\documentclass[preprint2]{emulateapj}
\let\svthefootnote\thefootnote

\usepackage[T1]{fontenc}
\usepackage{mathrsfs}

\def\Sec{\hbox{${}^{\prime\prime}$\llap{.}}}
\def\deg{\hbox{${}^\circ$}} \def\min{\hbox{${}^{\prime}$}}
\def\sec{\hbox{${}^{\prime\prime}$}}
\def\lae{\mathrel{<\kern-1.0em\lower0.9ex\hbox{$\sim$}}}
\def\gae{\mathrel{>\kern-1.0em\lower0.9ex\hbox{$\sim$}}}
\def\msun{${\cal M}_\odot$}

\def\nd{\nodata}

\def\sex{{\tt SExtractor}}

\DeclareMathAlphabet{\mathpzc}{OT1}{pzc}{m}{it}
\shorttitle{The NGVS Galaxy Luminosity Function }

\shortauthors{Ferrarese et al.}
\begin{document}

\title{The Next Generation Virgo Cluster Survey (NGVS). XIII. The Luminosity and Mass Function of Galaxies in the Core of the Virgo Cluster and the Contribution from Disrupted Satellites$^{\dag}$}
\let\thefootnote\relax\footnote{$^{\dag}$Based on observations obtained with MegaPrime/MegaCam, a joint project of the Canada France Hawaii Telescope (CFHT) and CEA/DAPNIA. CFHT  is operated by the National Research Council (NRC) of Canada, the Institut National des Science de l'Univers of the Centre National de la Recherche Scientifique (CNRS) of France and the University of Hawaii.\vskip.3in}
\addtocounter{footnote}{-2}\let\thefootnote\svthefootnote

\author{
Laura Ferrarese\altaffilmark{1}, 
Patrick C\^ot\'e\altaffilmark{1},
R\'uben S\'anchez-Janssen\altaffilmark{1},
Joel Roediger\altaffilmark{1},
Alan W. McConnachie\altaffilmark{1},
Patrick R. Durrell\altaffilmark{2},
Lauren A. MacArthur\altaffilmark{1,3},
John P. Blakeslee\altaffilmark{1},
Pierre-Alain Duc\altaffilmark{4},
S. Boissier\altaffilmark{5},
Alessandro Boselli\altaffilmark{5},
St\'ephane Courteau\altaffilmark{6},
Jean-Charles Cuillandre\altaffilmark{4},
Eric Emsellem\altaffilmark{7},
S.D.J. Gwyn\altaffilmark{1},
Puragra Guhathakurta\altaffilmark{8},
Andr\'es Jord\'an\altaffilmark{9}, 
Ariane Lan\c{c}on\altaffilmark{10},
Chengze Liu\altaffilmark{11},
Simona Mei\altaffilmark{12,13},
J. Christopher Mihos\altaffilmark{14},
Julio F. Navarro\altaffilmark{15}, 
Eric W. Peng\altaffilmark{16,17},
Thomas H. Puzia\altaffilmark{9}, 
James E. Taylor\altaffilmark{18},
Elisa Toloba\altaffilmark{8,19} \& 
Hongxin Zhang\altaffilmark{9,16}
}
\affil{$^1$National Research Council of Canada, 5071 West Saanich Road, Victoria, BC, V9E 2E7, Canada}
\affil{$^2$Department of Physics and Astronomy, Youngstown State University, Youngstown, OH, USA}
\affil{$^3$Department of Astrophysical Sciences, Princeton University, Princeton, NJ 08544, USA}
\affil{$^4$CEA/IRFU/SAP, Laboratoire AIM Paris-Saclay, CNRS/INSU, Universit\'e Paris Diderot, Observatoire de Paris, PSL Research University, F-91191 Gif-sur-Yvette Cedex, France}
\affil{$^5$Aix Marseille Universit\'e, CNRS, LAM (Laboratoire d'Astrophysique de Marseille) UMR 7326, 13388, Marseille, France}
\affil{$^6$Department of Physics, Engineering Physics and Astronomy, Queen's University, Kingston, ON, Canda}
\affil{$^7$European Southern Observatory, Karl-Schwarzchild-Str. 2, 85748, Garching, Germany}
\affil{$^8$UCO/Lick Observatory, University of California, Santa Cruz, 1156 High Street, Santa Cruz, CA 95064, USA}
\affil{$^9$Instituto de Astrof\'{\i}sica, Pontificia Universidad Cat\'olica de Chile, 7820436 Macul, Santiago, Chile} 
\affil{$^{10}$Observatoire Astronomique, Universit\'e de Strasbourg \& CNRS UMR 7550, 11 rue de l'Universit\'e, 67000 Strasbourg, France}
\affil{$^{11}$Center for Astronomy and Astrophysics, Department of Physics and Astronomy,  and Shanghai Key Lab for Particle Physics and Cosmology, Shanghai Jiao Tong University, Shanghai 200240, China}
\affil{$^{12}$GEPI, Observatoire de Paris, CNRS, 5 Place Jules Jannssen - 92195 Meudon, France}
\affil{$^{13}$Universit\'e Paris Denis Diderot, 75205, Paris Cedex 13, France}
\affil{$^{14}$Department of Astronomy, Case Western Reserve University, Cleveland, OH}
\affil{$^{15}$Department of Physics and Astronomy, University of Victoria, Victoria, BC, Canada}
\affil{$^{16}$Department of Astronomy, Peking University, Beijing 100871, China}
\affil{$^{17}$Kavli Institute for Astronomy and Astrophysics, Peking University, Beijing 100871, China}
\affil{$^{18}$Department of Physics and Astronomy, University of Waterloo, Waterloo, Ontario N2L 3G1, Canada}
\affil{$^{19}$Department of Physics, Texas Tech University, Box 41051, Lubbock, TX}

\begin{abstract}
We present measurements of the galaxy luminosity and stellar mass function in a 3.71 deg$^2$ (0.3 Mpc$^2$) area in the core of the Virgo cluster, based on $u^*griz$ data from the Next Generation Virgo Cluster Survey (NGVS). The galaxy sample --- which consists of 352 objects brighter than $M_g =-9.13$ mag, the 50\% completeness limit of the survey --- reaches 2.2 mag deeper than the widely used Virgo Cluster Catalog (VCC) and at least 1.2 mag deeper than any sample previously used to measure the luminosity function in Virgo. Using a Bayesian analysis, we find a best-fit faint end slope of $\alpha = -1.33 \pm 0.02$ for the $g-$band luminosity function; consistent results are found for the stellar mass function as well as the luminosity function in the other four NGVS bandpasses. We discuss the implications for the faint-end slope of adding 92 ultra compact dwarfs galaxies (UCDs) --- previously compiled by the NGVS in this region --- to the galaxy sample, assuming that UCDs are the stripped remnants of nucleated dwarf galaxies. Under this assumption, the slope of the luminosity function (down to the UCD faint magnitude limit, $M_g = -9.6$ mag) increases dramatically, up to $\alpha = -1.60 \pm 0.06$ when correcting for the expected number of disrupted non-nucleated galaxies. We also calculate the total number of UCDs and globular clusters that may have been deposited in the core of Virgo due to the disruption of satellites, both nucleated and non-nucleated. We estimate that $\sim 150$ objects with $M_g \lesssim -9.6$ mag and that are currently classified as globular clusters, might, in fact, be the nuclei of disrupted galaxies. We further estimate that as many as 40\% of the (mostly blue) globular clusters in the Virgo core might once have belonged to such satellites; these same disrupted satellites might have contributed $\sim$ 40\% of the total luminosity in galaxies observed in the core region today. Finally, we use an updated Local Group galaxy catalog to provide a new measurement of the luminosity function of Local Group satellites, $\alpha = -1.21 \pm 0.05$, which is only 1.7$\sigma$ shallower than measured in the core of the Virgo cluster.
\end{abstract}


\section{Introduction}
\label{sec:intro}

The processes by which gas is retained and cooled within dark matter halos are central to our understanding of how galaxies form and evolve. Starting with the work of Rees \& Ostriker (1977) and White \& Rees (1978), it has been recognized that the shape of the galaxy luminosity function, $\phi(L)$, is sensitive to such processes. At high luminosities, $\phi(L)$ is characterized by a sharp, exponential cutoff that is commonly interpreted as the signature of strongly suppressed cooling in massive halos, as can be achieved (albeit under rather extreme conditions) by thermal conduction, high energy expulsion of gas via superwinds, or feedback from Active Galactic Nuclei (see Benson et al. 2003a; Croton et al. 2006 and references therein). At low luminosities, $\phi(L)$ is usually parameterized as a power-law, although significant uncertainties remain as to the exact value of its slope, $\alpha$, and its possible dependence on environment and galaxy type. Even whether a single, luminosity independent, slope is an adequate description of the data is still subject to debate (see reference below and, for the mass function, Li \& White 2009; Baldry et al 2012). 

In the Local Group, where Milky Way satellites as faint as  $M_V \sim -2$ mag can be detected (e.g. McConnachie 2012; Laevens et al. 2015a, 2015b; Martin et al. 2015; Koposov et al. 2015; Kim \& Jerjen 2015; Kim et al. 2015a,b; Bechtol et al. 2015; Drlica- Wagner et al. 2015), $\alpha$ has generally been found to be of order $-1.0$ to $-1.1$ (Pritchet \& van den Bergh 1999; Trentham et al. 2005; Koposov et al. 2008; McConnachie et al. 2009, and references therein; Chiboucas et al. 2009), although difficulties in assessing completeness corrections --- which can be large in the case of the Milky Way satellites --- as well as small number statistics, are notable complications (e.g. Tollerud et al. 2008). These issues aside, there seems to be little doubt that the measured slope is far shallower than that predicted by $\Lambda$CDM cosmological simulations for the mass function of dark matter halos ($\alpha \sim -1.9$, e.g. Springel et al. 2008). This discrepancy has come to be known as the ``missing satellites'' problem (Klypin et al. 1999; Moore et al. 1999). Reconciling observations and simulations has spurred significant activity on the theoretical front,  including modifying the power spectrum on small scales or changing the properties of the dark matter particles themselves. Within the $\Lambda$CDM framework, a favored scenario is the suppression of star formation in low-mass halos due to reheating of the intergalactic medium by re-ionization (e.g. Bullock et al. 2000; Benson et al. 2002; Somerville 2002; Governato et al. 2007; Okamoto et al. 2008), although additional reheating of cold disk gas to the halo temperature caused by supernovae and winds in star-forming galaxies also appears necessary (Benson et al. 2003a). 

It is important to recognize that the Local Group is only one of many possible environments, and that the universality of the galaxy luminosity function remains very much open to debate. In the dynamically unevolved M81 group, deep observations (reaching $M_R \sim -10$ , or $M_V \sim -9.5$ mag; Chiboucas et al. 2009; Chiboucas et al. 2013) point to a faint end slope of $-1.28 \pm 0.06$. This is consistent with what found for  the more dynamically evolved Cen A Group ($-1.23^{+0.04}_{-0.10}$,  based on data reaching $M_B \sim -10$, or $M_V \sim -10.8$ mag; Karachentsev et al. 2002; Chiboucas et al. 2009) and for the NGC 5846 and NGC 1407 groups ($\alpha \sim -1.35$ to $-1.3$ but at somewhat brighter limiting magnitudes, $M_R \sim -12$ mag or $M_V \sim -11.5$ mag; Mahdavi et al. 2005; Trentham et al. 2006). At the extreme end of the evolutionary scale, deep observations (reaching $M_B \sim -9.2$ mag) of the fossil group NGC 6482 also point to a slope $\alpha = -1.32 \pm 0.05$ (Lieder et al. 2013). While steeper than measured in the Local Group (but still significantly shallower than predicted by $\Lambda$CDM models), the slopes listed above suggest an apparent lack of environmental dependence in the luminosity function of galaxy groups spanning a significant range of evolutionary states (Chiboucas et al. 2009).

However, not all studies agree with the findings above. Steeper slopes --- consistent, in fact, with $\Lambda$CDM predictions --- have also been reported. Krusch et al. (2006) targeted five nearby Hickson compact groups, to a depth of $M_B \sim -13$ mag ($M_V \sim -13.8$ mag), finding a steep upturn in the luminosity function starting at $M_B \sim -15$ mag, with a faint-end slope of $\alpha \sim -1.7$. A similar upturn, and a slope $-1.9 \lesssim \alpha \lesssim -1.6$, was measured by Gonzalez et al. (2006) for the composite luminosity function of nearby groups and poor clusters identified in SDSS data and for which membership was assessed based on colors and morphology. The reported upturn did not appear to depend on group mass, richness, or environment.

The picture becomes increasingly more complicated as one moves up the mass scale, from groups to clusters. Although there is general agreement that the luminosity function in galaxy clusters shows a broad trough around $M_B \sim -17$ mag, followed by a steepening at fainter magnitudes (e.g., Virgo: Sandage et al. 1985; Tully et al. 2002; Trentham \& Tully 2002; Trentham \& Hodgkin 2002; Rines \& Geller 2008; McDonald et al. 2009; Coma: Tully et al. 2002; Yamanoi et al. 2012 and references therein), there is no consensus on the precise value of the faint-end slope. This is true not only when comparing different clusters, but also when comparing slopes measured for the {\it same} cluster by different teams (see, e.g. Table 4 of Yamanoi et al. 2012 or Table 1 of De Propris et al. 2003). De Propris et al. (2003) measured a faint-end slope of $\alpha = -1.28 \pm 0.03$ for the composite luminosity function of 60 low-redshift clusters ($z < 0.11$) observed as part of the 2dF Galaxy Redshift Survey: a slope consistent with that measured by Chiboucas et al. (2009, 2013) for nearby galaxy groups.  However, Popesso et al. (2005) measured a faint-end slope of $-2.1 \lesssim \alpha \lesssim -1.6$ in a sample of X-ray selected galaxy clusters noting that, in all SDSS bands, the luminosity function shows a steep upturn at $M_g \sim -16$ mag. In Coma, Adami et al. (2007), reaching a magnitude of $M_R = -9.3$, reported a faint-end slope of $\alpha \sim -2$, while Yamanoi et al. (2012) measured an even more extreme value, as steep as $\alpha \sim -3$ (but see their Table 4 for different results). 

Given this situation, it is difficult to characterize any dependence of the faint-end slope on environment. For instance, based on a comparison of results for several environments, from the Local Group to the Coma cluster, Tully et al. (2002) argued that the steepness of the faint-end slope correlates with the cluster richness --- a claim supported by Trentham et al. (2005) based on a composite luminosity function for both groups and clusters. At the same time, De Propris et al. (2003) found no evidence that $\phi(L)$ depends on either cluster richness, velocity dispersion, or amount of substructure for their cluster samples. The situation with the {\it field} luminosity function is equally problematic. De Propris et al. (2003) found their cluster luminosity function to be slightly steeper than measured for the field using 2dF data ($\alpha \sim -1.2$), whereas Blanton et al. (2005) found that the luminosity function of field galaxies (down to $M_r \sim -13$ and for galaxies within 150 Mpc) shows an upturn at $M_r \sim -18$ mag, with a faint-end slope as steep as $\alpha \sim -1.5$ after accounting for incompleteness (but see Blanton et al. 2001, 2003). Most recently, Klypin et al. (2015) measured a slope of $\alpha \sim -1.3$ for field galaxies within 10 Mpc of the Milky Way.

This paper targets what is arguably the most extreme and unique environment in the local universe --- the core of the Virgo cluster. Our analysis is based on high-quality imaging from the Next Generation Virgo Cluster Survey (NGVS, Ferrarese et al. 2012). At a distance of 16.5 Mpc (Mei et al. 2007; Blakeslee et al. 2009)  and with a gravitating mass of ${\cal M}_{200}= 5.5\times10^{14}$ \msun~(Durrell et al. 2014; McLaughlin~1999), the Virgo cluster is the dominant mass concentration in the local universe, the center of the Local Supercluster, and the largest concentration of galaxies within $\sim 35$ Mpc. Not surprisingly, it has long been a prime target for studies of the luminosity function in cluster environments. The landmark study of Sandage, Binggeli \& Tammann (1985), which was based on the Virgo Cluster Catalog (VCC, Binggeli, Sandage \& Tammann 1985), reached a high level of completeness for galaxies as faint as $M_B \sim -13$ and mean surface brightness within an effective radius of $\langle\mu(B)\rangle_e \sim 25.3$ mag arcsec$^{-2}$. Surveying an area of 140 deg$^2$, these authors found a faint-end slope of $\alpha = -1.3$ when including galaxies of all Hubble types, and as steep as $\alpha \sim -1.45$ when the sample was restricted to just early types (regular and dwarf ellipticals).  However, Virgo has not been immune from the controversies noted above, and subsequent investigations, some of them reaching fainter magnitudes but at the expense of a reduced spatial coverage, found quite discordant slopes, ranging widely from $\alpha \sim -1.0$ to $-2.2$  (Impey et al. 1988; Phillipps et al. 1998; Trentham \& Hodgkin 2002; Trentham \& Tully 2002; Sabatini et al. 2003; Rines \& Geller 2008). 

Such wide range in the measured slopes can arise, at least in part, from technical issues, such as the detection techniques, the criteria adopted in the determination of cluster membership, the completeness corrections, or even the methodology used and the assumptions made in fitting the faint end slope -- all of which differ amongst the various studies. However, slope differences could also have a physical origin, for instance they could reflect true variations of the luminosity function as a function of stellar population (for surveys conducted in different bands), environment (reflected in the field size/location), and/or galaxy mass and morphological mix (which can be sampled differently according to the survey's depth, as well as field placement).  The NGVS offers a unique opportunity to explore the Virgo luminosity function using a homogeneous, deep (reaching $M_g = -9.13$ mag at 50\% completeness), high resolution (0\Sec55 median seeing full-width-half-maximum in the $i-$band), panchromatic ($u-$ to $z-$band) dataset covering the entire cluster (104 deg$^{2}$), with the added benefit of a well understood completeness function. The luminosity function of the galaxy population across the entire NGVS area, including a discussion of environmental dependencies from the core to the outskirt of the cluster, will be presented in an upcoming contribution (Ferrarese et al., in preparation). In this paper, we focus on the densest region at the dynamical center of Virgo: a $3.71$ deg$^{2}$ (0.3 Mpc$^2$) area surrounding M87. This Virgo ``core region'', described in Ferrarese et al. (2016), has special significance as it is located at the bottom of the cluster's potential well, and is by far the region with the highest density of baryonic substructures  -- galaxies, globular clusters, and Ultra Compact Dwarfs (UCDs). Observationally, it was the first region targeted as part of the NGVS and, being the region most extensively studied in the literature, it provided a testbed for designing and optimizing the algorithms used for the detection and characterization of low mass galaxies throughout the cluster (Ferrarese et al. 2016).

The properties of the galaxies in the Virgo core --- including their structural properties, color-magnitude relations, shapes and nucleation properties --- are examined in a series of companion papers (e.g., Grossauer et al. 2015; S\'anchez-Janssen et al. 2016a,b; Roediger et al. 2016; C\^ot\'e et al. 2016). Here we discuss the luminosity and mass function of galaxies residing in the core of the cluster. We start our analysis by considering all baryonic structures whose galactic nature is undisputed, including `extreme' populations such as compact ellipticals and `Ultra Diffuse Galaxies' (UDGs, van Dokkum et al. 2015; Koda et al. 2015; Mihos et al. 2015). Capitalizing on the sample of $\sim$ one hundred UCDs assembled using NGVS data in the Virgo core region (Zhang et~al. 2015;  Liu et al. 2015a), we further extend our study and incorporate UCDs in our galaxy sample, under the (not unanimously accepted) assumption that UCDs represent the nuclei of low-mass galaxies stripped within the cluster environment. Finally, we make use of a detailed investigation of the nucleation fraction of Virgo galaxies to estimate the total number of satellites expected to have been disrupted, and their contribution to the amount of diffuse light, and the present-day UCD and globular cluster population in the core of Virgo.

This paper is organized as follows: \S\ref{sec:data} describes the data and the Markov Chain Monte Carlo Bayesian analysis method used to fit the luminosity and stellar mass functions; \S\ref{sec:lfvirgo} discusses the galaxy luminosity function in the Virgo core based on NGVS data, and compares it with previous measurements (\S\ref{sec:lfvirgo2}). The possible contribution of UCDs to the luminosity function is considered in \S\ref{sec:lfucds}, while \S\ref{sec:lflg} compares the Virgo luminosity function to that measured for the Local Group. The Virgo core stellar mass function is presented in \S\ref{sec:mfvirgo}, and a summary is given in \S\ref{sec:summ}.
Throughout this paper, we assume a distance modulus to the Virgo cluster of $(m-M)_0$ = 31.09 mag, corresponding to a distance of 16.5 Mpc (Mei et al. 2007; Blakeslee et al. 2009).

\newpage

\section{Data and Methodology}
\label{sec:data}

Using the 1 deg$^2$  MegaCam instrument at the Canada-France-Hawaii Telescope (CFHT), the NGVS surveyed the Virgo cluster out to the virial radius of its two main substructures -- for a total areal coverage of 104 deg$^{2}$ -- in the $u^*,g,i,z$ passbands. A subset of the cluster, including the core region discussed in this paper, was also imaged in the $r-$band. Thanks to a dedicated data acquisition strategy and processing pipeline, the NGVS reached a 10$\sigma$  point-source depth of $g \sim 25.9$ mag and a surface brightness limit of $\mu_g \sim 29$ mag arcsec$^{-2}$ (2$\sigma$ above the mean sky level), at sub-arcsecond spatial resolutions.  Details about the survey design, motivation, and specifications can be found in Ferrarese et al. (2012).

The core region discussed in this paper was the first to be imaged as part of the survey. It comprises four MegaCam pointings and its boundaries are defined as:

\begin{center}
$
\begin{array}{rcrcl}
12^{h}26^{m}20^{s} & \leq & \alpha({\rm J2000}) & \leq & 12^{h}34^{m}10^{s} \nonumber\\
11^{\deg}30^{\min}22^{\sec} & \leq & \delta({\rm J2000}) & \leq & 13^{\deg}26^{\min}45^{\sec} \nonumber\\
\end{array}
$
\end{center}

\noindent giving a total areal coverage of $\Omega = 3.71$ deg$^{2}$ (0.3 Mpc$^2$)\footnote{This is slightly smaller than the nominal 4 deg$^2$ covered by four MegaCam pointings, both because the pointings overlap slightly (by design) and because the outer edges of the mosaic -- which are not as well sampled by the adopted dithering pattern and have lower signal-to-noise ratio -- have  been excluded from the analysis.}. In terms of clustercentric radius, this area samples Virgo's A subcluster out to  $\sim 20\%$ of its virial radius (McLaughlin~1999). 

The most crucial (and most difficult) step for a reliable estimation of the luminosity function is membership assignment: indeed the inclusion of interlopers is likely responsible for some extremely steep faint-end slopes reported in the literature (see \S~\ref{sec:lfvirgo2}). Our analysis takes advantage of the robust membership assignments made possible by the panchromatic nature, surface brightness sensitivity and exceptional spatial resolution of the NGVS data. A catalog of 404 galaxies in the core region --- all {\it bona-fide} Virgo members ---  is presented in Ferrarese et al. (2016), to which we refer the reader for a detailed description of the galaxy detection procedures and membership criteria. Briefly, the detection algorithm \sex~(Bertin et al. 2002) is run, using optimally designed masks, on images that are pre-processed to isolate faint, low surface objects (such as low-mass galaxies belonging to Virgo) by minimizing contamination from compact, high surface brightness interlopers (such as foreground stars, Virgo globular clusters, and high redshift galaxies). Photometric redshifts and structural parameters (based on two-dimensional modelling of the light distribution) are measured for each of the $\sim 70,000$ objects detected, and are used to define a multi-parameter space that provides maximal separation between known (including but not limited to spectroscopically confirmed) Virgo members and background galaxies (spectroscopically confirmed as such and/or detected in control fields located at three virial radii, or $\approx 16$\deg, from the core of Virgo). Each of the $\sim 1500$ galaxies deemed to be possible members of the cluster based on their location within this multi-dimensional space is then visually inspected, producing the final catalogue of 404 objects described in Ferrarese et al. (2016), where a number of validation tests (including cross-correlations with catalogues of known Virgo members and spectroscopic confirmation of some newly identified members) are also discussed.

The 404 galaxies in our sample span the magnitude range  $-22.2 < M_g < -7.4$ mag. Of these galaxies, 250 had been detected in previous surveys and reconfirmed in our analysis, while 154 are new NGVS detections\footnote{A number of galaxies reported in the existing literature as Virgo member were deemed to be either spurious detections or background objects based on NGVS data, see Appendix A of Ferrarese et~al. (2016).}. Ferrarese et al. (2016) also discusses the membership `probability' (a term that we apply very loosely): of the 404 galaxies, 173 are classified as {\tt certain} members of the cluster, 103 as {\tt likely}, and 128 as {\tt possible}. Figure~1 plots galaxy photometric errors against their integrated magnitudes, measured in the $g-$band from a curve of growth analysis and corrected for Galactic extinction (following Schlegel et al. 1988; see Tables 4 and 5 of Ferrarese et~al. 2016)\footnote{The extinction coefficients adopted are $A(g)/E(B-V) = 3.560$,  $A(u)/E(B-V) = 4.594$, $A(r)/E(B-V) =  2.464$, $A(i)/E(B-V) = 1.813$ and $A(z)/E(B-V) =  1.221$}.

\begin{figure}
\epsscale{1.00}
\plotone{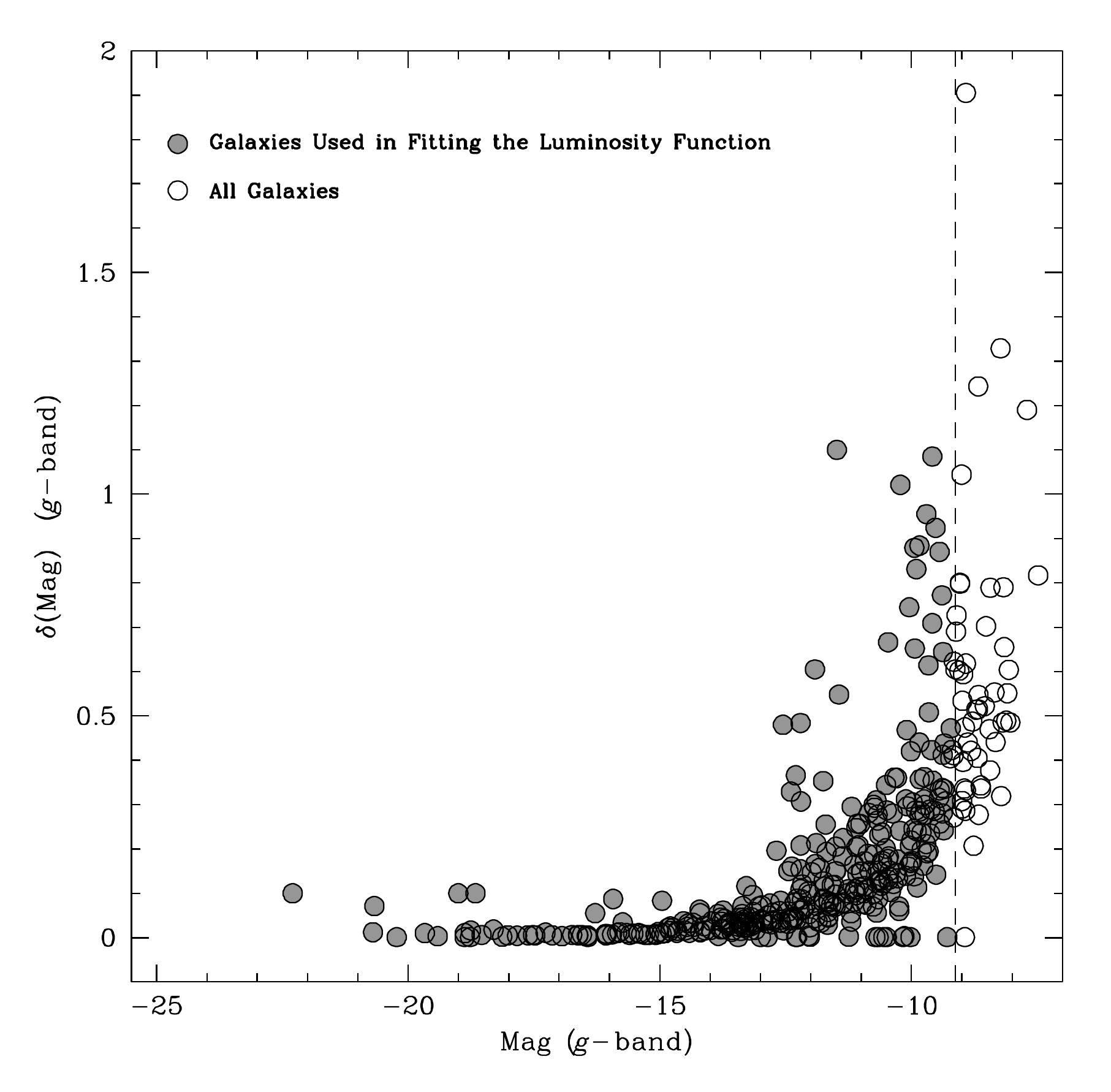}
\caption{Dependence of $g-$band errors on absolute magnitude $M_g$, based on a curve of growth analysis and corrected for Galactic extinction,  for the full sample of 404 galaxies in the core of the Virgo cluster (Ferrarese et~al. 2016). Galaxies plotted as gray filled circles show the sample of 352 galaxies brighter than $M_g = -9.13$~mag ($g \sim 22$, before correcting for Galactic extinction) used to measure the core luminosity and stellar mass functions.}
\label{fig1}
\end{figure}

Ferrarese et~al. (2016) presents an extensive set of simulations that can be used to estimate the detection efficiency as a function of the galaxy photometric and structural parameters: i.e., total magnitude, S\'ersic index, effective radius, axial ratio, and surface brightness at, and averaged within, one effective radius. In the simulations, galaxies were randomly generated to loosely populate the scaling relations expected for Virgo members, and assumed to have surface brightness profiles described by S\'ersic models. The detection efficiency does not differ appreciably between the four core region fields -- not surprising since the frames are very similar in terms of object density, depth and image quality. Marginalizing over all effective radii and effective surface brightnesses (or, equivalently, central or average surface brightness) covered in the simulations, the $g-$band magnitude at which completeness falls below 50\% is found to be $M_g = -9.13$. To avoid large incompleteness corrections, we have excluded all galaxies fainter than this limit (correction for Galactic extinction is applied after this magnitude cut) when computing the $g-$band luminosity function. This reduces the sample from our original 404 galaxies to 352, of which 106 are new detections. All but five of the galaxies excluded in this way were classified as {\tt possible} cluster members; the remaining five are {\tt likely} members. In Figure~1, the 352 galaxies used to calculate the core luminosity function are shown as filled gray circles. In the other four NGVS bands ($u^*,r,i$ and $z$), where the survey reaches comparable depth as in the $g-$band, we have computed the luminosity function in each band using the 50\% limit appropriate to that band. The number of galaxies in this case varies from 319 in the (shallower) $z-$band, to 354 in the (deeper) $r-$band (see Table~1). All but a few of the galaxies are early-type, both in morphology and in colour: for instance, fitting the $g~vs~u^*-g$ color-magnitude relation with a low order polynomial reveals $\sim 17$ objects bluer than the ridge line by more than one standard deviation. For this reason, we will not discuss the luminosity function for galaxies divided by morphology or colour, noting only that excluding the blue outliers in the color-magnitude relation has no effects on our results.

\begin{deluxetable*}{llc|ll|cl|l}
\tabletypesize{\footnotesize}
\tablewidth{0pc}                        
\tablenum{1}
\tablecaption{Markov Chain Monte Carlo (MCMC) Fitting Results for the Virgo Cluster Core\label{tab1}}
\tablehead{
\multicolumn{1}{c}{Filter} &
\multicolumn{1}{c}{$No.$} &
\multicolumn{1}{c}{$M_{\rm lim}$ (mag)} &
\multicolumn{1}{|c}{$M^*$ (mag)} &
\multicolumn{1}{c}{$\alpha$} &
\multicolumn{1}{| c}{$M^*$ (fixed) (mag)} &
\multicolumn{1}{c}{$\alpha$} &
\multicolumn{1}{|c}{Notes} \\
\multicolumn{1}{c}{(1)} &
\multicolumn{1}{c}{(2)} &
\multicolumn{1}{c}{(3)} &
\multicolumn{1}{|c}{(4)} &
\multicolumn{1}{c|}{(5)} &
\multicolumn{1}{c}{(6)} &
\multicolumn{1}{c}{(7)} &
\multicolumn{1}{|c}{(8)} 
}
\startdata
$u^*$ & 339 & $-$8.36 & $-21.5^{+1.5}_{-1.4}$ & $-1.338^{+0.019}_{-0.017}$ & $-$19.17  & $-1.332^{+0.018}_{-0.019}$& Galaxies only\\
$g$ & 352 & $-$9.13 & $-23.3^{+1.1}_{-1.1}$ & $-1.346^{+0.017}_{-0.016}$ & $-$20.48  & $-1.330^{+0.018}_{-0.016}$& Galaxies only\\
$r$ & 354 & $-$9.54 & $-23.5^{+1.1}_{-1.2}$ & $-1.337^{+0.016}_{-0.015}$  & $-$21.12  & $-1.323^{+0.018}_{-0.017}$& Galaxies only\\
$i$ & 336 & $-$9.98 & $-24.0^{+0.1}_{-1.2}$ & $-1.350^{+0.018}_{-0.017}$  & $-$21.46  & $-1.330^{+0.019}_{-0.018}$& Galaxies only\\
$z$ & 319 & $-$10.34 & $-24.1^{+1.0}_{-1.1}$ & $-1.357^{+0.017}_{-0.017}$ & $-$21.73  & $-1.348^{+0.019}_{-0.018}$& Galaxies only\\
&&&&&&&\\
\hline
&&&&&&&\\
$g$ & 418 & $-$9.60  & $-23.4^{+1.1}_{-1.2}$ & $-1.446^{+0.018}_{-0.017}$ & $-$20.48  & $-1.425^{+0.017}_{-0.020}$ &Galaxies + UCDs\\
$g$ & 110 & $-$15.6 & $-22.5^{+1.2}_{-1.4}$ & $-1.562^{+0.082}_{-0.074}$ & $-$20.48  & $-1.528^{+0.076}_{-0.074}$ &Galaxies + UCDs Progenitors
\enddata
\tablecomments{The columns list the filter in which the fit is performed (column 1), the number of galaxies (column 2) and the 50\% completeness faint magnitude limit of the sample (column 3). The best-fit $M^*$ and $\alpha$ (in a Schechter formalism) are given in cols. 4 and 5 for the case in which $M^*$ is allowed to vary, and cols. 6 and 7 for the case where $M^*$ is fixed to the value measured in the field by Smith et al. (2009).  The last two rows show the results of the fits obtained when the UCD sample from Liu et al. (2015a) and their progenitor galaxies are added to the sample (see \S\ref{sec:lfucds}).}                        
\end{deluxetable*}

In the next section, the simulations presented in Ferrarese et~al. (2016) are also used to correct the luminosity function for completeness. The completeness correction we use is only a function of total magnitude (i.e. it is marginalized with respect to all other parameters). However, surface brightness is also a determining factor when detecting galaxies, and the question arises as to whether at any given magnitude, there might be populations of low surface brightness galaxies that could elude detection altogether. While we cannot speculate on, nor correct for galaxies not known to exist, a population of `Ultra Diffuse Galaxies' (UDGs) has recently been detected in Coma and Virgo (van Dokkum et al. 2015; Koda et al. 2015; Mihos et al. 2015). In Coma, these UDGs have average surface brightness within an effective radius $23.5 < \langle\mu(g)\rangle_e < 26.0$ mag arcsec$^{-2}$ and sizes $1.5 < r_e < 4.6$ kpc, while the three Virgo objects discovered by Mihos et al. (2015) are more extreme: $26.85 < \langle\mu(g)\rangle_e < 27.15$ mag arcsec$^{-2}$ and $2.8 < r_e < 9.5$ kpc. As discussed in Ferrarese et al. (2016), the NGVS detection rate of galaxies drops abruptly as a function of surface brightness: at $\langle\mu(g)\rangle_e \sim 27$ mag arcsec$^{-2}$ the detection rate is 50\%, and is essentially zero at $\langle\mu(g)\rangle_e \sim 29$ mag arcsec$^{-2}$. The NGVS is over 80\% sensitive to the analogues of the UDGs detected in Coma by van Dokkum et al. (2015) -- indeed several examples are included in the Virgo core sample as well as in the full Virgo catalogue to be discussed in an upcoming contribution. Only one of the Mihos et al. 2015 galaxies was detected by our automated algorithm (consistent with the expected 50\% detection rate at the surface brightnesses of those galaxies) while the other two were detected in the following by-eye inspection. We conclude that all known galaxy types are represented in our luminosity function, and no obvious selection biases are present that depend on the galaxy surface brightness.

We use a Bayesian approach in fitting the luminosity function. This approach works directly with discrete data: i.e., the magnitudes and their associated errors for the galaxies that make up our sample (Figure 1). By not binning, we circumvent potential complications associated with the choice of the bin size, avoid the loss of information resulting from collapsing discrete data within a single bin, and retain information on the error distribution for each individual galaxy. Although these advantages are negligible for very large samples (i.e., in the limit of very small bin sizes), a Bayesian approach is preferred when dealing with smaller datasets. In our case, the optimal bin size for a histogram representation of the $g-$band luminosity function would be $2(IQR)N^{-1/3} = 0.97$ mag, where $IQR$ is the interquartile range of the magnitude distribution and $N$ is the total number of objects (Izenman 1991). This would result in our 352 galaxies being divided into 14 bins, containing between 1 and 63 objects each --- clearly a rather coarse representation of the data. 

Our Bayesian approach is based on a numerical maximization of the likelihood function, $\mathscr{L}$, which we define as 

\begin{equation}
\begin{array}{lcrcl}
\log_{10}(\mathscr{L}[M(d_1...d_n)|\Theta]) = \sum\limits_{i=1}^n \log_{10} \int\Phi(M, \Theta)\\
 \times \mathscr{C}(M) \times \mathscr{G}[M(d_i),\sigma_{M[d_i]}] \rm{d} M
\end{array}
\end{equation}

\noindent where $M$ is the absolute magnitude; $\Phi(M, \Theta)$ is the adopted functional form for the luminosity function, expressed as a function of $M$ and having free parameter(s) $\Theta$; $\mathscr{C}(M)$ is the completeness function, i.e., the fraction of galaxies recovered as a function of $M$; and $\mathscr{G}[M(d_i),\sigma_{M[d_i]}]$ is a Gaussian with center and dispersion equal to the magnitude $M$ and associated error $\sigma$ for the $i^{th}$ of the $n$ datapoints, $d_i$. In our particular application, $\Phi(M, \Theta)$ is described as a Schechter (1976) function,

\begin{equation}
\begin{array}{lcrcl}
\Phi(M,\Theta){\rm d}M = \\
\Phi^* \times [10^{0.4(M^* - M)}]^{(\alpha + 1)}{\rm e}^{-10^{0.4(M^* - M)}}{\rm d}M, \\
\end{array}
\end{equation}

\noindent with free parameters $\Theta = \{M^*,\alpha\}$. In this formalism, $\Phi^*$ is not treated as a free parameter, but is constrained by the data as the ratio of the total number of objects to the integral of  $\Phi(M, \Theta) \times \mathscr{C}(M)$ estimated over the magnitude range spanned by the data. The completeness function $\mathscr{C}(M)$ is discussed in Ferrarese et~al. (2016)  and represented for our purposes as a fourth order polynomial.

The uncertainties in $M^*$ and $\alpha$ are defined as the standard deviation in the posterior probability function:

\begin{equation}
\begin{array}{rcrcl}
\mathscr{P}[\Theta | M(d_i)] \propto \mathscr{L}[M(d_i)|\Theta] \times p(\Theta) \\
\end{array}
\end{equation}

\noindent where $p(\Theta)$ is the ``prior'' function: i.e., a function that encapsulates any prior knowledge about the parameters. We adopt flat priors for both $M^*$ and $\alpha$ over the range $-2.2 \leq \alpha \leq -0.8$ and $-26.5 \leq M^* \leq -14.5$ mag. The posterior distribution of the parameters $\Theta = \{M^*,\alpha\}$ is sampled using the affine-invariant Markov Chain Monte Carlo (MCMC) method implemented in the Python {\tt EMCEE} package (Foreman-Mackey et al. 2013). We use a set of 200 ``walkers'' initialized in a small Gaussian space centered on a previously computed maximum likelihood solution, and run {\tt EMCEE} for 500 steps. A value of each parameter is computed for each walker at each of the 500 steps. Typically this MCMC ``chain'' converges to a stable solution after $50-100$ steps, which we discard when exploring the posterior distributions.  

\section{The Luminosity Function in the Core of the Virgo Cluster}
\label{sec:lfvirgo}

All galaxies in the core region have been observed and measured in five MegaCam bandpasses: $u^*,g,r,i$ and $z$. We begin by discussing the MCMC results obtained using the $g-$band photometry, shown in Figures 2~and~3 and tabulated in Table~1. For visualization purposes, the MCMC best fits are plotted against the differential luminosity function shown in histogram form although, as noted above, it is the {\it individual} datapoints (shown by the vertical red lines), not the histogram, that have been fitted. Also for visualization purposes, we plot the luminosity function against the completeness-corrected data (shown by the filled gray histogram) down to $M_g = -9.13$ mag, the magnitude beyond which the completeness falls below 50\%, while MCMC compares the raw data to the luminosity function corrected for completeness (see \S\ref{sec:data}). The green line in Figure~2 shows the best MCMC fit (with 1$\sigma$ confidence limits on $\alpha$) when both $M^*$ and $\alpha$ are treated as free parameters. Meanwhile, the posterior probability density functions are shown in the left panel of Figure 3. It is immediately apparent that the data do not constrain $M^*$, for the obvious reason that the cluster core contains very few bright galaxies. 

\begin{figure}
\epsscale{1.0}
\plotone{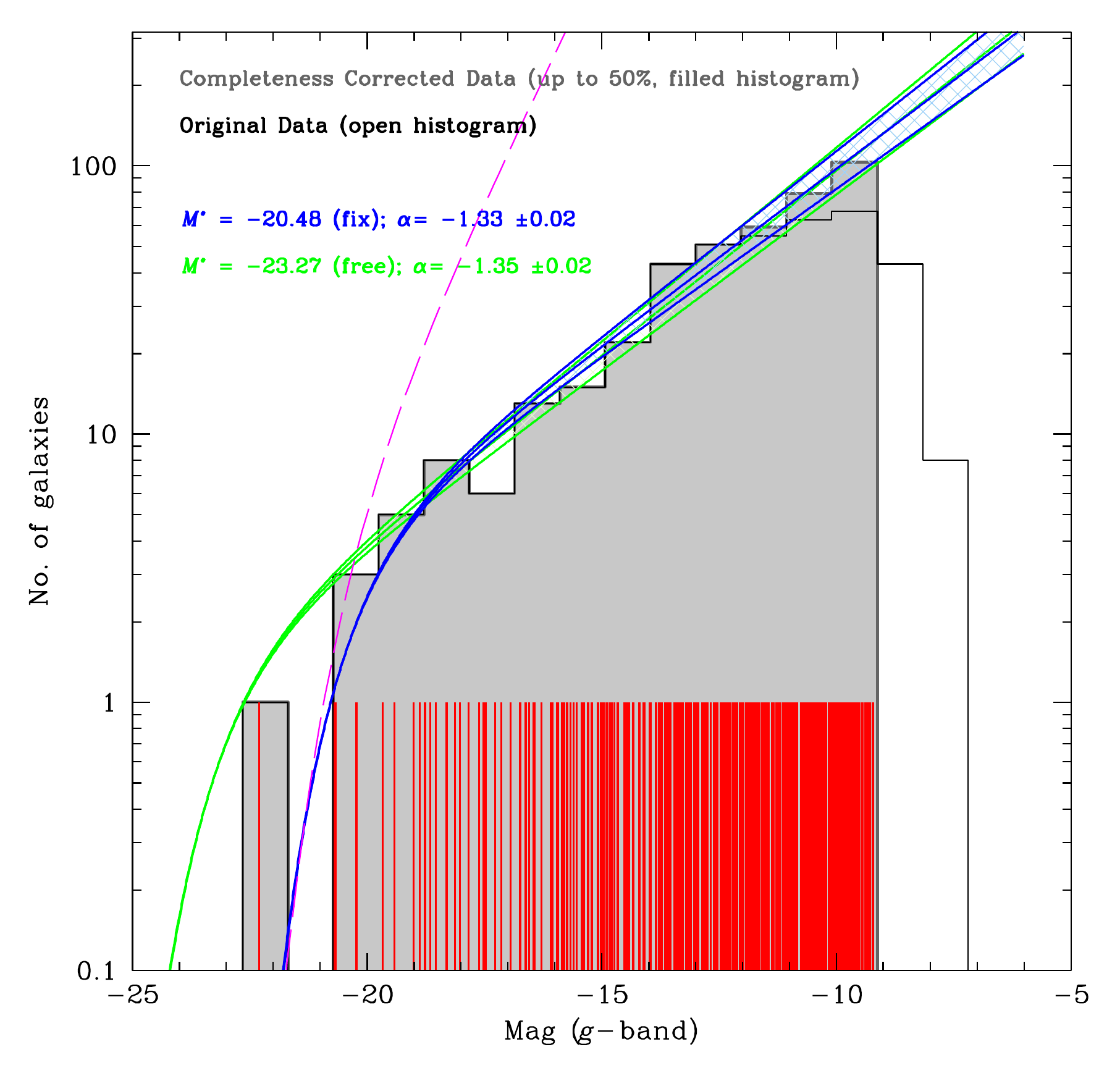}
\caption{Schechter fits to the luminosity function of galaxies in the core of the Virgo cluster, based on $g-$band photometry. The gray histogram shows the differential luminosity function found using galaxies brighter than $M_g = -9.13$ mag (corresponding to the 50\% completeness limit of the data before correcting for Galactic extinction) and corrected for incompleteness. The open black histogram shows the differential luminosity function constructed from the original data. The vertical red lines show the individual magnitudes for the 352 galaxies used in the analysis. The green line, along with 1$\sigma$ confidence limits on $\alpha$, shows the best MCMC fit to the individual data when both $M_g^*$ and $\alpha$ are allowed to vary. The blue line (also with 1$\sigma$ confidence limits) assumes $M_g^* = -20.48$ mag, the value measured for low-redshift field galaxies by Smith et al. (2009). Finally, the magenta dashed line represents a Schechter function with $\alpha = -1.9$, the slope of the $\Lambda$CDM halo mass function. All magnitudes are corrected for Galactic extinction.}
\label{fig2}
\end{figure}

\begin{figure*}
\epsscale{1.0}
\plottwo{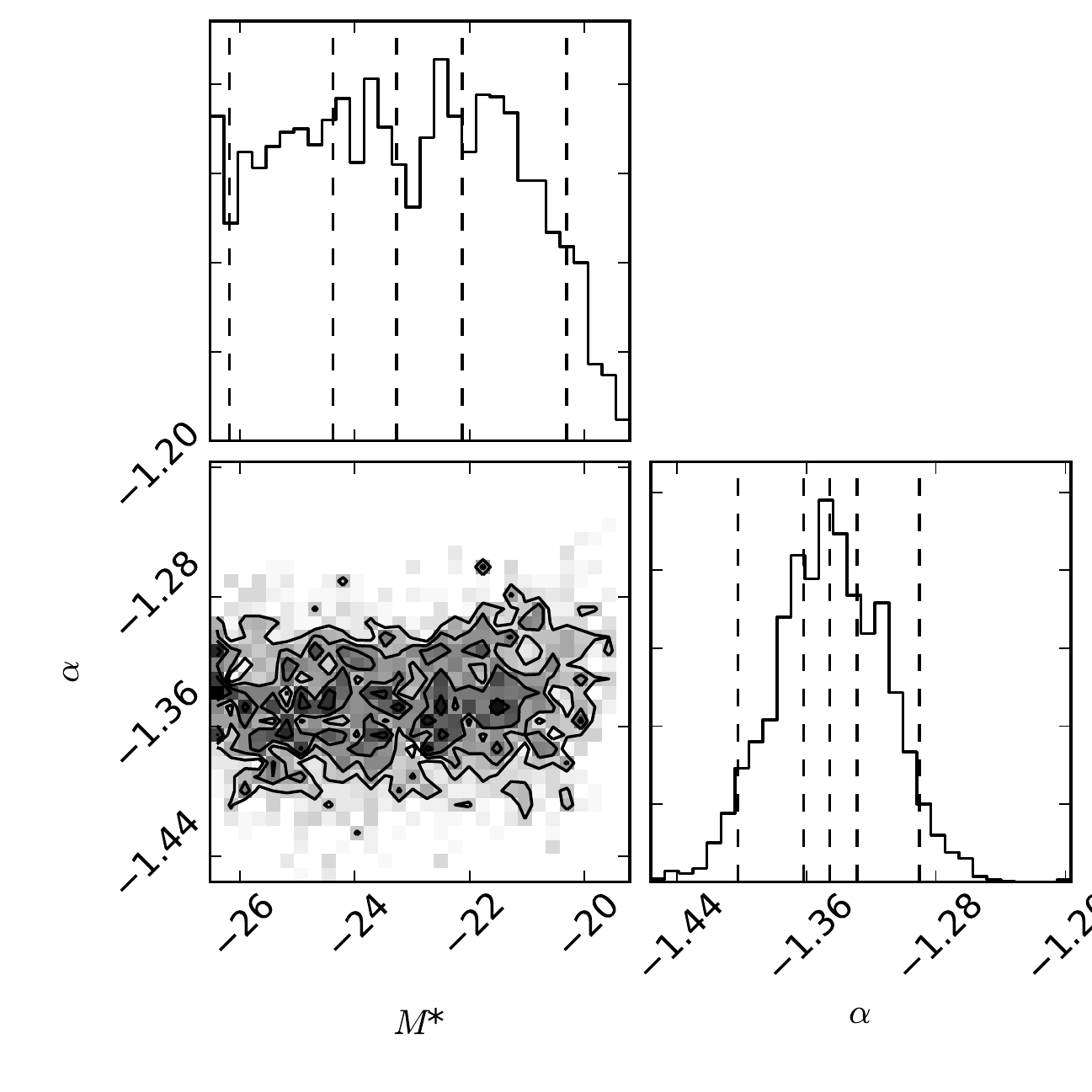}{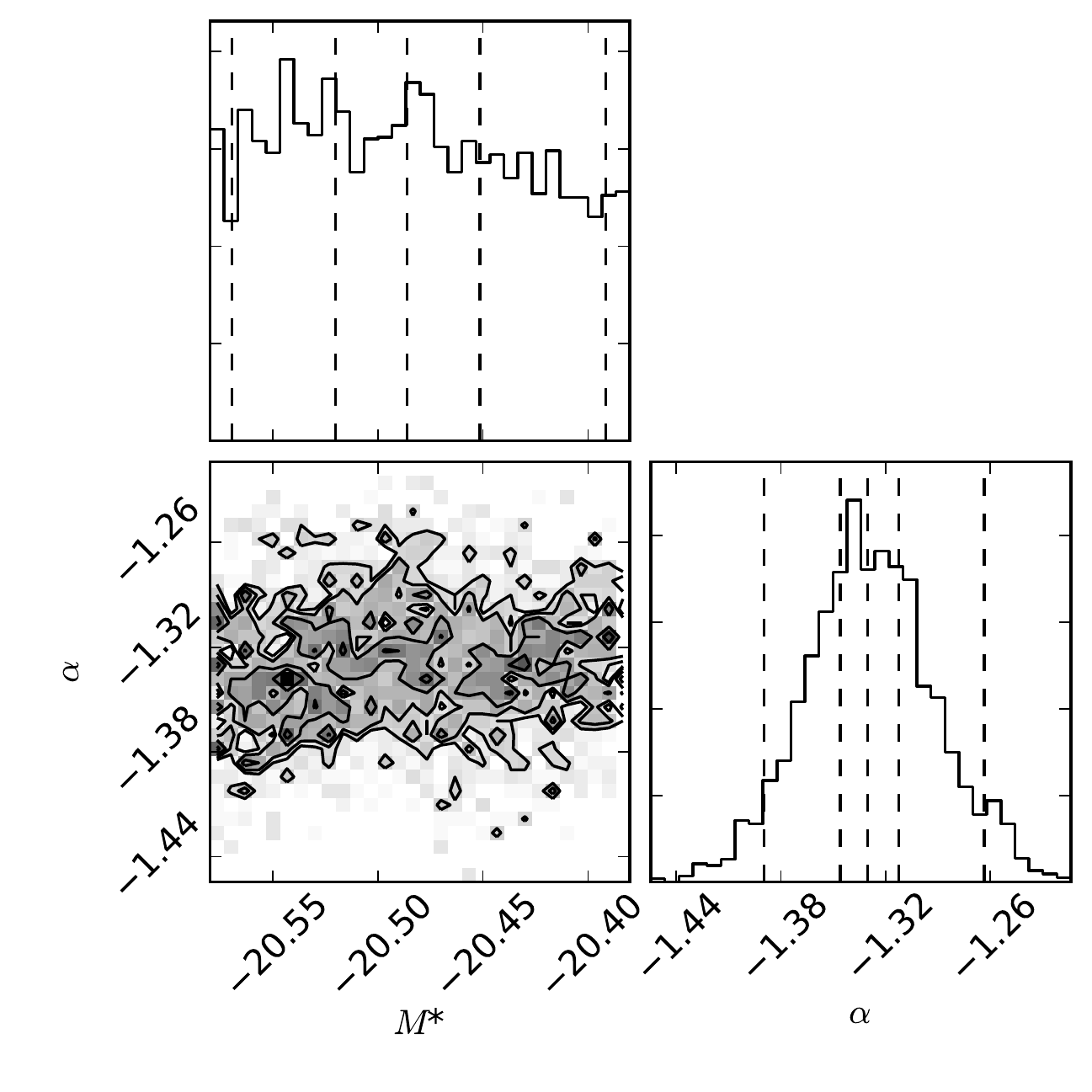}
\caption{Posterior probability density functions for $M_g^*$ and $\alpha$ for the restricted sample of 352 galaxies in the Virgo core (see Figure 2). Results are shown for two cases: when $M_g^*$ is allowed to vary ({\it left panel}) and when constrained to a very narrow range around $M^* = -20.48$ mag ({\it right panel}). From the left panel, it is clear that $M_g^*$ is essentially unconstrained by the data, owing to the small number of bright galaxies in the core. At the same time, $\alpha$ is tightly constrained and largely independent of $M_g^*$ due to the wide range in magnitude spanned by the sample ($\approx 13$~mag).}
\label{fig3}
\end{figure*}

In spite of this, $\alpha$ is well constrained by the data thanks to the wide magnitude range spanned by our sample: e.g., we measure a best-fit slope of $\alpha = -1.35 \pm 0.02$ in the $g-$band. To test the sensitivity of $\alpha$ on $M^*$, we also fix $M^*$ to the value measured for field galaxies by Smith et al. (2009). These authors measured the luminosity function of a sample of $\sim 40\,000$ low redshift ($z\sim0.1$) field galaxies using $r-$band SDSS photometry, finding $M_r^* -5\log h = -20.40 \pm 0.04$ mag, or $M_r^* = -21.12$ mag for $H_0 = 72$ km~s$^{-1}$~Mpc$^{-1}$, a value that is broadly consistent with the one measured for the luminosity function of nearby clusters (e.g. Mines \& Geller 2008; Yagi et al. 2002). To convert this value to the NGVS bands, we use the average colors measured for the six NGVS galaxies in the core region brighter than $r = 11.4$ mag: $(u^*-r) = 1.95 \pm 0.07$, $(g-r) = 0.64 \pm 0.05$, $(i-r) = -0.34 \pm 0.03$ and $(z-r) = -0.61 \pm 0.08$ mag. We then run the fits by setting the prior on $M^*$ to be non-zero only within 0.1 mag of the  value measured by Smith et al. (2009), converted to the corresponding NGVS band using the colors listed above. The result for the $g-$band (for which the Smith et al. 2009 value corresponds to $M^* = -20.48$ mag) is shown by the blue line in Figure~2 and the posterior probability density function is shown in the right panel of Figure~3. The best-fit slope in this case is $\alpha = -1.33 \pm 0.02$, fully consistent with the value obtained when $M^*$ is unconstrained. Table~1 also lists the faint-end slopes measured using $u^*, r, i$ and $z$ photometry. All slopes are insensitive to the value of $M^*$ and are consistent with each other to within $1\sigma$. 

Starting with the work of Sandage et al. (1985), it has been noted that: (1) the galaxy luminosity function in Virgo displays a ``dip'' at $M_g \sim -17$ mag and/or; (2) it does not follow a single Schechter function (Trentham \& Tully 2002; Trentham \& Hodgkin 2002; Rines \& Geller 2008; McDonald et al. 2009). A similar observation has been made for the luminosity function of field galaxies: i.e., a broad and shallow trough appears at a similar magnitude (Blanton et al. 2005; Trentham et al. 2005; Smith et al. 2009; see also Loveday 1997). In rich galaxy clusters, the luminosity function also appears to show a steepening at faint magnitudes (Trentham 1998; Tully et al. 2002; Yagi et al. 2002; Popesso et al. 2006; Agulli et al. 2014; although the faint-end slope measurements are not consistent among clusters). Finally, a dip followed by an upturn is seen in at least some galaxy groups (e.g., the M81 group, Chiboucas et al. 2013), although the luminosity function shape is generally not well determined due to small number statistics (Trentham et al. 2005; Chiboucas et al. 2009). 

\begin{deluxetable}{llll}
\tabletypesize{\footnotesize}
\tablewidth{0pc}                        
\tablenum{2}
\tablecaption{MCMC Fitting Results Within Restricted Magnitude Ranges ($g-$band)\label{tab2}}
\tablehead{
\multicolumn{1}{c}{No.} &
\multicolumn{1}{c}{Sample} &
\multicolumn{1}{c}{$M^*$ (mag)} &
\multicolumn{1}{c}{$\alpha$} \\
\multicolumn{1}{c}{(1)} &
\multicolumn{1}{c}{(2)} &
\multicolumn{1}{c}{(3)} &
\multicolumn{1}{c}{(4)}
}
\startdata
71 & $M\leq-14$ & $-$20.48  &   $-1.266^{+0.056}_{-0.057}$ \\
116 & $M\leq-13$ & $-$20.48  & $-1.381^{+0.038}_{-0.040}$ \\
170 & $M\leq-12$ & $-$20.48  & $-1.372^{+0.032}_{-0.033}$ \\
226 & $M\leq-11$ & $-$20.48  & $-1.337^{+0.027}_{-0.024}$ \\
295 & $M\leq-10$ & $-$20.48  & $-1.332^{+0.018}_{-0.017}$ \\
46 & $-17.4\leq M \leq -14.4$ & $-$20.48  & $-1.39^{+0.11}_{-0.11}$ \\
292 & $-14.4\leq M \leq -9.13$ & $-$20.48  & $-1.263^{+0.034}_{-0.035}$ 
\enddata
\tablecomments{All results refer to the luminosity function measured in the $g-$band. The columns list the number of galaxies (column 1), the magnitude range spanned by the fitted sample (column 2), the assumed $M^*$ (column 3, fixed to the value measured in the field by Smith et al. 2009) and the faint end slope $\alpha$ (column 4).}      
\end{deluxetable}

\begin{deluxetable*}{lrlllll}
\tabletypesize{\footnotesize}
\tablewidth{0pc}                        
\tablenum{3}
\tablecaption{Summary of Faint-End Slope Measurements for the Virgo Cluster\label{tab3}}
\tablehead{
\multicolumn{1}{c}{Reference} &
\multicolumn{1}{c}{Area} &
\multicolumn{1}{c}{Mag. Limit} &
\multicolumn{1}{c}{$\alpha$} &
\multicolumn{1}{c}{No.} &
\multicolumn{1}{c}{Region targeted} &
\multicolumn{1}{c}{Comments}  \\
\multicolumn{1}{c}{(1)} &
\multicolumn{1}{c}{(2)} &
\multicolumn{1}{c}{(3)} &
\multicolumn{1}{c}{(4)} &
\multicolumn{1}{c}{(5)} &
\multicolumn{1}{c}{(6)} &
\multicolumn{1}{c}{(7)}
}
\startdata
Sandage et al. 1988 & 140 & $M_B < -11$ & $-1.30$                     & 1647 & Entire Cluster & All galaxies, $M_B < -11.7$\\
                                  &       &                       &  $-1.40$                     & 1210 & Entire Cluster & Early-Types, $M_B < -11.7$\\
Impey et al. 1988 & 7.7 &  $M_B < -11$ & $-1.7$                            & 137 & Entire Cluster & Dwarf galaxies from Sandage et al. (1985)\\
Phillipps et al. 1998 & 3.4 &  $M_R < -11$ & $-2.18\pm0.12$         & $\sim 4000$ & Core and Outskirt & Dwarf galaxies, $-16 \leq M_R \leq -11.5$\\
Trentham \& Tully 2002 & 0.76 &  $M_R < -10$ & $-1.03$              & 99 & Core & $-16 \leq M_R \leq -10$\\
Trentham \& Hodgkin 2002 & 24.9 &  $M_B < -11$ & $-1.35$         & 449 & Core to outskirt & $M_B > -18$\\
 &  &   & $-1.7$                                                                                 & \nd  & Core to outskirt & $-17 \leq M_B \leq -14$\\
 &  &   & $-1.1$                                                                                 & \nd  & Core to outskirt & $-14 \leq M_B \leq -12$\\
Sabatini et al. 2003 & 14 &  $M_B < -10.5$ & $-1.6 \pm 0.1$         & 241 & Core to outskirt & $-14.5 \leq M_R \leq -10.5$\\
 &  &   &                                                           $-1.4 \pm 0.2$          & \nd & Core & \\
 &  &   & $-1.8\pm0.2$                                                                      & \nd & Outskirt & \\
Rines \& Geller 2008 & 35.6 &  $M_r < -13.5$ & $-1.28 \pm 0.06$ & 484 & Entire Cluster & $M_r < -13.5$\\
Lieder et al. 2012 & 3.7 & $M_V < -13.0$ & $-1.50\pm0.17$           & 216 & Core &  $-18.8 \leq M_V \leq -13$\\
This paper & 3.7 & $M_g < -9.13$ & $-1.33 \pm 0.02$                    & 404 & Core & $M_g < -9.13$
\enddata
\tablecomments{The columns list the literature reference (column 1); the area covered by the survey, in deg$^2$ (col 2, but note that only sources within the core region are considered in the comparison shown in Figures 4 and 5); the magnitude limit and faint-end slope quoted in the original paper (columns 3 and 4); the number of galaxies used in the fit (col. 5); the region of the cluster targeted (col. 6) with `core' referring to a region within the area discussed in this paper, `entire cluster' referring to the entire region within (at least) one virial radius, and `core to outskirt' referring to strip(s) extending from the core outwards; and the type of galaxy/magnitude range over which the luminosity function was fitted (col. 7). With the exception of Rines \& Geller (2008), who required spectroscopic confirmation, membership criteria in all other studies are based on morphological and photometric properties (although the details vary significantly between studies), supplemented by velocities, when available, and often (but not always) followed by visual confirmation. Note that  Impey et al. (1988) base their luminosity function on the VCC catalogue; however, while Sandage et al. (1985) completeness correction was applied at $M_B> -13$, Impey et al. (1988) estimate the incompleteness in the VCC sample to set in at $M_B> -14$ based on the 137  galaxies detected within their survey.}
\end{deluxetable*}

Marginal evidence for a dip in the NGVS Virgo core luminosity function at $M_g \sim -17$ mag can be seen in Figure~2, although the statistics are poor. To address point (2) above, i.e. examine the stability of the faint-end slope measurements, we performed two tests. First, we ran a series of fits restricting the magnitude range of the data. The results, which are summarized in Table~2, demonstrate that cutting the sample at progressively brighter magnitudes (from $M_g \leq -10$ to $M_g \leq -14$ mag) does not appreciably change the slope. The second test was carried out by fitting separately a bright and a faint sample.  Trentham \& Hodgkin (2002) noted that the slope of the Virgo luminosity function for a sample restricted to $-17 < M_B < -14$ mag is  $\alpha \sim -1.7$, significantly steeper than derived for fainter galaxies ($-14 < M_B < -12$, $\alpha \sim -1.1$); indeed a `step' at $\sim -14$ is suggested by the NGVS data.  Assuming $(B-g) \sim 0.4$ (the average difference between the $B-$band magnitudes in the VCC and the $g-$band magnitudes measured by the NGVS for the galaxies in common, see Ferrarese et al. 2016) and fitting the 46 NGVS galaxies with $-17.4 \leq M_g \leq -14.4$ mag gives $ \alpha = -1.39 \pm 0.11$, while fitting the 292 galaxies with magnitudes $-14.4 \leq M_g \leq -9.13$ gives $ \alpha = -1.26 \pm 0.03$. Thus, we find that the slope for the faint sample is indeed shallower than that for the brighter sample, but the difference is at the 1$\sigma$ level and not nearly as pronounced as reported by Trentham \& Hodgkin (2002). A more thorough investigation of this issue will be presented in a future NGVS contribution that examines the luminosity function over the entire cluster and benefits from greatly reduced statistical uncertainties.

\subsection{Comparison with Previous Work}
\label{sec:lfvirgo2}

Before proceeding, we pause to put our results into the context of previous luminosity function measurements in Virgo. The relevant studies, which are discussed in detail below,  are summarized in Table 3, while a detailed comparison between the literature and the NGVS data is presented in Appendix A of Ferrarese et al. (2016).

Figures 4 and 5 compare previous luminosity functions to that from the NGVS. In both figures, the gray filled histogram represents the raw NGVS data while the dashed histogram shows the completeness corrected luminosity function, up to the point where completeness drops below 50\%. Colored histogram bars show data from the literature\footnote{We are unable to show the data from Phillipps et al. (1998), who did not publish their catalogue, and Impey et al. (1988), who did not publish the photometry for their entire sample.}. Figure 4 shows the original published photometry, using appropriate color transformations, if needed, while Figure 5 uses the NGVS photometry for galaxies in common between the published and the NGVS samples. In all cases, we plot only those galaxies that lie within the boundaries of the core region defined in this paper. With the exception of the VCC and EVCC, none of the other surveys covered the core region in its entirety, so we show counts without ({\it left panel}) and with ({\it right panel}) the application of a correction factor equal to the ratio of the NGVS areal coverage to the area covered by each survey (under the assumption that galaxies are distributed uniformly across the area covered by each survey). Note that this approach may artificially increase the number of very bright  galaxies.   

\begin{figure*}
\epsscale{1.0}
\plottwo{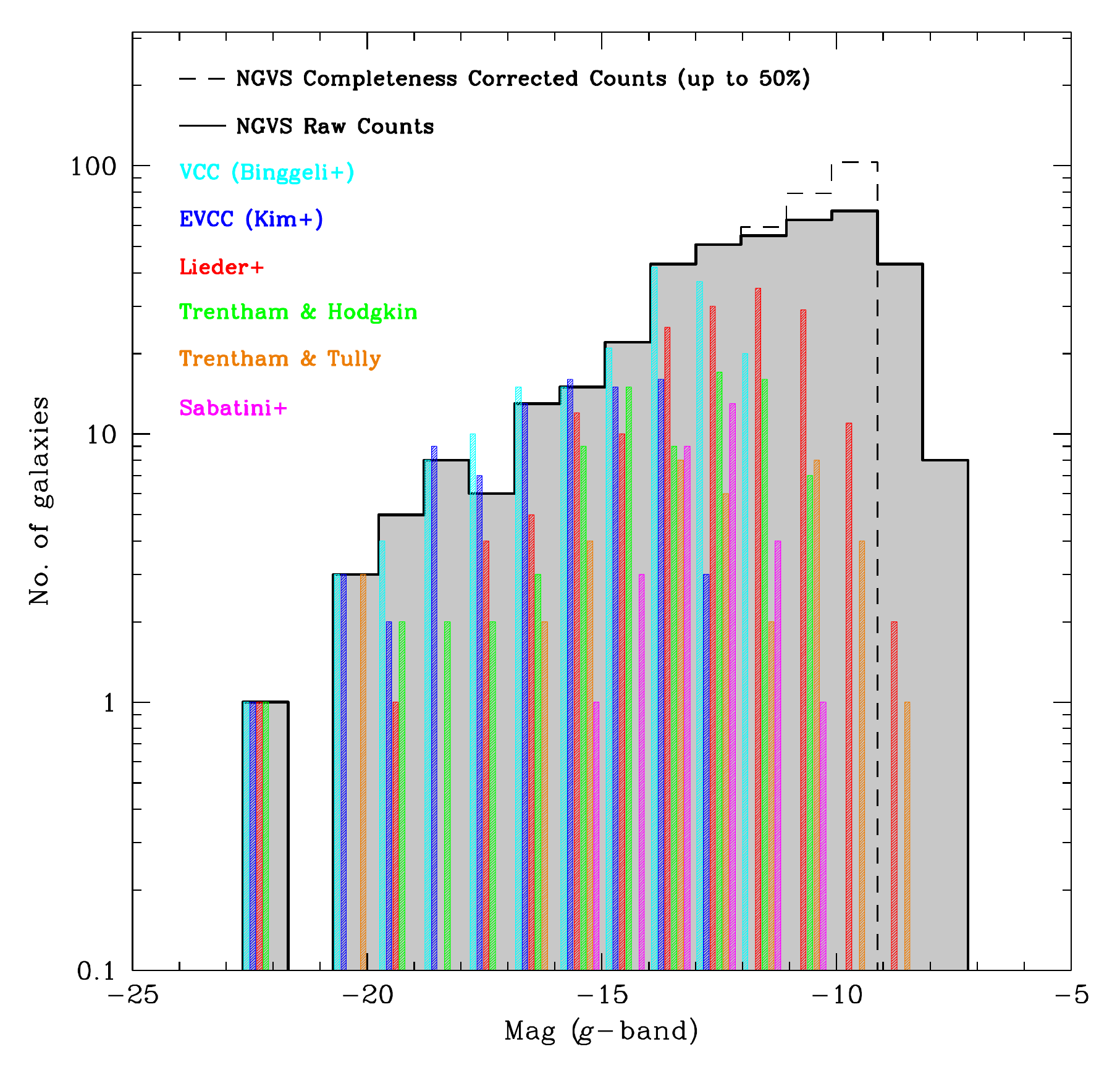}{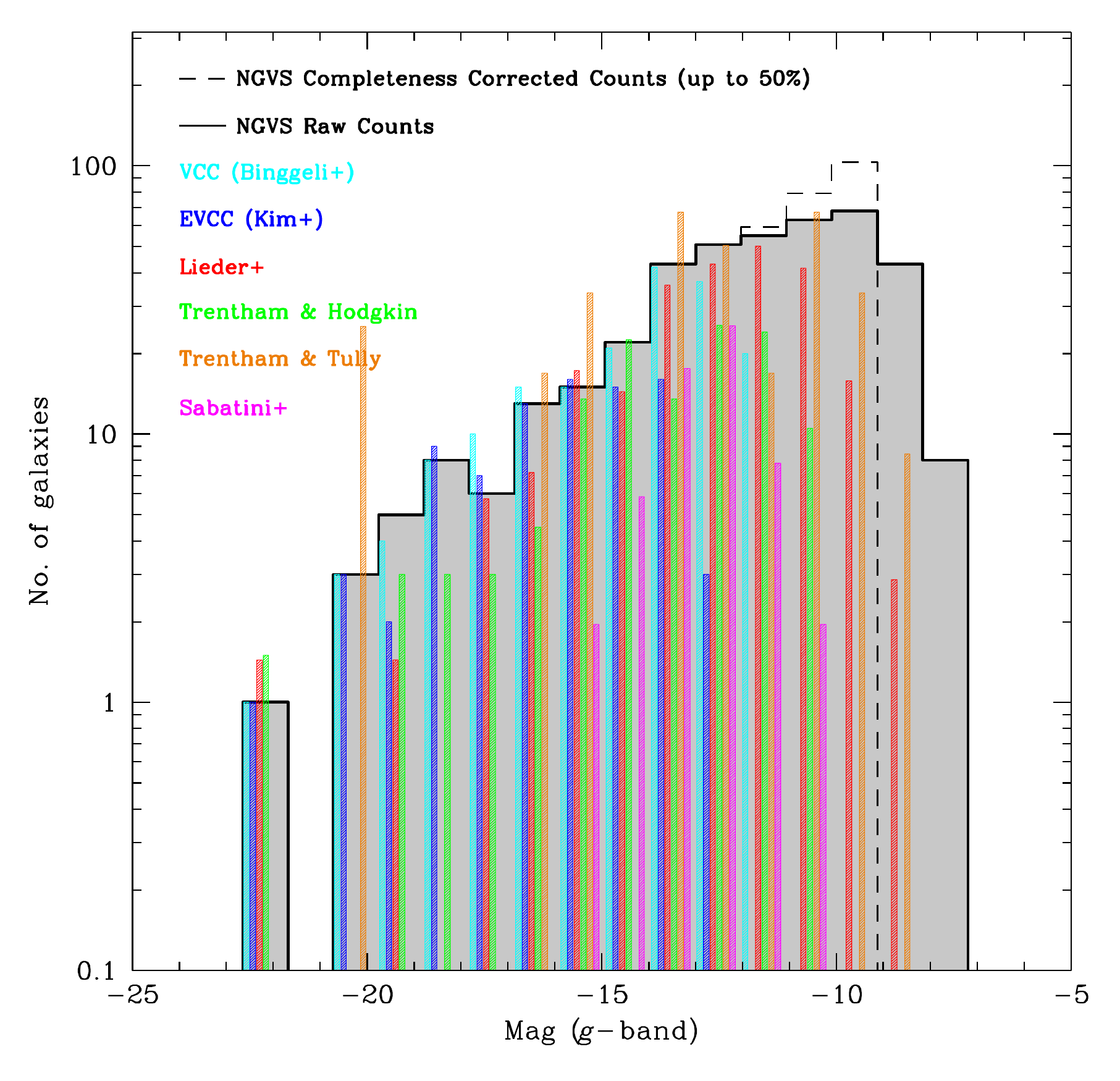}
\caption{Comparison between the NGVS luminosity function of the core of Virgo, and that derived in independent previous published works. The NGVS data are shown by the gray histogram, representing the raw counts, and the black dashed histogram, representing the luminosity function corrected for completeness, up to a maximum correction of 50\% (corresponding to $M_g \le -9.13$ mag). The cyan, blue, red, green, orange and magenta histograms represent data from the VCC, EVCC, Lieder et al. (2012), Trentham \& Hodgkin (2002), Trentham \& Tully (2002), and Sabatini et al. (2005), respectively, but only for detections within the core region discussed in this paper. The literature data have been corrected for Galactic extinction, transformed to the $g-$band using a constant color term (except for the EVCC, for which data are tabulated in the $g-$band), and translated to absolute magnitude using a Virgo distance modulus of $(m-M)_0$ = 31.09 mag. In the left panel, the counts from the literature are presented as in the various works, while in the right panel, they have been scaled proportionally to the fractional area covered within the core by the various surveys. The VCC and the EVCC catalogs cover the entire core region and therefore appear the same in both panels.}
\label{fig5}
\end{figure*}

\begin{figure*}
\epsscale{1.0}
\plottwo{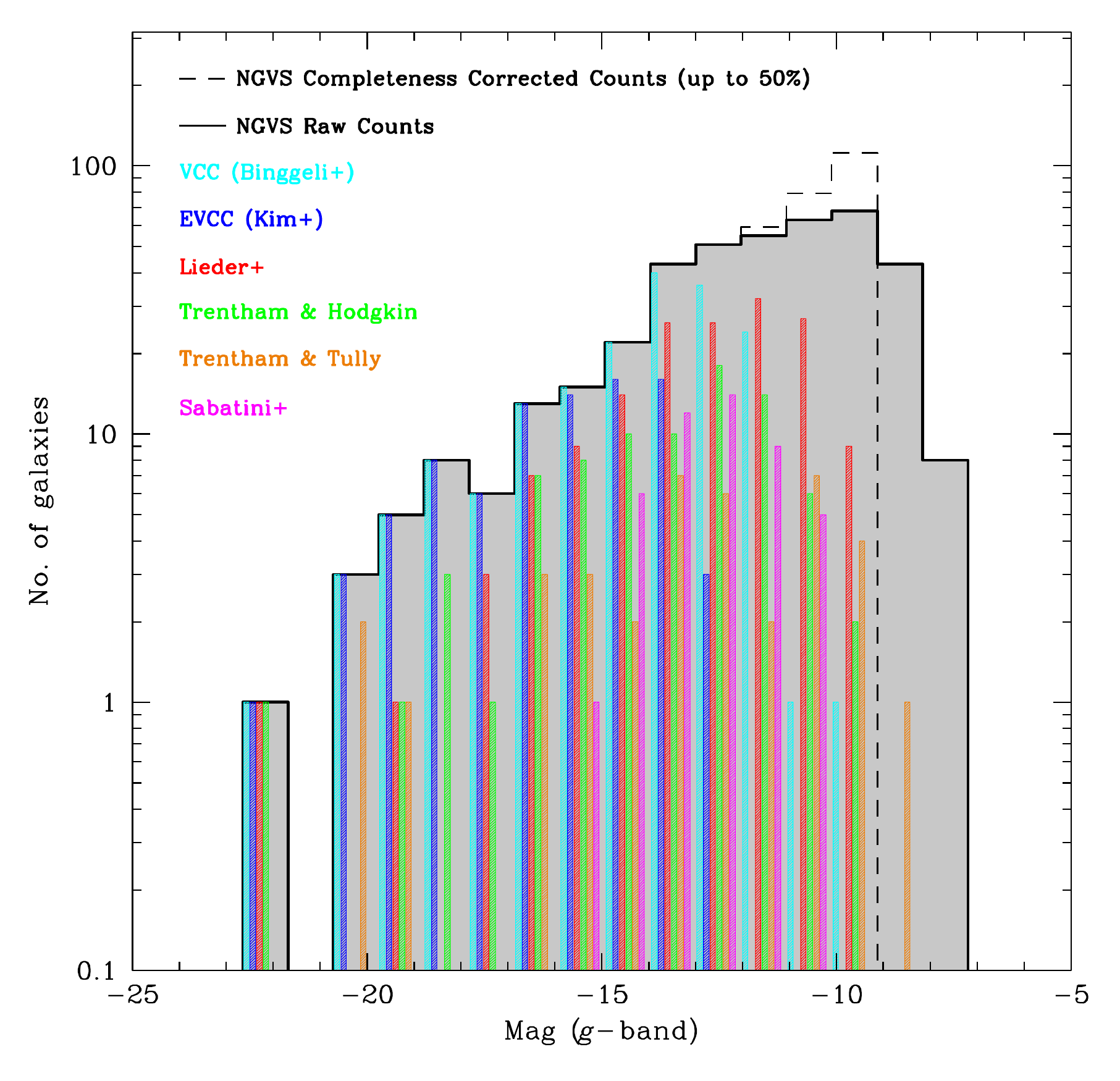}{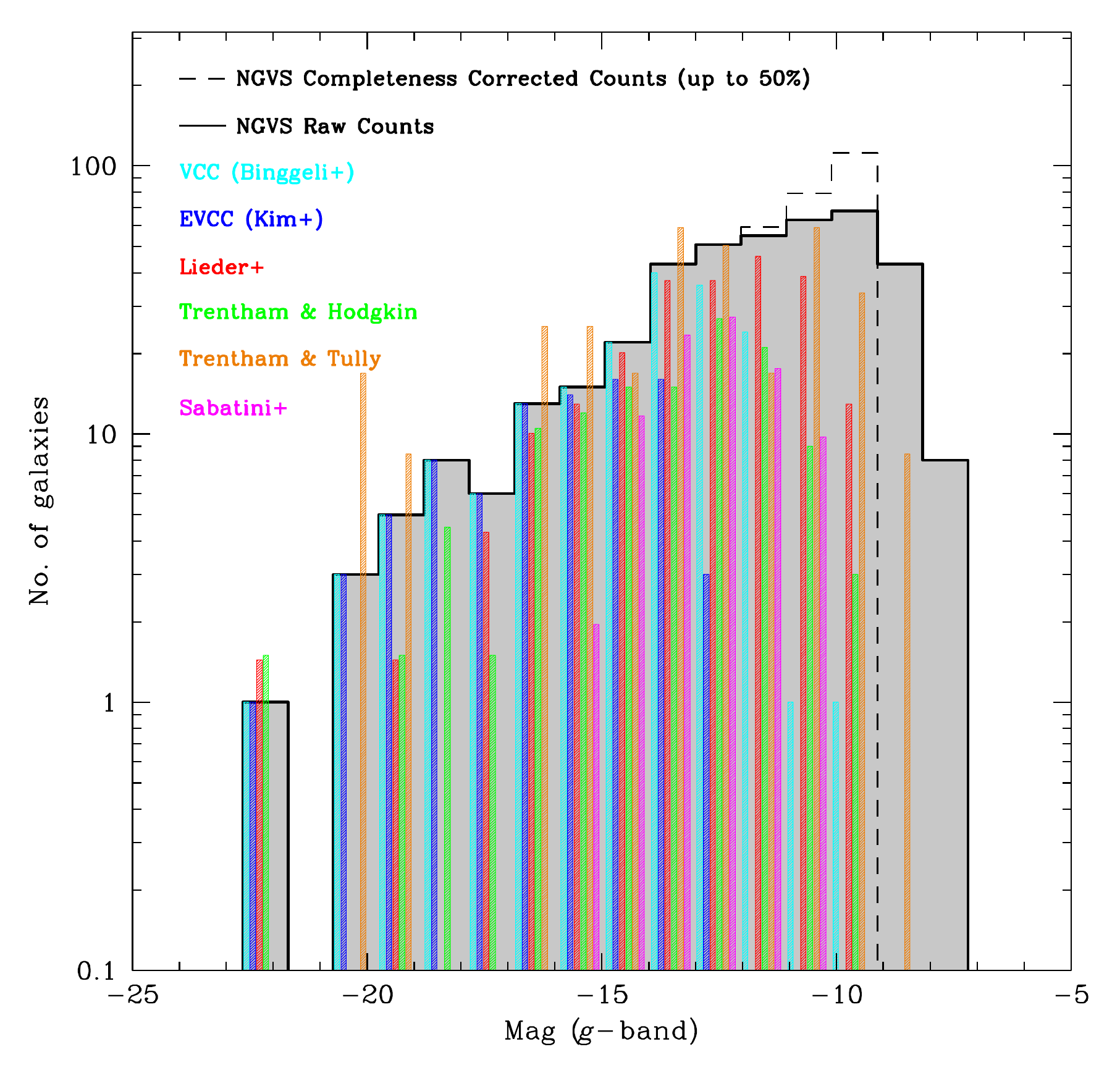}
\caption{The same as Figure 4 except that the galaxies previously reported in the literature have now been matched to the NGVS catalog: i.e., galaxies deemed to be spurious in the NGVS membership analysis (see Ferrarese et~al. 2016) have been removed and magnitudes have been plotted using the NGVS $g-$band magnitudes, corrected for Galactic extinction. On the left, the counts from the literature are as presented in the various works, while on the right, they have been scaled proportionally to the fractional area covered within the core by the various surveys.}
\label{fig6}
\end{figure*}

As mentioned in \S\ref{sec:intro}, the VCC (Binggeli, Sandage \& Tammann 1985) formed the basis of the first secure determination of the luminosity function in the Virgo cluster (Sandage, Binggeli \& Tammann 1985). These authors divided their sample according to morphological type, motivated by the observation that the dip seen at $M_B \sim -17.5$ implies that a single Schechter function cannot adequately reproduce the data. At the faint end (i.e., down to the detection limit of the survey, $M_B \sim -11$ mag) where their sample is dominated by early-type galaxies, they found a slope $\alpha \sim -1.35$\footnote{Note that, fainter than $M_B = -13$, these authors applied a completeness correction to the VCC data.}. More recently, the VCC analysis has been extended to even wider fields by Kim et al. (2014), who used the SDSS Data Release 7 to spectroscopically identify cluster members over an area of 725 deg$^2$. Their catalog extends to $M_r \sim -12$ mag --- the nominal spectroscopic limit of the SDSS --- but the authors warn that incompleteness in the SDSS spectroscopic data sets in at significantly brighter magnitudes, $M_r \sim -14.5$. Although the authors do not quote a faint-end slope, this is not expected to be significantly different from that of Sandage et al. (1985), given the overlap between the two samples. 

On the other hand, two studies that followed the VCC by pushing to fainter magnitudes and/or lower surface brightnesses argued for considerably steeper slopes. First, Impey et al. (1988) published a deeper galaxy catalog based on photographically amplified UK Schmidt plates covering a 7.7 deg$^2$ region (partially overlapping the NGVS core region). The authors discovered a number of galaxies with $M_B < -11$ mag, including some objects in the magnitude range where the VCC was supposedly complete. By applying a more aggressive completeness correction to the VCC data, the authors reported a slope of $\alpha \sim -1.7$, although they did caution that both large uncertainties in the data and observational biases and selection effects could conspire towards a steepening of the derived slope. A decade later, Phillipps et al. (1998) surveyed an area of 3.4 deg$^2$, partially overlapping with the cluster core, to a completeness limit $M_R \sim -11$ mag. In the range $-16 < M_R < -11.5$, the authors found an extremely steep luminosity function, with slope $\alpha \sim -2.18 \pm 0.12$, and suggested that the actual slope might be even steeper, $\alpha \sim -2.5$, at fainter magnitudes.  

Based on the NGVS data, it seems quite likely that both the Impey et al. (1988) and the Phillipps et al. (1998) slopes were severely overestimated. Within the core region, Impey et al. (1988) identified eight galaxies that were not cataloged in the VCC. None of these is recovered in the (deeper and more sensitive) NGVS images (see Ferrarese et~al. 2016): although Impey et al. (1988) do not list their magnitudes, these galaxies are located in uncrowded regions and must be brighter than $B \sim -11$ mag (the faint-end limit of their survey), conditions under which the NGVS is essentially complete. Eliminating these spurious detections would  bring the Impey et al. (1988) slope in line with the VCC value. Meanwhile, the Phillipps et al. (1999) survey did not include color information, without which identification of background interlopers, especially at faint magnitudes, becomes problematic. Indeed, it has previously been suggested that the Phillipps sample was severely contaminated by background sources (Trentham \& Tully 2002; Trentham \& Hodgkin 2002). Over the range $-13 \leq M_g \leq -12$ mag, where the NGVS is essentially complete (see Ferrarese et~al. 2016), a slope $\alpha = -2.0$ predicts 130 times more galaxies than are detected in the NGVS: i.e., we would expect to detect over 7000 galaxies when, in fact, we find just 58.  We note that depth is unlikely to be responsible for the difference in slope between our study and  Impey et al. (1998) and Phillipps et al. (1998), since restricting the NGVS sample to magnitudes brighter than the NGVS 50\% completeness limit does not affect our result (Table 2).   

Much shallower slopes --- shallower, in fact, than the original Sandage et al. (1985) value --- were found in two contemporaneous studies by Trentham \& Tully (2002) and Trentham \& Hodgkin (2002). The first study surveyed a 0.76 deg$^2$ area in the core of the cluster,  while the second covered two elongated regions extending North and West of the core, for a total coverage of 24.9 deg$^2$. Both surveys were estimated to be complete down to $M_B \sim -11$~mag and both found that a single Schechter function did not provide a good fit to the entire luminosity function (which is not surprising since both relied on the VCC at the bright end of the sample). In the range $-16 < M_R < -10$ (corresponding to $-14.5 \lesssim M_B \lesssim -8.5$, using the transformation provided by the authors ), Trentham \& Tully (2002) found a shallow slope of $\alpha \sim -1.03$. Based on an inspection of the left panels of Figures 4 and 5, this shallow slope might be due to some residual incompleteness in their sample for galaxies fainter than $M_g \sim -12$~mag ($M_R \sim -12.7$~mag).

Trentham \& Hodgkin (2002) found an average slope for galaxies fainter than $M_B = -18$~mag of $\alpha = -1.35$, which is consistent with the value of Sandage et al. (1985). As mentioned above, however, they emphasized that the slope displays a significant magnitude dependence: i.e., it steepens to $\alpha \sim -1.7$ for $-17 < M_B < -14$ mag, while at fainter magnitudes, $-14 < M_B < -12$~mag, it flattens to $\alpha \sim -1.1$.  

Finally,  Rines \& Geller (2008) used radial velocities from the SDSS Data Release 6 to spectroscopically confirm cluster members over an area of 35.6 deg$^2$ centered on M87. Down to $M_r \sim -13.5$~mag, they reported a luminosity function slope of $\alpha \sim -1.28 \pm 0.06$, which is consistent with the value found here. 

Since the surveys discussed above generally sample different regions within the cluster, it is fair to ask whether the differences in the measured slopes might be due, at least in part, to an environmental dependence of the luminosity function. The two studies that, to date, have explored a possible environmental dependence for the luminosity function (using optical data) appear to indicate a shallower slope in the core region with respect to the outer parts of the cluster. Sabatini et al. (2003) analyzed a 14 deg$^2$ strip that partially includes the core region; fitting the core region in the range $-14.5 < M_B < -10.5$ mag (beyond which the authors estimated their sample to be severely incomplete), led to a faint-end slope of $\alpha = -1.4 \pm 0.2$, while fitting the outer regions of the cluster seemed to indicate a steeper slope of $\alpha = -1.8 \pm 0.2$. A similar trend was reported by Lieder et al. (2012), who covered a region of $\sim 3.71$ deg$^2$, overlapping with most of the core region discussed in this paper, and extending further in the North-West direction. Over the range $-18.8 \lesssim M_V \lesssim -13.0$ mag, where the authors estimated their sample to be complete, the faint-end slope was measured to be $\alpha = -1.50 \pm 0.17$. However, restricting the sample to the field surrounding M87 yielded a marginally shallower slope, $\alpha = -1.31 \pm 0.08$. Finally, in the ultraviolet, however, Boselli et al. (2016)  used GALEX data covering an area of more than 300 deg$^2$, finding no significant differences in the overall UV luminosity function from the core to the cluster outskirts (although we note that their survey is shallower than those discussed so far, reaching early type galaxies with $g-$band magnitude $\sim -15$).  In a future paper, we will revisit the question of a possible environmental dependence using the full sample of NGVS detections distributed across the entire cluster.

\section{The Connection to Ultra Compact Dwarfs and Globular Clusters}
\label{sec:lfucds}
UCDs have traditionally been defined (somewhat {\it ad hoc}) as compact stellar system with properties intermediate between those of compact, low-mass galaxies and globular clusters (e.g., Hilker et al. 1999; Drinkwater et al. 2000; Phillips et al. 2001). With masses in the range $2\times10^6 \lesssim {\cal M} \lesssim 10^8$ \msun, magnitudes $-13 \lesssim M_V \lesssim -9$, and half light radii $10 \lesssim r_h \lesssim 100$ pc, they are as faint as the faintest dwarf galaxies known in the Local Group, but are about one order of magnitude more compact. Their origin is still somewhat obscure (e.g. Mieske et al. 2002, 2012; Fellhauer \& Kroupa 2002, 2005; Br{\"u}ns \& Kroupa 2012), but in recent years, the hypothesis that at least some of them might be the stripped remnants of nucleated early-type galaxies has gained significant support (Bekki et al. 2001; Drinkwater et al. 2003; Ha{\c s}egan et al. 2005; Paudel et al. 2010; Pfeffer \& Baumgardt 2013; Seth et al. 2014; Janz et al. 2015; Liu et al. 2015ab; Zhang et al. 2015). In view of these studies, it is interesting to explore how the galaxy luminosity function changes when UCDs are added to the sample. We provide here some simple calculations on this issue, although we emphasize at the outset that there are numerous assumptions and simplifications that underlie such calculations. We further emphasize that while it is likely that at least some UCDs might not have a galactic origin -- e.g. they might simply be massive globular clusters -- for simplicity (and lack of better constraints) in our calculations we assume {\it all} UCDs to originate from stripped nucleated galaxies. 

\begin{figure*}
\epsscale{1.0}
\plottwo{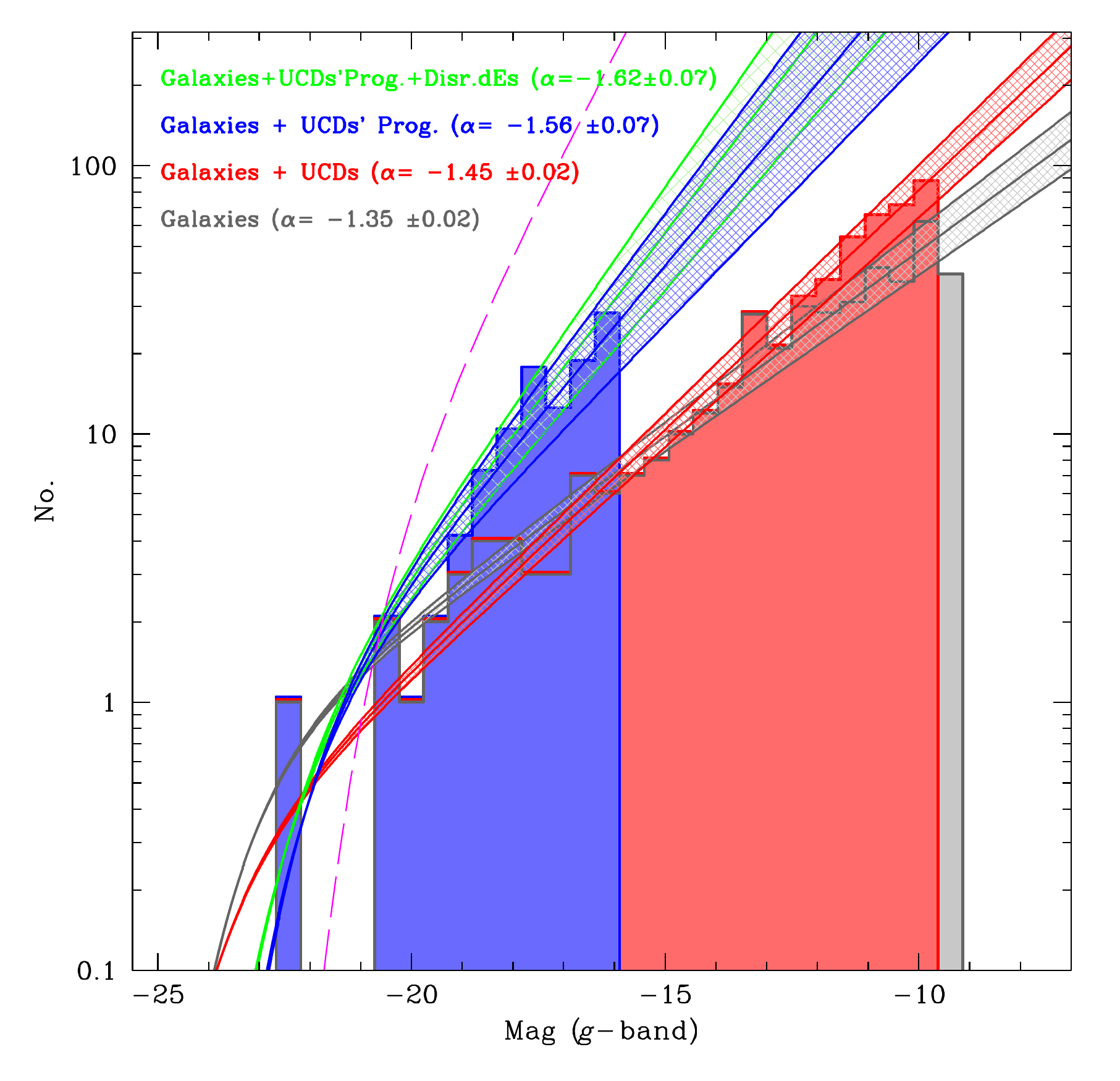}{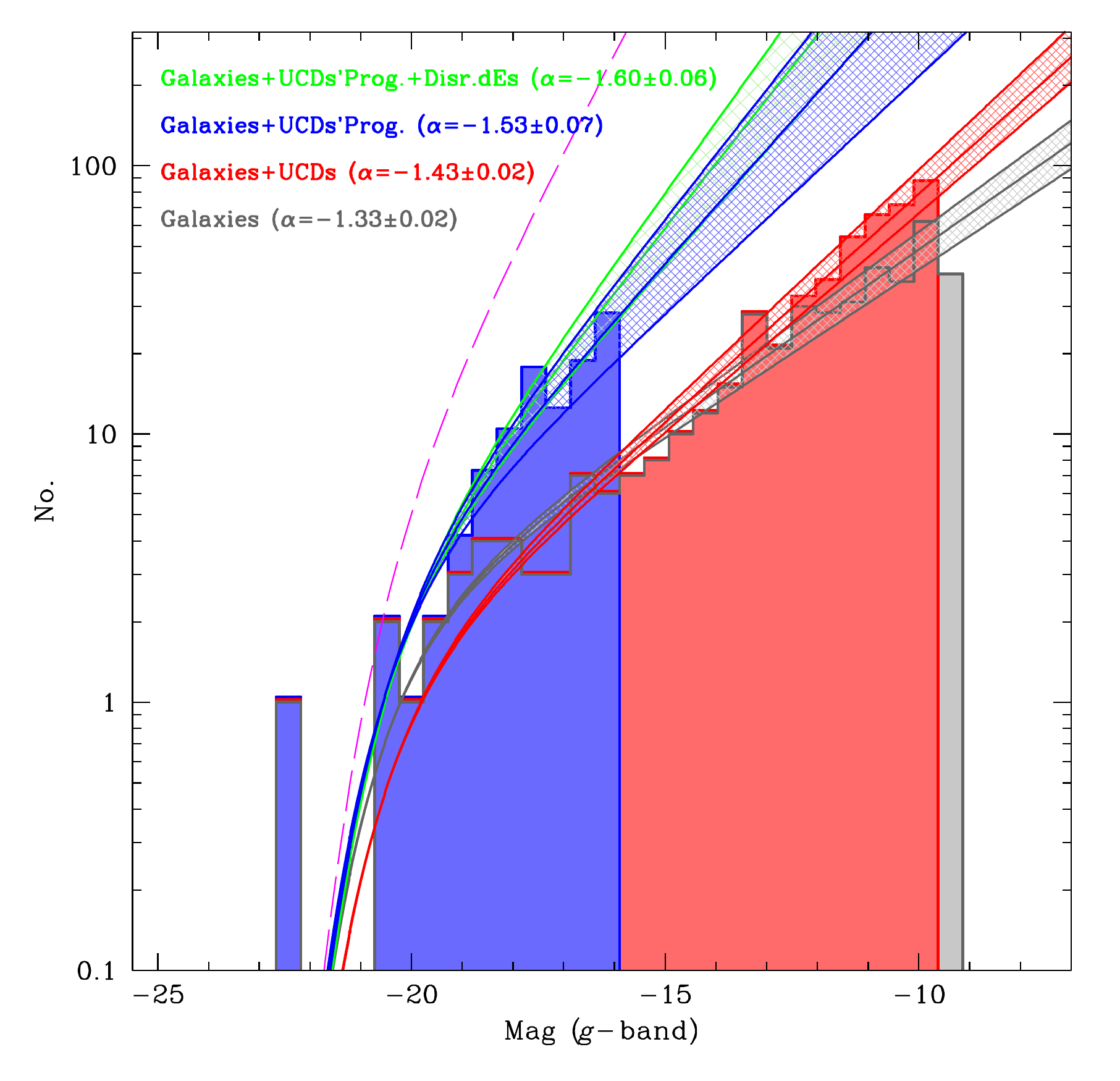}
\caption{The NGVS luminosity function of galaxies in the Virgo core region (gray histogram). The red histogram shows the galaxy sample after including magnitudes for the UCD sample of Liu et al (2015a), corrected for Galactic extinction, while the blue histogram shows the results of including magnitudes for the UCD {\it progenitors}, whose magnitude are estimated using the relation between nuclear and total galaxy magnitude discussed in S\'anchez-Janssen et al. (2016b). This second sample thus assumes that the UCDs are the surviving nuclei of tidally stripped galaxies (note that this does not include a correction to account for {\it non-nucleated} galaxies that might have been totally disrupted). Formal Schechter fits to the three samples are shown by the gray, red and blue curves, respectively. The green curve shows the luminosity function derived when disrupted non-nucleated galaxies (whose numbers are estimated from the observed number of UCDs, combined with the present-day ratio of nucleated to non-nucleated galaxies) are also added to the sample. In the right panel, $M_g^*$ is held fixed while on the left, $M_g^*$ is allowed to vary. The magenta dashed line represents a Schechter function with $\alpha = -1.9$, the slope of the $\Lambda$CDM halo mass function.}
\label{fig7}
\end{figure*}

Our departure point is the catalog of confirmed and possible UCDs surrounding M87 published by Liu et al. (2015a) using NGVS data. A thorough search for UCDs in the core of the cluster (the same region considered in this work), yielded a total of 92 objects with magnitudes in the range $-12.1 \leq M_g \leq -9.6$.  UCDs were identified on the basis of location in the $u^*iK_s$ diagram (Mu\~noz~et al. 2014), magnitude, half-light radius and surface brightness. It is important to realize that the faint magnitude limit of the UCD sample, $g \sim 21.5$~mag, was artificially imposed during the selection process (see Liu et~al. 2015a for details). 

In analyzing the ``corrected'' luminosity function, we follow two approaches meant to provide a rough idea of the magnitude of the expected change in the faint-end slope of the luminosity function due to the inclusion of UCDs. In the first approach, we simply add the magnitude of each UCD to the galaxy sample. This approach is justified if, for instance, UCDs are the remnants of primordial compact galaxies (Drinkwater et al. 2004) and/or if their evolutionary history is correctly reflected in the cosmological simulations that predict the mass function of dark matter halos. We then use MCMC to refit the luminosity function down to $M_g \leq -9.6$ (i.e. the UCD faint magnitude limit, $\sim 0.5$ mag brighter than the galaxy's 50\% completeness limit) and find a modest --- but statistically significant --- increase in the faint-end slope, from $\alpha = -1.33 \pm 0.02$ (for the galaxies alone) to $\alpha = -1.42 \pm 0.02$ when $M^*$ is fixed to $-20.48$ mag. A similar steepening is found if $M^*$ is allowed to vary (see Table~1). 

In the second approach, we assign to each UCD {\it the magnitude the presumed UCD host galaxy would have if stripping had not taken place}.  Following earlier work from C\^ot\'e et al. (2006), S\'anchez-Janssen et al. (2016b) provide a fit to the relation between nuclear and galaxy magnitudes for the sample of nucleated early-type galaxies in the core of Virgo, based on NGVS data. For galaxies hosting nuclei of magnitude comparable to the observed UCDs, the nuclear luminosities are found to be proportional to the $\sim0.5$ power of the host galaxy luminosity: we use this relation to estimate the initial magnitude of the UCD progenitor\footnote{Note that the term `progenitor' does not imply the object to be unevolved. Rather, the `progenitor's magnitude' is the magnitude the host galaxy would have, at the present time, had it not been subjected to the extreme stripping processes that have lead to the observed UCD, but had instead followed the evolution of a `typical' nucleated galaxy.}~under the assumption that the UCD, viewed at the present time, corresponds to the original nucleus. The error on the UCD progenitor magnitude is taken to be equal to the scatter in the relation of S\'anchez-Janssen et al. (2016b), $\sigma(M_g) \simeq$ 1.1 mag. Once the correction is applied, the (pre-stripping) UCD progenitors have magnitudes in the range $-18.8 \leq M_g \leq -14.5$ mag, corresponding to stellar masses of $1.0\times10^{10} \geq {\cal M} \geq 1.1\times10^{8}$ \msun~ (fainter and less massive progenitors are excluded due to the selection criteria imposed on the UCD sample). An MCMC fit to the luminosity function of the combined UCD progenitors and galaxy population down to  $M_g \leq -15.6$ mag (${\cal M} = 4.3\times10^8$ \msun, which we take as the completeness limit of the sample, and is equal to the magnitude of the faintest of UCD progenitors, from which we subtract the scatter in the S\'anchez-Janssen relation), leads to a faint-end slope of $\alpha = -1.53 \pm 0.07$ when $M^*$ is fixed. The luminosity functions of the combined galaxy and UCD sample are shown in Figure 6, both in the case in which $M^*$ is allowed to vary, or held fixed. We note that, taking our argument even further to correct, besides UCDs, also compact ellipticals (which, like UCDs, might result from extreme tidal stripping, e.g. Guerou et al. 2015) would not affect our results. The core region, in fact, includes only two such systems (VCC 1297 == NGC4486B, and EVCC813); the nucleated nature of VCC 1297 is difficult to assess in view of the steep surface brightness profile of the galaxy (e.g. C\^ot\'e et al. 2006; Ferrarese et al. 2016) while the nucleus of EVCC813 is at the bright end of the distribution of observed UCDs, and the addition of its progenitor would not appreciably alter the luminosity function shown in Figure 6.

There is an additional correction that should be applied to the above estimate. UCD progenitors are, by definition, nucleated galaxies. However, the nucleation fraction, $\eta$, (i.e., the ratio of nucleated galaxies to galaxies of all types), is known to vary as a function of magnitude (Binggeli \& Cameron 1991; C\^ot\'e et~al 2006; Turner et al. 2012; den Brok et al. 2014; S\'anchez-Janssen et al. 2016b). In the magnitude range spanned by the UCD progenitors, $\eta$, varies from $\sim 94$\% at $M_g = -18.8$ mag to $\sim 52$\% at $M_g = -14.5$ mag. If we assume that for every UCD (i.e., every disrupted nucleated galaxy), $(1/\eta-1)$  non-nucleated galaxies existed that are now completely disrupted, and correct the luminosity function accordingly, then the faint-end slope steepens to $\alpha = -1.60 \pm 0.06$ (when the value of $M^*$ is held fixed, or $\alpha = -1.62 \pm 0.07$ when it is allowed to vary).

A few conclusions can be drawn from the above analysis. First,  the inclusion of UCDs can have a significant impact on the faint-end slope of the luminosity function. However, even in the most extreme case ($\alpha = -1.60 \pm 0.06$, obtained by adding to the galaxy sample the magnitudes of the presumed UCD progenitors and correcting for the nucleation fraction) the faint-end slope is still shallower than the slope of the $\Lambda$CDM mass function at the low-mass end ($\alpha \sim -1.9$). Moreover, if the goal is to compare the observed galaxy luminosity function to the present-day dark matter halo mass function, it is not entirely clear whether UCDs should be included at all. Whether UCDs are still surrounded by a dark matter halos is subject to debate (Ha{\c s}egan et al. 2005; Mieske et~al. 2008; Chilingarian et al. 2008; Goerdt et al. 2008; Frank et al. 2011; Seth et al. 2014). If they are not, then the simulations presumably also predict the infall halos to be completely disrupted, and therefore not to contribute to the present-day halo mass function, in which case UCDs should not be included in the comparison. A second conclusion is that, because UCD numbers appear to scale with the underlying total gravitating mass (Liu et al. 2015a), including UCDs will likely lead to an increase in the apparent environmental dependence of the faint-end slope, with steeper slopes associated with deeper potential wells. We plan to test this hypothesis in a future contribution that uses the galaxy sample over the full cluster. 

\subsection{The Rate of Satellite Disruption}
\label{sec:lfucds2}

Whether the simulations predict the total disruption of their progenitors halo or not, UCDs can still be used to provide a rough estimate of the expected number of disrupted halos under our initial assumption: i.e., that  the original UCD progenitor magnitude can be inferred from the nuclear-to-total galaxy magnitude relation of present-day early-type galaxies. 

In the mass range spanned by the UCD progenitors in our sample, $1.0\times10^{10} \geq {\cal M} \geq 1.1\times10^{8}$ \msun, there are 92 UCDs in the core of Virgo, and 49 galaxies, of which 33 are nucleated.  At a minimum, therefore, for every galaxy-hosting halo that survives to the present time, we expect 92/49 = 1.9 halos to have been disrupted -- in fact, since UCD can only originate from nucleated galaxies, that estimate is likely closer to 3 (92/33). In other words, in this scenario, $\sim 70$\% of all infall halos that reach maximum mass after entering the cluster core will not survive to the present day. This number is likely a lower limit since, as mentioned above, not all galaxies are nucleated and therefore not all galaxies, when disrupted, would leave a recognizable remnant behind in the form of a UCD. Assuming, as done before, that for every UCD (i.e., every disrupted nucleated galaxy), $(1/\eta-1)$ non-nucleated galaxies existed that are now completely disrupted, the fraction of infall halos that reach maximum mass and do not survive to the present-day increases from $\sim 70$\% at a stellar masses ${\cal M} = 1.0\times10^{10}$ \msun, to 82\% at ${\cal M} = 1.1\times10^{8}$ \msun.  Using high resolution simulations from the Aquarius (Springel et al. 2008) and Phoenix (Gao et al. 2012) projects, Han et al. (2016) estimate that $\sim 45\%$ of infall haloes are disrupted, lower than our average value of 70\%. In addition, Han et al. (2015)  find that the disruption rate of satellites is largely independent on halo mass down to the resolution of the simulations ($\sim 10^6$ \msun), while since the observed fraction of nucleated to non-nucleated galaxies decreases as a function of magnitude (at the low-mass end), our assumptions necessarily imply that the disruption rate should increase at low masses. We stress, however, that our rough calculation assumes that nucleated and non-nucleated galaxies share the same spatial distribution and orbital properties, an hypothesis whose validity is still under active investigation (Binggeli et al. 1987; C\^ot\'e et al. 2006; Lisker et al. 2007; S\'anchez-Janssen et al. 2016b).

\subsection{Globular Clusters and Intracluster Light from Disrupted Satellites}
\label{sec:lfucds3}

Finally, it is interesting to explore the possible contribution of UCDs and their progenitors to the globular cluster population in the Virgo core. If UCDs are the surviving nuclei of tidally stripped galaxies, then it is expected that the globular cluster population associated with those galaxies now resides in the core of the cluster.  Indeed, it has been recognized for some time that the accretion and disruption of low-mass galaxies is likely one of the most important mechanisms by which galaxies assemble their globular cluster systems (e.g., C\^ot\'e et al. 1998, 2002; Beasley et al. 2002; Jord\'an et al. 2004; Tonini 2013). Using the extensive NGVS imaging data in the Virgo core region, which allows us to assemble complete and homogenous samples of galaxies, UCDs and globular clusters in this environment, we explore this scenario by noting that the globular cluster `specific frequency' $S_N = N_{GC}\times10^{0.4(M_V+15)}$ --- defined as the number of globular clusters $N_{GC}$ normalized to a $V-$band magnitude $M_V = -15$ --- varies as a function of galaxy luminosity, but is $S_N \sim 3$ (albeit with large scatter) in the magnitude range occupied by the presumed progenitors of the existing UCDs (see, e.g., Peng et al. 2008 or Harris et al. 2013). For this choice of $S_N$, the UCD progenitors are expected to have contributed a total of roughly $1950$ globulars to the core of the cluster. Most of these clusters come from the brightest progenitors: e.g., a $M_g =-18.8$ mag progenitor will have contributed $\sim 120$ globulars, while a $M_g =-14.5$ mag progenitor is expected to have contributed only $\sim 2$ to 3 globulars. In addition, most accreted clusters are expected to belong to the blue globular cluster population (C\^ot\'e et al. 1998; Peng et al. 2006; Tonino 2013).

There is a second way in which the stripping of galaxies can affect the present-day population of globular clusters (or objects identified as such). As mentioned above, a faint magnitude cutoff of $M_g = -9.6$  was imposed on the UCD population selected from the NGVS (Liu et al. 2015a). This was solely out of necessity: i.e., one of the criteria used in selecting UCDs is their larger size (compared to globulars), and sizes cannot be measured reliably fainter than $M_g \simeq -9.6$ mag using NGVS data (Liu et~al. 2015a). Fainter objects are therefore excluded {\it a priori} when selecting UCDs, but they do exist in the core region and, when detected, would invariably be classified as globular clusters. The progenitors of such faint UCDs are expected to be fainter than the progenitors of the UCDs identified by Liu et al (2015a). To estimate their numbers, we can proceed in two separate ways. The first, probably incorrect method, is to assume that the luminosity function of the UCD progenitors and galaxies has a Schechter-like form, with a slope of $\alpha = -1.53 \pm 0.07$  (see Table 1), all the way down to the 50\% completeness limit of the NGVS, $M_g = -9.13$ mag. The integral of this luminosity function over the range $-14.5 \leq M_g \leq -9.13$ mag then yields the total number of existing galaxies plus disrupted UCD progenitors: $2150^{+946}_{-638}$, where the error reflects the uncertainty in the slope of the luminosity function. From this number, we need to subtract the number of existing galaxies in the same magnitude range: correcting the number of observed galaxies (297) for completeness gives 384 galaxies that have survived to the present day. This implies that $\sim 1766^{+946}_{-638}$ galaxies have been disrupted. Using the relation between nuclear and galaxy magnitude from S\'anchez-Janssen et al. (2016b), the nuclei of galaxies in the $-14.5 \leq g \leq -9.13$ mag range have magnitudes $-9.6 \leq g \leq -6.7$. This falls squarely in the magnitude range dominated by globular clusters, whose luminosity function can be described as a Gaussian with turnover magnitude $M_{g,T} \simeq -7.22$~mag and dispersion $\sigma \simeq 1.28$ mag (see, e.g., Villegas et al. 2010). The progenitors of these galaxies are also expected to contribute their globular cluster populations to the core of Virgo. The globular cluster specific frequency is not well quantified for early-type galaxies in this magnitude range (e.g. Georgiev et al. 2010), but assuming $S_N = 3$ would lead to a total of $\sim 280$ additional globular clusters. 

As already noted, the above arguments are likely incorrect at some level, since there is no reason to believe that the combined luminosity function of UCD progenitors and galaxies is well described by a Schechter function all the way down to the completeness limit of the survey --- a factor 150 beyond the magnitude of the progenitor of the faintest UCD currently detected. In fact, there are reasons to believe that this luminosity function is {\it not} well described by a Schechter function. All UCD progenitors are, by definition, nucleated, but the present-day fraction of nucleated galaxies decreases dramatically as galaxies become fainter, falling to $\sim 1$\% at  $M_g = -9.13$ mag, the 50\%  completeness limit of the survey. If the same nucleation fraction applies to galaxies that have been disrupted (which it might not), then we would expect the luminosity function of the UCD progenitors to have a bell-shaped distribution, following the galaxy luminosity function at the bright end and declining sharply at the faint end. 

\begin{figure}
\epsscale{1.2}
\plotone{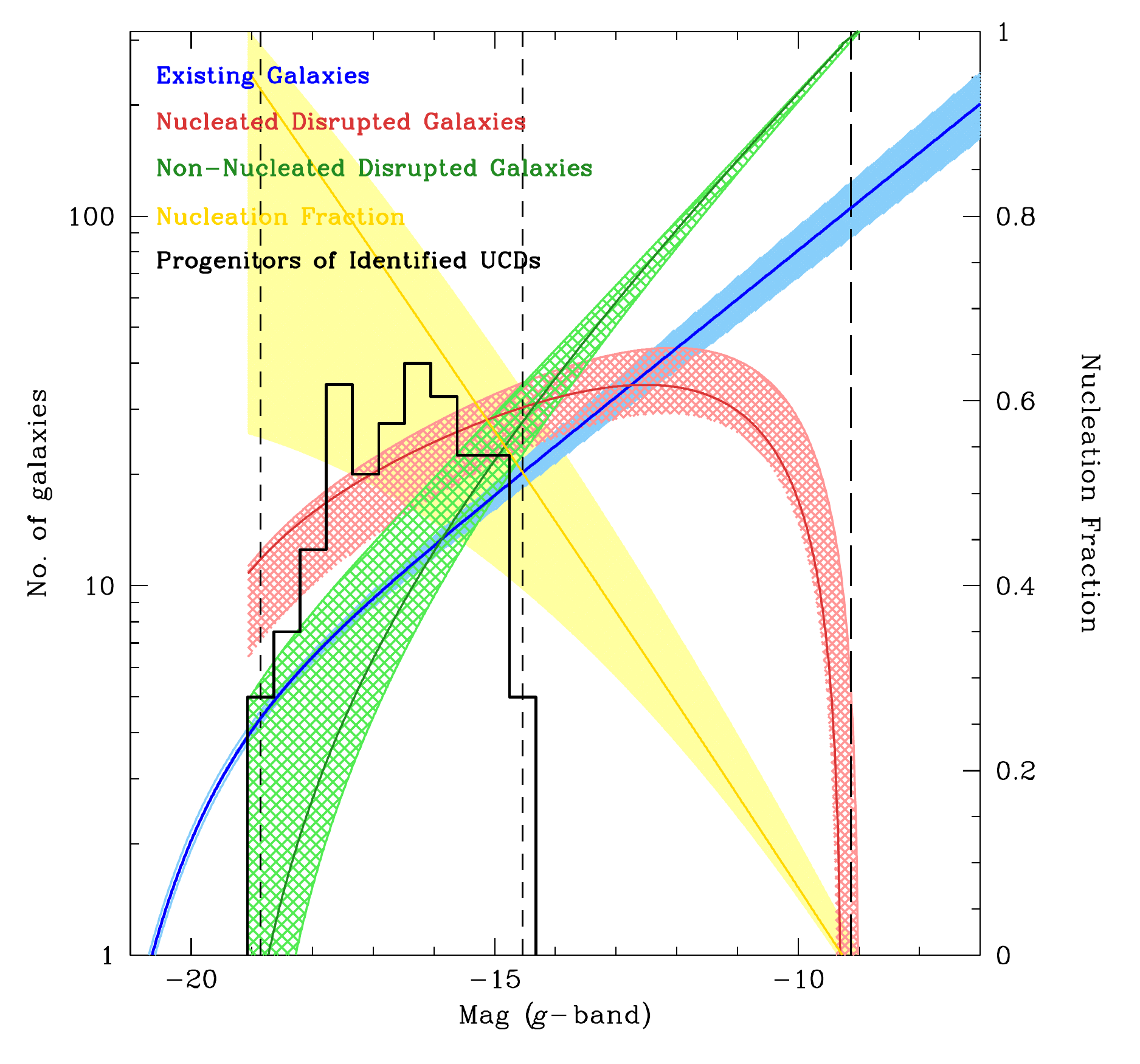}
\caption{Schematic representation of the luminosity function of the current sample of galaxies in the core of Virgo (blue curve). The red and green curves show, respectively,  the nucleated and non-nucleated galaxies that are presumed to have been disrupted, estimated from the observed galaxy luminosity function combined with the nucleation fraction observed at the present time (shown by the yellow wedge, to be read on the scale to the right). The confidence intervals account for the errors on the nucleation fraction. The black histogram shows the luminosity function of the presumed UCD progenitor galaxies, based on the sample of UCDs detected in the Virgo core region (Liu et al. 2015a). The vertical short-dashed lines identify the magnitudes of the brightest and faintest of the progenitors of the observed UCDs, while the long-dashed line shows the 50\% completeness limit of the (observed) galaxy population. See \S\ref{sec:lfucds3} for details.}
\label{fig8}
\end{figure}

\begin{deluxetable*}{lllll}
\tabletypesize{\footnotesize}
\tablewidth{0pc}                        
\tablenum{4}
\tablecaption{UCDs and Globular Estimates from Disrupted Galaxies\label{tab4}}
\tablehead{
\multicolumn{1}{c}{Magnitude Range} &
\multicolumn{1}{c}{Galaxy Type} &
\multicolumn{1}{c}{No. of Galaxies} &
\multicolumn{1}{c}{No. of UCDs} &
\multicolumn{1}{c}{No. of Associated GCs} \\
\multicolumn{1}{c}{$g-$band} &
\multicolumn{1}{c}{} &
\multicolumn{1}{c}{} &
\multicolumn{1}{c}{}  \\
\multicolumn{1}{c}{(1)} &
\multicolumn{1}{c}{(2)} &
\multicolumn{1}{c}{(3)} &
\multicolumn{1}{c}{(4)} &
\multicolumn{1}{c}{(5)} 
}
\startdata
$-$18.8 to $-$14.5 & Nucleated        & 92                           & 92                          & 1950                         \\
$-$14.5 to $-$9.13 & Nucleated        & $148^{+49}_{-24}$ & $148^{+49}_{-24}$& $77\pm16$   \\
$-$18.8 to $-$9.13 & Nucleated        & $240^{+49}_{-24}$ & $240^{+49}_{-24}$ & $2027\pm16$  \\
$-$18.8 to $-$14.5 & Non-Nucleated & $43^{+28}_{-11}$   & 0 & $491^{+724}_{-164}$ \\
$-$14.5 to $-$9.13 & Non-Nucleated & $650^{+26}_{-49}$  & 0 & $125^{+16}_{-18}$ \\
$-$18.8 to $-$9.13 & Non-Nucleated & $693^{+54}_{-60}$ & 0 &  $616^{+740}_{-182}$ \\
$-$18.8 to $-$9.13 & All & $933^{+103}_{-84}$ & $240^{+49}_{-24}$ &  $2643^{+756}_{-198}$ 
\enddata
\tablecomments{Column 1 lists the magnitude range over which the total number of disrupted satellites (column 3), their nuclei (present-day UCDs, column 4) and globular clusters (column 5) is calculated. Whether the galaxies are nucleated, non-nucleated, or include both categories is listed in column 2. By definition, non-nucleated galaxies do not produce UCDs, but they do contribute globular clusters to the core of Virgo (the calculation assume a luminosity independent globular cluster specific frequency $S_N = 3$). The total number of nucleated disrupted galaxies in the magnitude range $-18.8 \leq M_g \leq -14.5$ is set equal to the number of observed UCDs from Liu et al. (2015a). All other numbers are calculated as described in \S\ref{sec:lfucds3}. For comparison, the core region contains 92 known UCDs, $\sim$ 11\,000 globular clusters (of which $\sim$ 7300 belonging to the blue population), and 346 galaxies with $-18.8 < M_g <  -9.13$. At face value, this implies that $\sim 150$ of the globular clusters in the core of Virgo are in fact UCDs, that as many as one third of the blue globular clusters once belonged to disrupted satellites, and that for every galaxy surviving to the present day, three times as many have been disrupted.}
\end{deluxetable*}

This is shown schematically in Figure 7. The blue line represents the observed galaxy luminosity function (see \S\ref{sec:lfvirgo}), parameterized as a Schechter function with $\alpha = -1.33 \pm 0.02$. The yellow wedge shows the constraints on the nucleation fraction (to be read on the scale to the right) measured for present-day galaxies in the core of Virgo (S\'anchez-Janssen et al. 2016b). We note that, as discussed in S\'anchez-Janssen et al. (2016b), the steep drop in nucleation fraction at faint magnitudes is real and not due to observational biases: throughout the entire magnitude range spanned by the NGVS galaxies, the survey's point source sensitivity is at least two magnitudes fainter than the average nuclear magnitude expected based on the observed correlation between the magnitude of the nucleus, and that of its host galaxy. The red curve is meant to represent the luminosity function of nucleated galaxies that have been disrupted. Under the assumption that the nucleation fraction at earlier times was the same as we measure today, this is equal to the galaxy luminosity function multiplied by the nucleation fraction and normalized to the number of progenitors of the observed UCDs in the magnitude range $-18.8 \leq M_g \leq -14.5$. The green curve represents the luminosity function of disrupted non-nucleated galaxies under the same assumption for the nucleation fraction. For comparison, the black histogram shows the inferred progenitors of the observed UCDs.  Qualitatively, the UCD progenitor histogram agrees well with the red curve, especially when one considers that the UCD sample might be incomplete at the faint end. 

Proceeding with this cartoon picture, we can estimate the total number of UCDs in the $-9.6 \leq M_g \leq -6.7$ mag range by integrating the red curve (i.e. the luminosity function of disrupted nucleated galaxies) over the range $-14.5 \leq M_g \leq -9.13$ mag (corresponding to the magnitudes of the progenitors). This yields a total of $148^{+49}_{-24}$ UCDs. Additionally, these galaxies would have deposited $77 \pm 16$ globular clusters in the core region, assuming $S_N = 3$. Adding to this the 1950 globular clusters expected to have been associated with the brighter UCD progenitors (see above) leads to a total of $2175^{+65}_{-40}$ objects belonging to the present-day globular cluster population that were once associated with disrupted nucleated galaxies. Of these, $148^{+49}_{-24}$ are, in fact, the surviving nuclei of those galaxies. 

Finally, we can estimate the overall contribution from disrupted, non-nucleated galaxies. By definition, these galaxies would not contribute UCDs but would still contribute globular clusters. Integrating the green curve over the range $-18.8 \leq M_g \leq -9.13$ mag leads to a total of $693^{+54}_{-60}$ (now disrupted) non-nucleated galaxies, contributing $616^{+740}_{-182}$ globular clusters to the present-day population. In total, the number of globular clusters contributed by disrupted galaxies of all types is then $2791^{+705}_{-222}$, of which $148^{+49}_{-24}$  (5\%) would be faint UCDs ``masquerading'' as globular clusters. Table 4 summarizes these estimates  for the number of UCDs and globular clusters that may have originated from disrupted galaxies. 

We can now compare this number to the total number of globular clusters present in the Virgo core with $-9.6 \leq M_g \leq -6.7$ mag, under the assumption that the clusters of the disrupted galaxies still reside in the core region. A clean globular cluster catalog can be derived from NGVS data down to $M_g = -7.49$ mag (Durrell et al. 2014; Peng et al. in preparation; note that this is over two magnitudes brighter than the 10$\sigma$ point source detection limit, the main limitation in identifying globular clusters being confusion with the stellar locus). To obtain the total numbers down to $M_g = -6.7$ mag, we assume a Gaussian luminosity function with parameters described in Villegas et al. (2010). This leads to a total of $\sim 11\,000$ globulars that are currently found within the core of Virgo. Liu et al. (2015a), meanwhile, found from their colors that UCDs appear to be more closely associated with the blue, rather than the red, globular cluster subpopulation. Using a double Gaussian fitted to the observed color distribution allows us to divide the observed globular cluster sample into blue and red components, for a total of 7300 and 2800 objects, respectively. It then follows that, in this picture, $(2791^{+705}_{-222})/7300 = 39 ^{+9}_{-3}$\% of the blue globular clusters were formerly associated with (mostly bright and nucleated) galaxies that have been disrupted. About 2\% of these ``globulars'' would correspond to the nuclei of disrupted nucleated galaxies.

Needless to say, these results could have important implications for the origin of the diffuse light in the cluster core (see, e.g., Cooper et al. 2015 and references therein). Constraints on the fraction of intracluster light in Virgo are both uncertain and model dependent, but range between 7 and 20\% (e.g., Durrell et al. 2002; Dolag et al. 2010; Murante et al. 2007; Purcell et al. 2007; Seigar et al. 2007; Gonzalez et al. 2007; Zibetti 2008; Feldmeier et al. 2004; Castro-Rodriguez et al. 2009). However, these estimates should probably be viewed as lower limits for the core, given the large amount of diffuse light within this region (e.g., Mihos et al. 2005; Rudick et al. 2010).
Integrating the luminosity function of disrupted nucleated and non-nucleated galaxies (the red and green curves in Figure 7)  gives ${\cal L}_{\rm dis,n} \sim 6.5^{+0.5}_{-2.3}\times10^{10}$ ${\cal L}_{\sun}$ and ${\cal L}_{\rm dis,nn} \sim 1.9^{+2.3}_{-0.5}\times10^{10}$ ${\cal L}_{\sun}$, respectively, for a combined luminosity from disrupted galaxies of ${\cal L}_{\rm dis} = (8.5 \pm 2.8)\times10^{10}$ ${\cal L}_{\sun}$. Only a fraction of this luminosity contributes to the intracluster light observed today,  while the remainder is likely to have been incorporated into the haloes of giant ellipticals: for comparison, the luminosity of M87 is ${\cal L}_{\rm M87} = (1.0 \pm 0.1) \times10^{11}$ ${\cal L}_{\sun}$ and our full sample of 404 galaxies in the core region have a collective luminosity of ${\cal L}_{\rm tot} \sim 2\times10^{11}$ ${\cal L}_{\sun}$ (Ferrarese et~al. 2016). Taken at face value, the above results would suggest that a very substantial population of disrupted galaxies --- two to three times as many as the $\sim 400$ galaxies that we presently observe in the cluster core --- might be responsible for up to $\sim 85\%$ of the present-day luminosity of M87 or, alternatively, $\sim 40\%$ of the aggregate luminosity in galaxies residing in this region.

\begin{deluxetable}{lrrl}
\tabletypesize{\footnotesize}
\tablewidth{0pc}                        
\tablenum{5}
\tablecaption{Local Group Sample\label{tab5}}
\tablehead{
\multicolumn{1}{c}{Name} &
\multicolumn{1}{c}{$(m-M)$ (mag)} &
\multicolumn{1}{c}{$M_V$ (mag)} &
\multicolumn{1}{c}{References}\\
\multicolumn{1}{c}{(1)} &
\multicolumn{1}{c}{(2)} &
\multicolumn{1}{c}{(3)} &
\multicolumn{1}{c}{(4)} 
}
\startdata
\multicolumn{4}{l}{{\it M31 System}}\\
Andromeda & \nd  & $-$21.20 $\pm$ 0.80 &  1,2 \\
Triangulum & 24.54 $\pm$ 0.06 & $-$18.84 $\pm$ 0.12 & 1,2 \\
NGC 205 & 24.58 $\pm$ 0.07 & $-$16.48 $\pm$ 0.12 &  1,2\\
M32 & 24.53 $\pm$ 0.21 & $-$16.43 $\pm$ 0.23 &  2,3,4\\
IC 10 & 24.50 $\pm$ 0.09 & $-$15.00 $\pm$ 0.23 &  2,3,5,6\\
NGC 185 & 23.95 $\pm$ 0.09 & $-$14.75 $\pm$ 0.13 & 1,2 \\
NGC 147 & 24.15 $\pm$ 0.09 & $-$14.65 $\pm$ 0.13 & 1,2 \\
Andromeda VII & 24.41 $\pm$ 0.10 & $-$13.21 $\pm$ 0.32 & 1,7,8 \\
Andromeda II & 24.07 $\pm$ 0.14 & $-$12.57 $\pm$ 0.21 &  1,7,8\\
Andromeda XXXII & 24.45 $\pm$ 0.14 & $-$12.25 $\pm$ 0.71 &  9\\
Andromeda I & 24.36 $\pm$ 0.07 & $-$11.86 $\pm$ 0.12 &  1,7,8\\
Andromeda XXXI & 24.40 $\pm$ 0.12 & $-$11.70 $\pm$ 0.71 &  9\\
Andromeda VI & 24.47 $\pm$ 0.07 & $-$11.47 $\pm$ 0.21 &  1,8\\
Andromeda XXXIII & 24.49 $\pm$ 0.18 & $-$10.34 $\pm$ 0.72 &  10\\
Andromeda XXIII & 24.43 $\pm$ 0.13 & $-$10.23 $\pm$ 0.52 &  11\\
Andromeda III & 24.37 $\pm$ 0.07 & $-$10.17 $\pm$ 0.31 &  1,7,8\\
LGS 3 & 24.43 $\pm$ 0.07 & $-$10.13 $\pm$ 0.12 &  1,12\\
Andromeda XXI & 24.59 $\pm$ 0.06 & $-$9.79 $\pm$ 0.60 &  13,14\\
Andromeda XXV & 24.55 $\pm$ 0.12 & $-$9.75 $\pm$ 0.51 &  11\\
Andromeda V & 24.44 $\pm$ 0.08 & $-$9.54 $\pm$ 0.22 &  1,8\\
\multicolumn{4}{c}{}\\
\multicolumn{4}{l}{{\it Milky Way System}}\\
MilkyWay &  \nd  &  $-$20.90 $\pm$  0.80  &  \\
LMC & 18.52 $\pm$ 0.09 & $-$18.12 $\pm$ 0.13 & 2,15 \\
SMC & 19.03 $\pm$ 0.12 & $-$16.83 $\pm$ 0.23 &  2,16\\
Canis Major & 14.29 $\pm$ 0.30 & $-$14.39 $\pm$ 0.85 & 17,18\\
Sagittarius dSph & 17.10 $\pm$ 0.15 & $-$13.50 $\pm$ 0.34 &  19,20,21,22\\
Fornax & 20.84 $\pm$ 0.18 & $-$13.44 $\pm$ 0.35 &  23,24\\
Sculptor & 19.67 $\pm$ 0.14 & $-$11.07 $\pm$ 0.52 &  24,25\\
Leo II & 21.84 $\pm$ 0.13 & $-$9.84 $\pm$ 0.33 &  24,26\\
\multicolumn{4}{c}{}\\
\multicolumn{4}{l}{{\it Additional Local Group Galaxies}}\\
NGC 6822 & 23.31 $\pm$ 0.08 & $-$15.21 $\pm$ 0.22 &  2,27\\ 
IC 1613 & 24.39 $\pm$ 0.12 & $-$15.19 $\pm$ 0.16 &  2,28\\  
WLM & 24.85 $\pm$ 0.08 & $-$14.25 $\pm$ 0.13 &  1,2\\ 
Pegasus dIrr & 24.82 $\pm$ 0.07 & $-$12.22 $\pm$ 0.21 &  1,2\\ 
Leo A & 24.51 $\pm$ 0.12 & $-$12.11 $\pm$ 0.23 &  2,29,30,31\\ 
Leo I & 22.02 $\pm$ 0.13 & $-$12.02 $\pm$ 0.33 &  24,32\\   
Cetus & 24.39 $\pm$ 0.07 & $-$11.29 $\pm$ 0.21 &  1,8\\ 
Aquarius & 25.15 $\pm$ 0.08 & $-$10.65 $\pm$ 0.13 & 1,33\\ 
Phoenix & 23.09 $\pm$ 0.10 & $-$9.89 $\pm$ 0.41 &  34,35\\ 
Tucana & 24.74 $\pm$ 0.12 & $-$9.54 $\pm$ 0.23 &  36,37    
\enddata                        
\tablecomments{{\scriptsize The Table lists all Local Group satellites (column 1) brighter than $M_g = -9.13$ mag, the 50\% completeness limit of the NGVS Virgo sample. Distance moduli, $V-$band absolute magnitudes and references are given in cols. 3, 4 and 5, respectively. In the text, we assume $g-V = 0.14$ mag to estimate the $g-$band luminosity function. References are as follows:
(1) McConnachie, A.~W.,  et al. 2005;
(2) de Vaucouleurs,  et al. 1991;
(3) Huchra, J.~P.,  et al. 1999;
(4) Grillmair, C.~J.,  et al. 1996;
(5) Tully, R.~B.,  et al. 2006;
(6) Sanna, N.,  et al. 2010;
(7) Kalirai, J.~S.,  et al. 2010;
(8) McConnachie, A.~W. \& Irwin, M.~J. 2006;
(9) Martin, N.~F.,  et al. 2013;
(10) Martin, N.~F.,  et al. 2013;
(11) Richardson, J.~C.,  et al. 2011;
(12) Lee, M.~G. 1995;
(13) Martin, N.~F.,  et al. 2009;
(14) Conn, A.~R.,  et al. 2012;
(15) Clementini, G.,  et al. 2003;
(16) Udalski, A.,  et al. 1999;
(17) Bellazzini, M.,  et al. 2006;
(18) Bellazzini, M.,  et al. 2004;
(19) Monaco, L.,  et al. 2004;
(20) Ibata, R.~A.,  et al. 1994;
(21) Mateo, M.,  et al. 1998;
(22) Majewski, S.~R.,  et al. 2003;
(23) Pietrzy{\' n}ski, G.,  et al. 2009;
(24) Irwin, M. \& Hatzidimitriou, D. 1995;
(25) Pietrzy{\' n}ski, G.,  et al. 2008;
(26) Bellazzini, M.,  et al. 2005;
(27) Gieren, W.,  et al. 2006;
(28) Bernard, E.~J.,  et al. 2010;
(29) Dolphin, A.~E.,  et al. 2002;
(30) Vansevi{\v c}ius, V.,  et al. 2004;
(31) Cole, A.~A.,  et al. 2007;
(32) Bellazzini, M.,  et al. 2004;
(33) McConnachie, A.~W.,  et al. 2006;
(34) Hidalgo, S.~L.,  et al. 2009;
(35) van de Rydt, F., et al. 1991;
(36) Bernard, E.~J.,  et al. 2009;
(37) Saviane, I.,  et al. 1996.}}                        
\end{deluxetable}

\section{A Comparison to the Luminosity Function in the Local Group}
\label{sec:lflg}
 
The number of known Local Group galaxies continues to rise in the era of deep, pan-chromatic, wide-field surveys such as the SDSS (York et al. 2000), Pan-STARRS (Kaiser et al. 2010), the Dark Energy Survey (DES; Diehl et al. 2014) and PAndAS (McConnachie et al. 2009). A complete catalog of Local Group galaxies discovered up to 2012 was compiled by McConnachie (2012)\footnote{An updated list is maintained online at {\tt http://www.astro.uvic.ca/$\sim$alan/Nearby\_Dwarf\_Database.html}}; since then, four new M31 satellites have been discovered: Andromeda  XXXI, Andromeda XXXII and Andromeda XXXIII from Pan-STARRS data (Martin et al., 2013a,b) and Andromeda  XXX from PAndAS data (Conn et al. 2012, Collins et al. 2013, Irwin et al., in prep.). More dramatically, 23 new ``ultra-faint'' satellites have since been discovered in the immediate vicinity of the Milky Way. Four of these detections were made using Pan-STARRS data (Laevens et al. 2015a, 2015b, Martin et al. 2015) while the remaining objects were discovered in the first and second year of DES (Koposov et al. 2015; Kim \& Jerjen 2015; Kim et al. 2015a,b; Bechtol et al. 2015; Drlica-Wagner et al. 2015). For our purposes --- where we seek a homogeneous comparison of the Local Group and Virgo luminosity functions, based on NGVS data --- we may limit the analysis to Local Group satellites brighter than $M_g = -9.13$ mag, the 50\% detection limit that applies to the Virgo sample. This reduces the Local Group sample to the galaxies listed in Table 5, divided according to whether they are M31 satellites (20), Galactic satellites (8), or members of the Local Group but not associated with either galaxy specifically (10), following the prescription of McConnachie (2012)\footnote{A virial radius of 300 kpc is assumed for both the Milky Way and M31. The zero-velocity radius of the Local Group is taken to be 1060 kpc.}. For comparison with Virgo, we assume  a colour term $g-V = 0.14$ mag to convert the $V-$band magnitudes in Table 5 to $g-$band magnitudes, although, as we will show, the slope is not sensitive to the exact value used for the conversion (or, equivalently, to the value of $M^*$). 

\begin{figure}
\epsscale{1.0}
\plotone{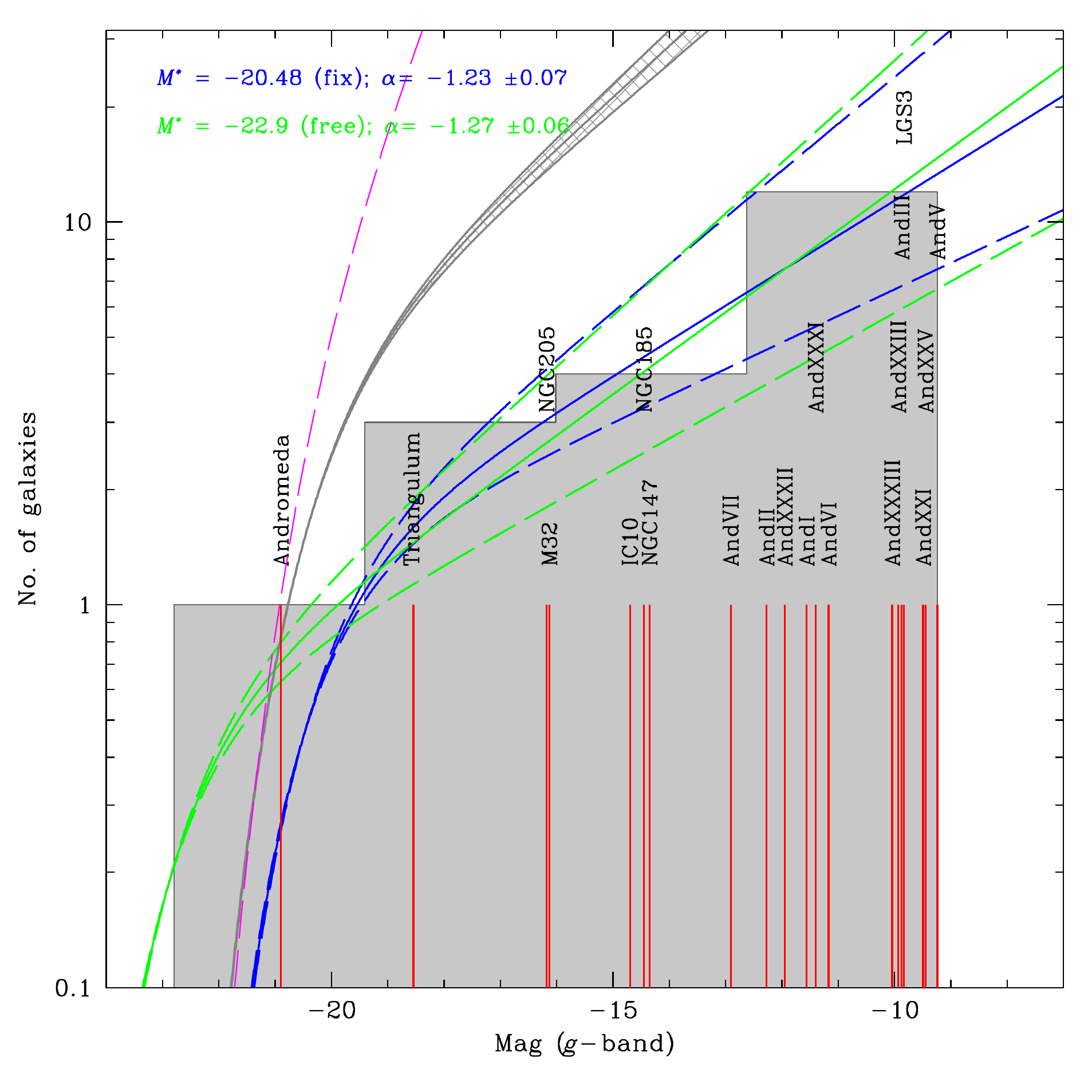}
\caption{Schechter fit to the luminosity function of Andromeda satellites brighter than $M_g = -9.13$ mag (the same magnitude limit as in our NGVS analysis of Virgo core galaxies, see Table 5). Galaxies are individually marked by the vertical red lines, and no correction for completeness has been applied. The green curve, with 1$\sigma$ confidence limits on $\alpha$, shows the best MCMC fit to the individual data when both $M_g^*$ and $\alpha$ are allowed to vary. The blue curve (again with 1$\sigma$ confidence limits) assumes $M_g^* = -20.48$ mag. The grey curve shows the best fit luminosity function measured in Virgo (the blue line in Figure 2), while the magenta dashed line represents a Schechter function with $\alpha = -1.9$, the slope of the $\Lambda$CDM halo mass function.}
\label{fig9}
\vskip .5in 
\epsscale{1.15}
\plottwo{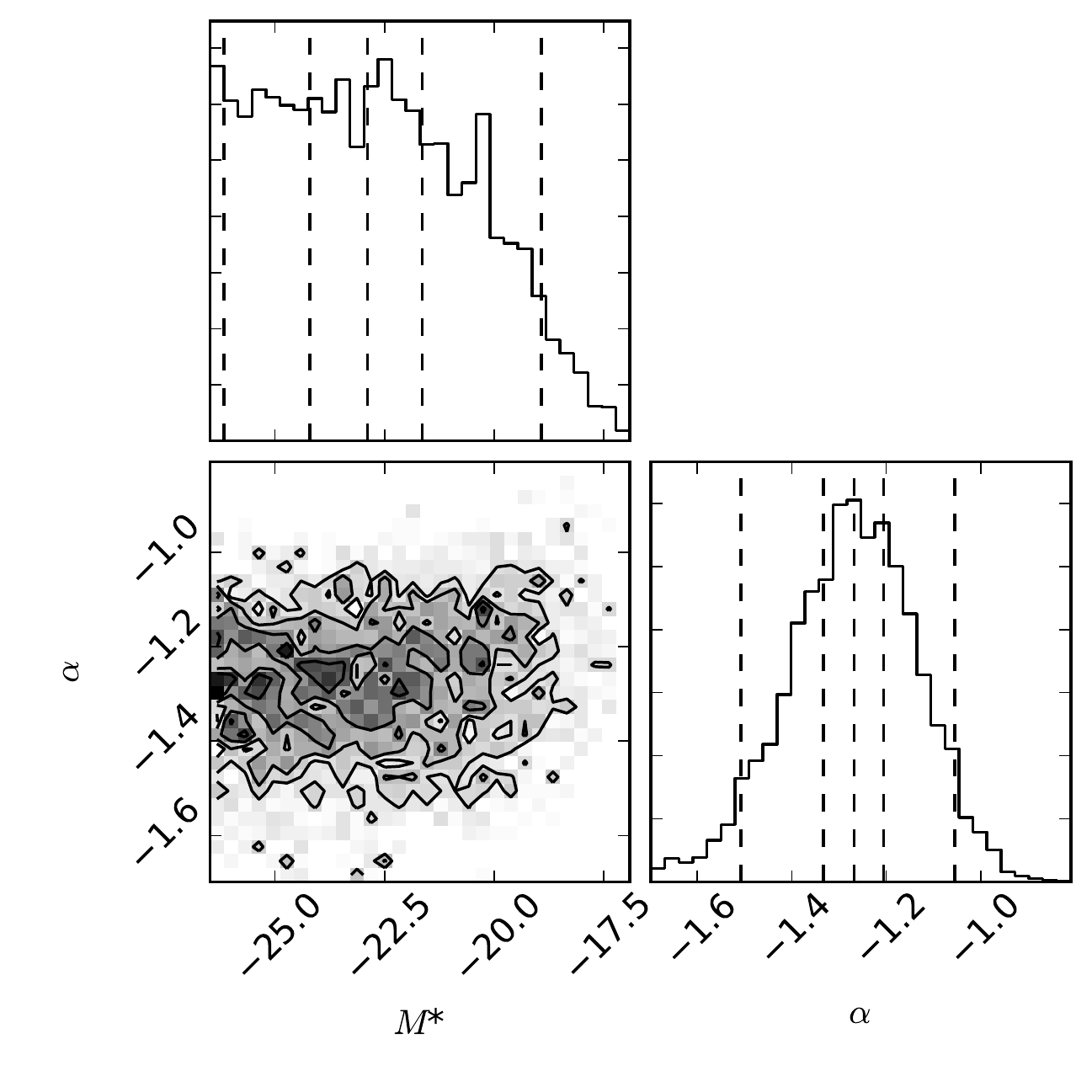}{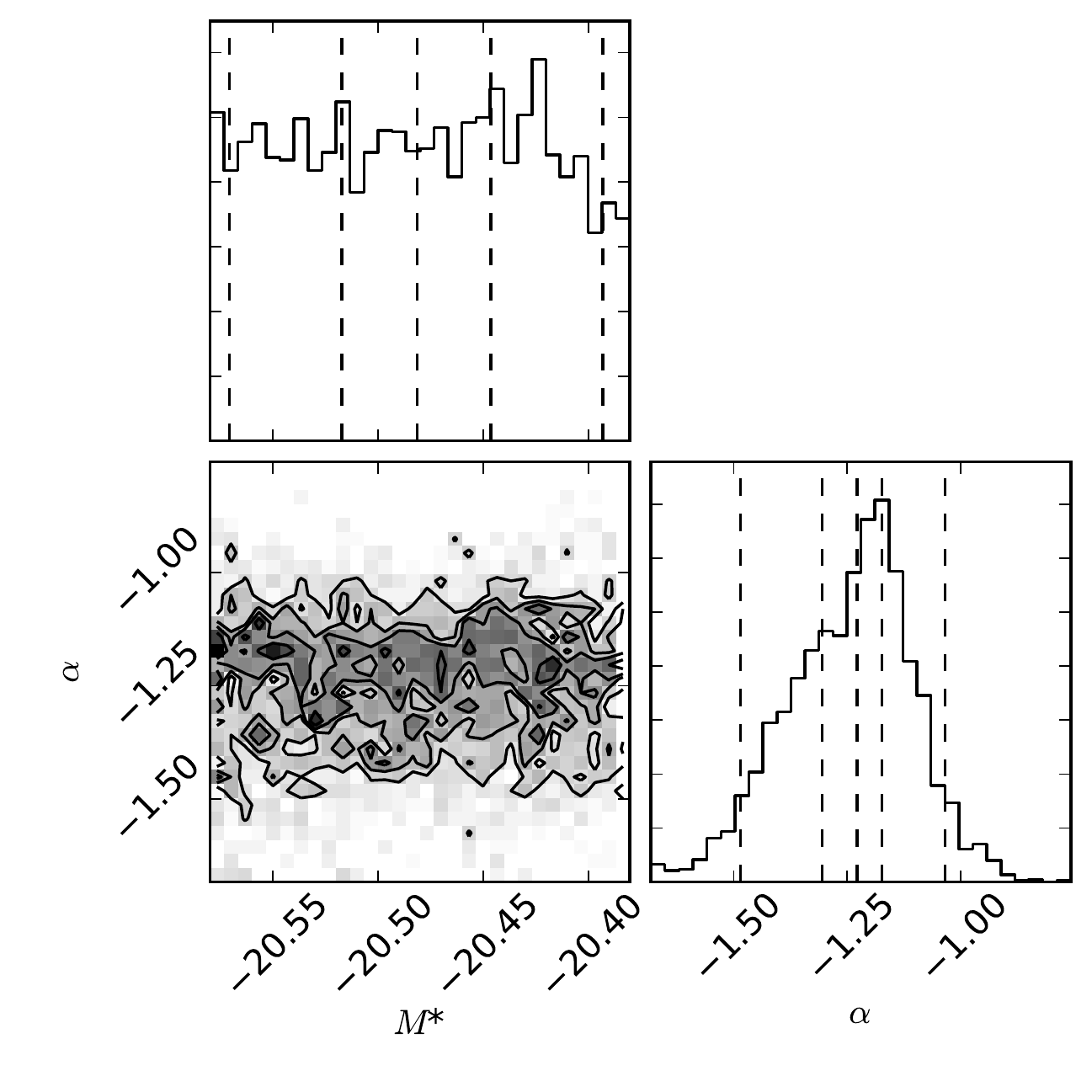}
\caption{Posterior probability density functions for $M_g^*$ and $\alpha$ for  M31 satellites brighter than $M_g = -9.13$ mag (Figure~8). Results are shown for two cases: when $M_g^*$ is allowed to vary ({\it left panel}) and when constrained to a very narrow range around $M^* = -20.48$ mag ({\it right panel}).}
\label{fig10}
\end{figure}

\begin{figure}
\epsscale{1.0}
\plotone{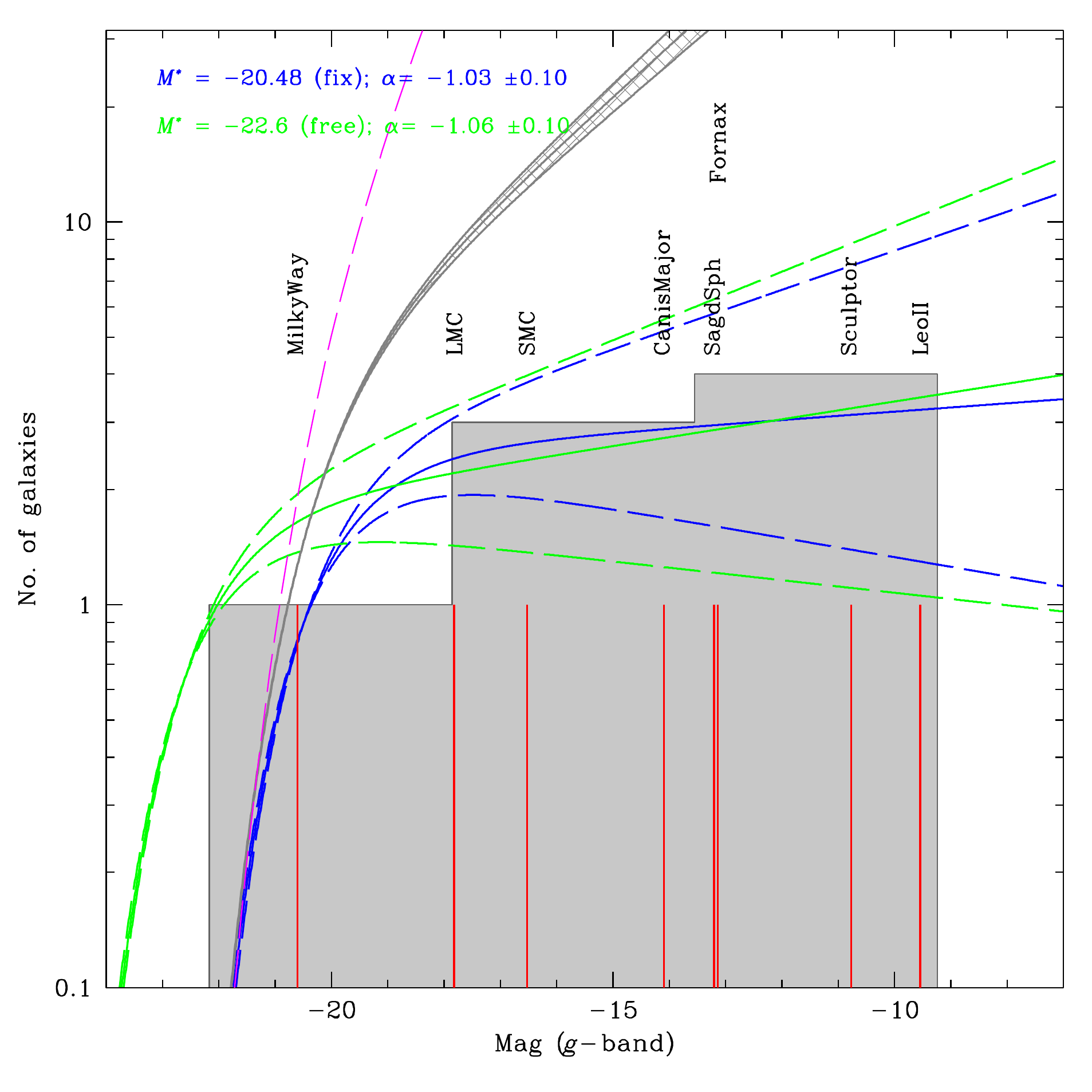}
\caption{Same as Figure 8, except for satellites of the Milky Way brighter than $M_g = -9.13$ mag (see Table 5).}
\label{fig11}
\vskip .5in 
\epsscale{1.15}
\plottwo{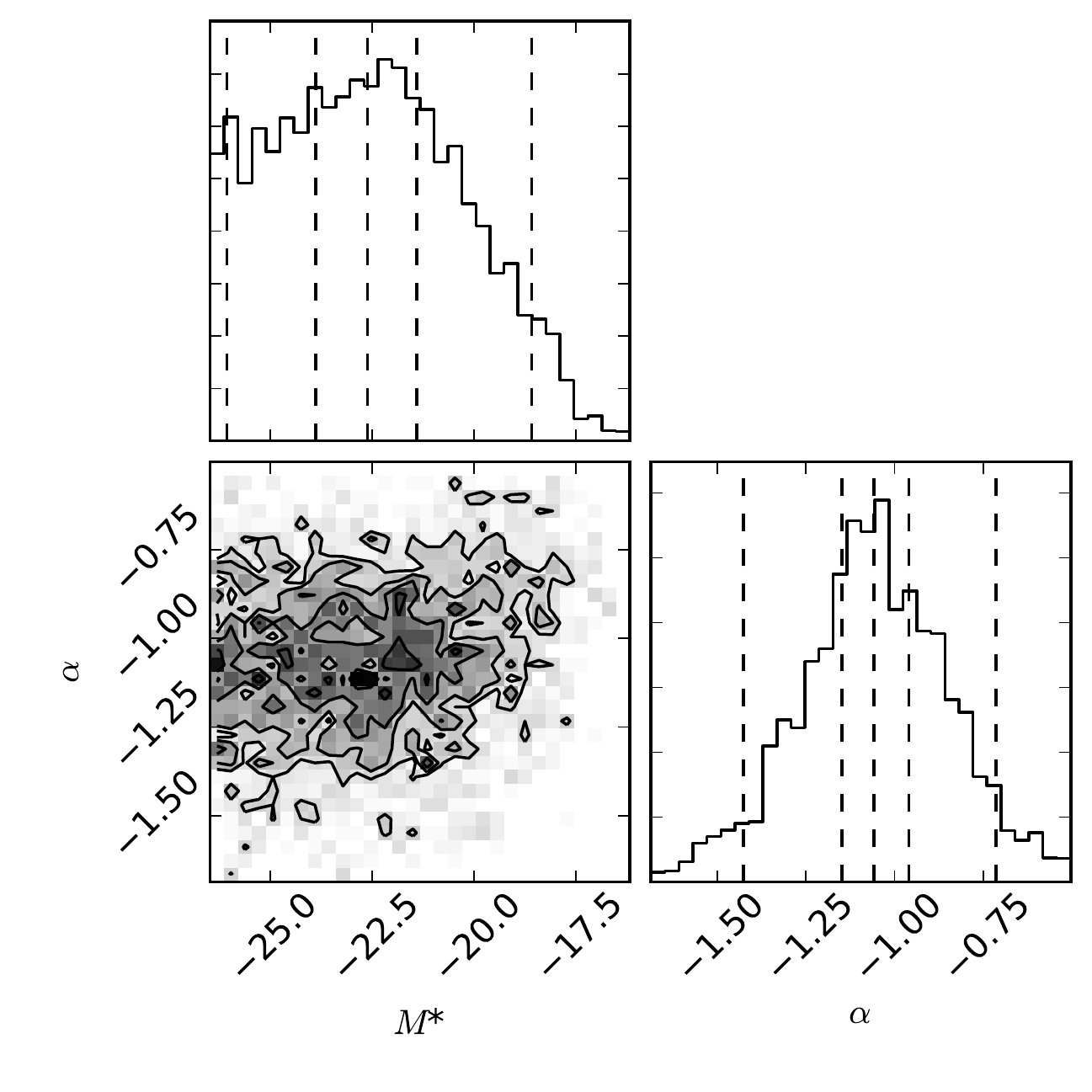}{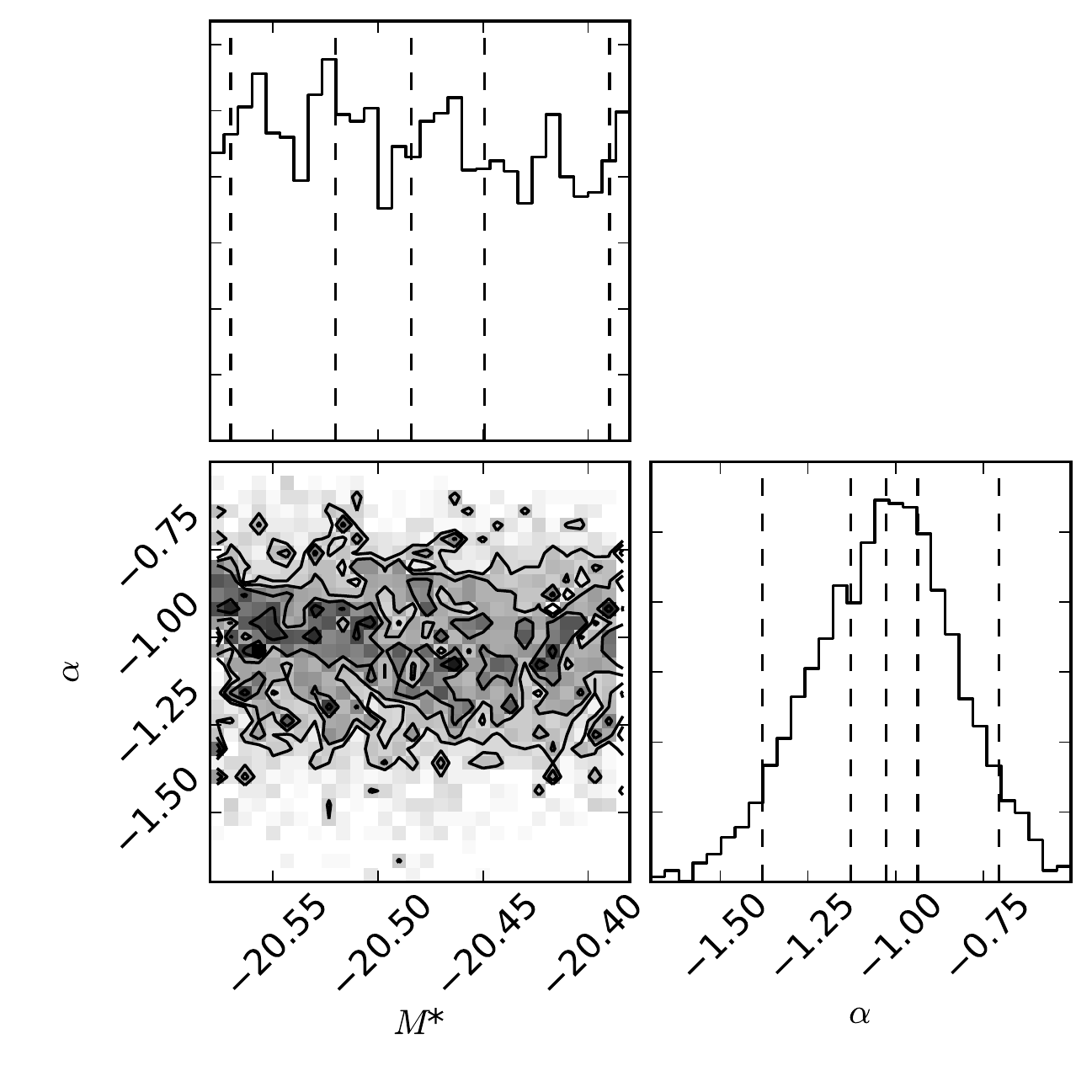}
\caption{Same as Figure 9, except for satellites of the Milky Way brighter than $M_g = -9.13$ mag (see Table 5).}
\label{fig12}
\end{figure}

\begin{figure}
\epsscale{1.0}
\plotone{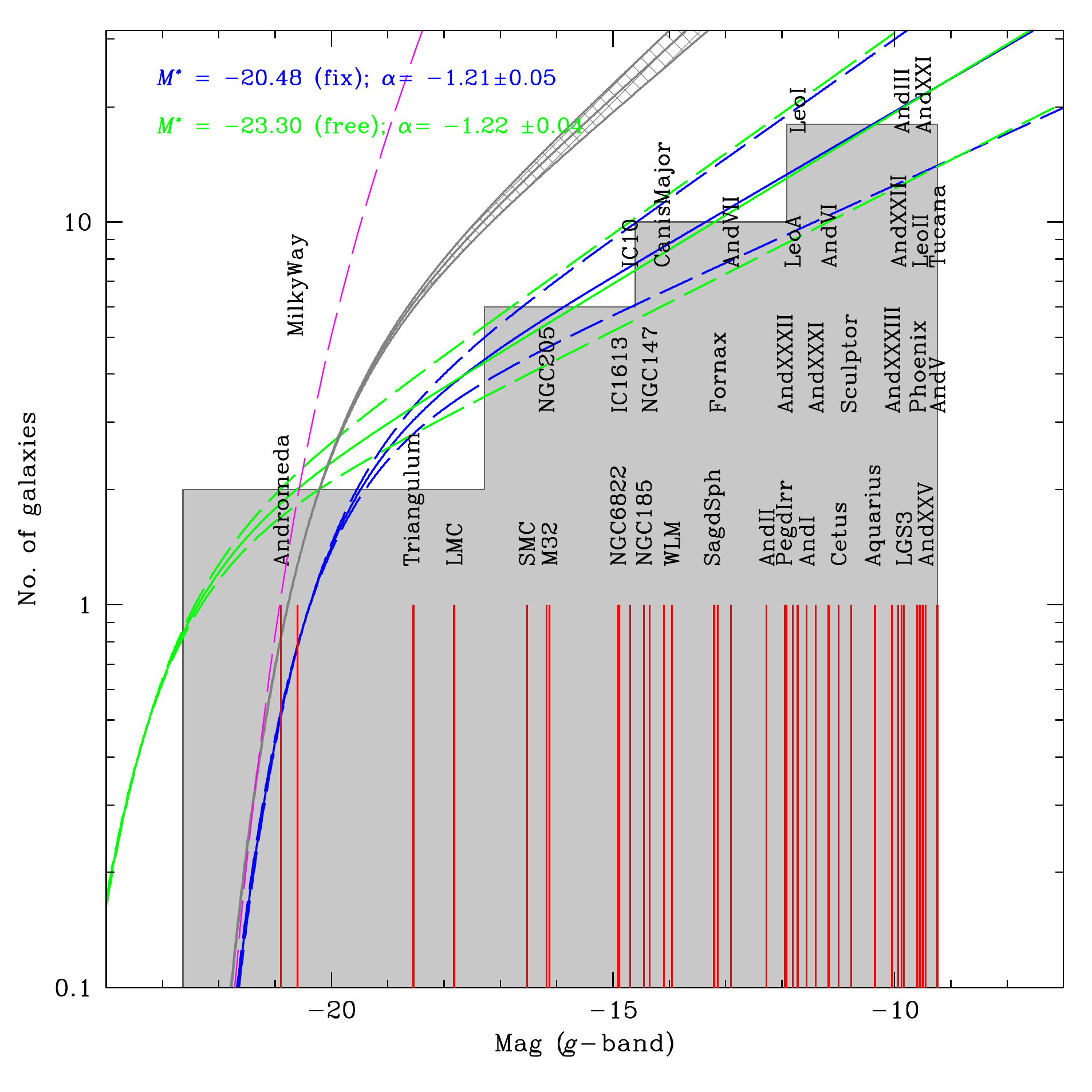}
\caption{As Figure 8, except for all Local Group galaxies brighter than $M_g = -9.13$ mag (see Table 5).}
\label{fig13}
\vskip .5in 
\epsscale{1.15}
\plottwo{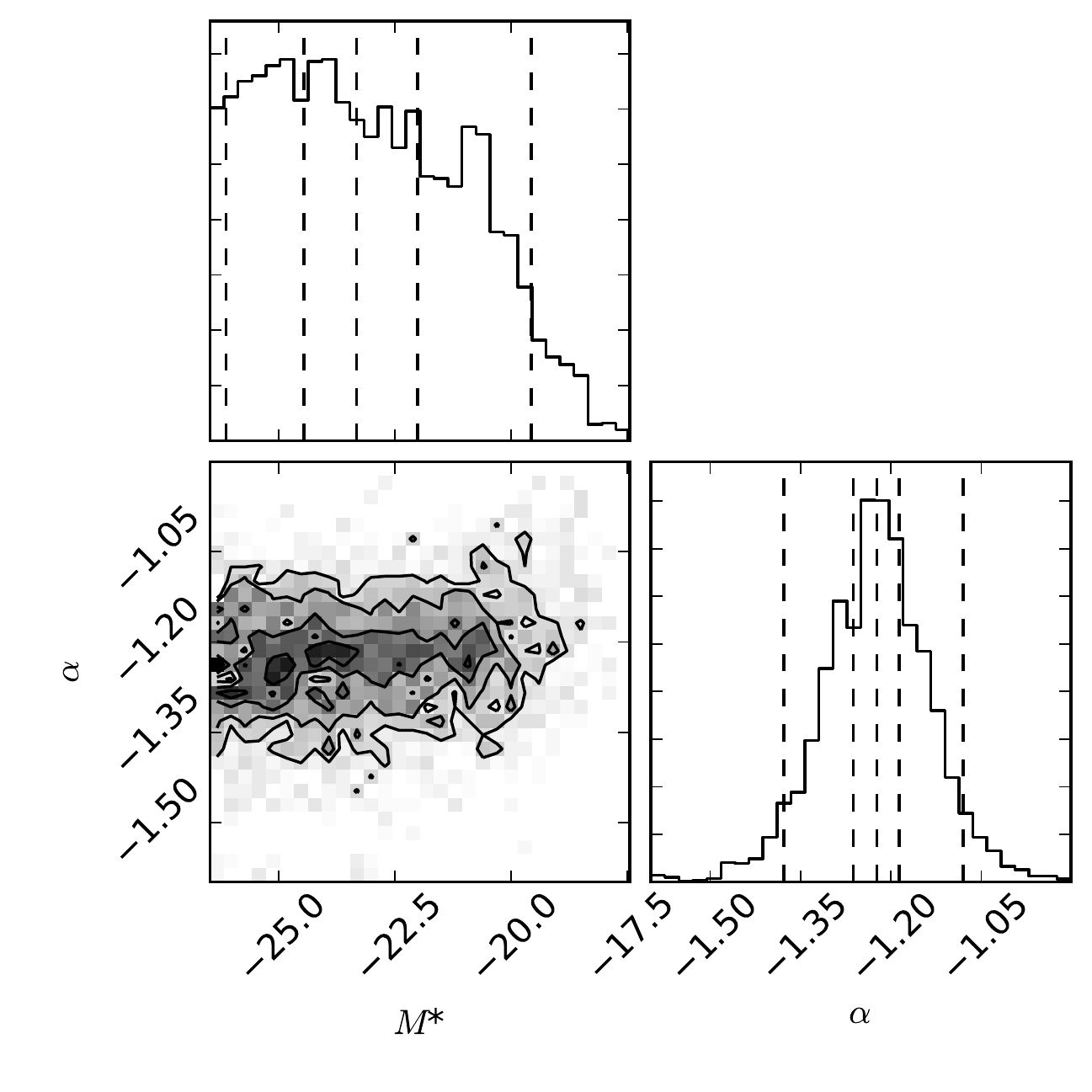}{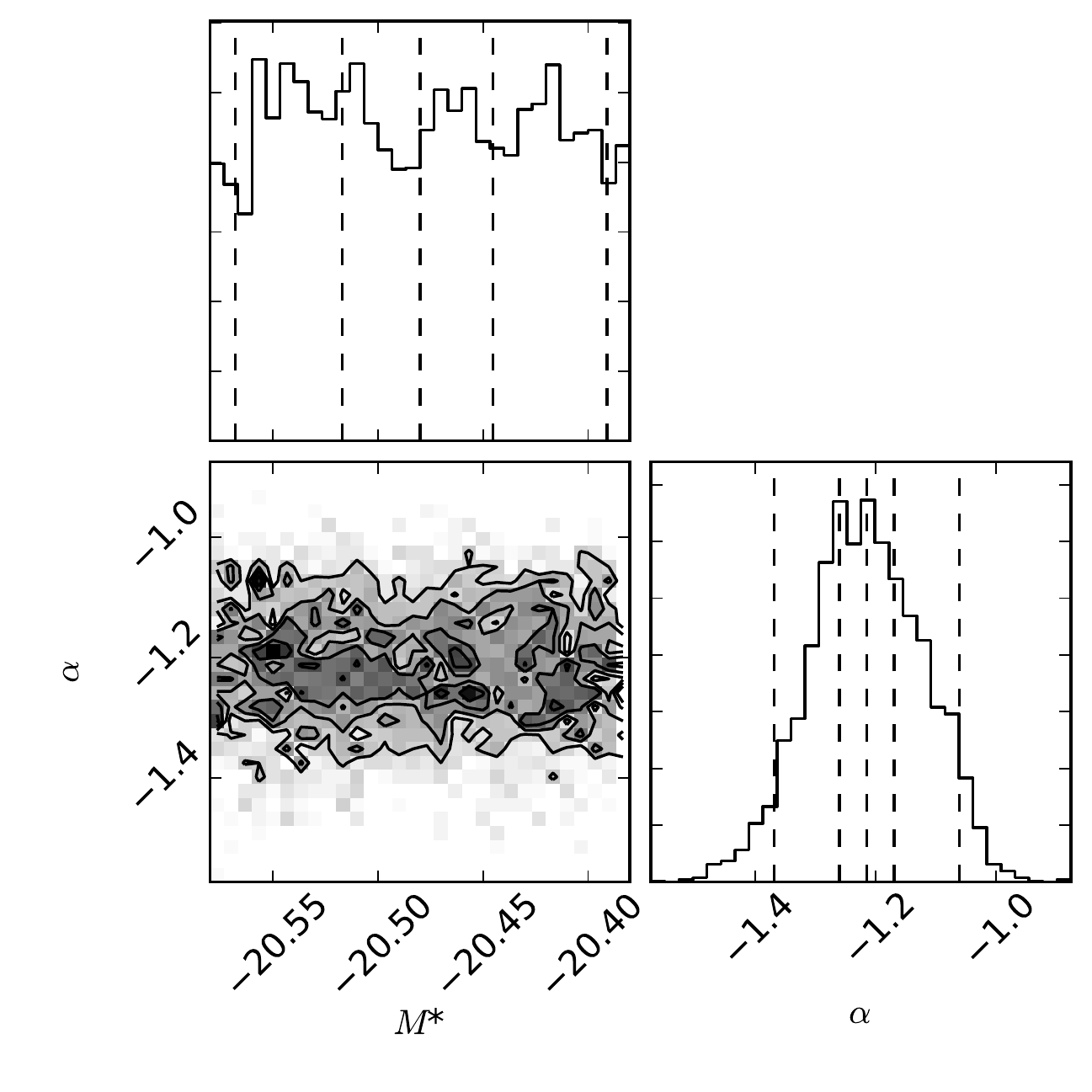}
\caption{As Figure 9, except for all Local Group galaxies brighter than $M_g = -9.13$ mag (see Table 5).}
\label{fig14}
\end{figure}

Unfortunately, completeness for the Local Group sample cannot be quantified as easily and reliably as it can for Virgo. Consider the PAndAS survey, which is complete to at least $M_V = -8$ mag ($M_g \sim -7.8$ mag; McConnachie et al. 2009). Although this limit is $\sim$ 1.3~mag fainter than for our NGVS analysis of Virgo, PAndAS covered Andromeda only out to 150 kpc (half a virial radius) and did not extend North of a Galactic latitude $b \sim -15\deg$ in order to avoid contamination from the Galactic disk. The Pan-STARRS 3$\pi$ survey has made it possible to extend the spatial coverage and reach an approximate completeness limit of $M_r \sim -9$ mag (Martin et al. 2013a,b).  However, the high Galactic latitude area --- which covers ~20\% of Andromeda's virial region --- remains problematic and faint satellites in that region would likely remain undetected. For the Milky Way satellites, completeness is even more problematic, with as many as 20\% to 50\% of Galactic satellites estimated to remain undiscovered, most likely behind the Galactic disk (e.g., Irwin 1994; McConnachie \& Irwin 2006; Tollerud et al. 2008; Koposov et al. 2008). Needless to say, the combined (i.e., Local Group) sample inherits the problems of both subsamples, further compounded by the fact that a truly complete census would require an even larger survey that is sensitive to satellites lying at larger distances. Luckily, since our NGVS luminosity function analysis is based on galaxies with $M_g < -9.13$~mag, for the sake of comparison we can neglect the ``ultra-faint'' galaxies whose completeness is most uncertain for the Local Group and its subsystems.

\begin{deluxetable*}{ll|ll|ll}[H]
\tabletypesize{\footnotesize}
\tablewidth{0pc}                        
\tablenum{6}
\tablecaption{MCMC Fitting Results for the Local Group\label{tab6}}
\tablehead{
\multicolumn{1}{c}{Sample} &
\multicolumn{1}{c}{No.} &
\multicolumn{1}{|c}{$M^*$ (free) (mag)} &
\multicolumn{1}{c}{$\alpha$} &
\multicolumn{1}{|c}{$M^*$ (fix) (mag)} &
\multicolumn{1}{c}{$\alpha$} \\
\multicolumn{1}{c}{(1)} &
\multicolumn{1}{c}{(2)} &
\multicolumn{1}{|c}{(3)} &
\multicolumn{1}{c}{(4)} &
\multicolumn{1}{|c}{(5)} &
\multicolumn{1}{c}{(6)} 
}
\startdata
MW System &   8 & $-22.6^{+1.2}_{-1.3}$  & $-1.06^{+0.10}_{-0.09}$ & $-$20.48  & $-1.03^{+0.10}_{-0.10}$\\
M31 System &  20 & $-22.9^{+1.2}_{-1.3}$  & $-1.267^{+0.063}_{-0.065}$ & $-$20.48  & $-1.228^{+0.055}_{-0.077}$\\
Local Group &  38 & $-23.3^{+1.3}_{-1.1}$  & $-1.223^{+0.037}_{-0.039}$ & $-$20.48  & $-1.214^{+0.045}_{-0.045}$
\enddata
\tablecomments{Results of the MCMC fits to the $g-$band luminosity function of the Local Group (Table 5). The sample is divided in Milky Way galaxies, M31 galaxies, and all Local Group galaxies (as indicated in column 1). The sample size is listed in column 2. Cols. 3 and 4 list the MCMC $M^*$ ($g-$band) and faint-end slope $\alpha$ in the case in which both parameters are allowed to vary. Cols. 5 and 6 list the same parameters in the case when $M^*$ is held fixed to the value determined for local field galaxies by Smith et al. (2009).}
\end{deluxetable*}

We will not attempt to correct for completeness in our estimation of the Local Group luminosity function. Completeness corrections are most severe at fainter magnitudes (i.e., LMC or M32 analogs are exceedingly unlikely to have been missed), and our decision to not account for incompleteness means that the faint-end slopes reported here should, in general, be interpreted as lower limits. Figures 8 to 13 show the MCMC fits and posterior probability density functions for the Andromeda, Milky Way and all Local Group galaxies. In each case, the fits were performed twice --- first by allowing both $M_g^*$ and $\alpha$ to vary as free parameters, and then by fixing the value of $M_g^*$ to $-20.48$ mag. 

The results are tabulated in Table 6. As was the case for the Virgo analysis, $\alpha$ does not appear to be overly sensitive to the precise value of $M_g^*$. For the Andromeda system, the slope,  $\alpha_{\rm M31} = -1.23 \pm 0.07$ is significantly steeper than the value of $-0.98 \pm 0.07$ based on an earlier release of the PAndAS data (McConnachie et al. 2009)\footnote{The difference in slope between our estimate and that of  McConnachie et al. (which was based on galaxies spanning a comparable range in magnitude) is due entirely to the addition of new Andromeda satellites discovered since that study (i.e., 13 galaxies were used in McConnachie et al. 2009, as opposed to the sample of 20 galaxies considered here).}, or the $-1.13\pm 0.06$ measurement of Chiboucas et al. (2009). In fact, our slope for M31 is only $1\sigma$ shallower than measured in Virgo --- and the difference is likely to be even less significant considering that, as already mentioned, the M31 sample might still be somewhat incomplete even down to the (relatively) bright magnitudes considered here.

The luminosity function for the Milky Way system is more uncertain, both because of the smaller number of objects (only eight Galactic satellites are brighter than $M_g = -9.13$ mag), and because of large uncertainties in incompleteness. Our best-fit slope, $\alpha_{\rm MW} = -1.03 \pm 0.10$, is however consistent with the value determined by Koposov et al. (2008) which not only extended to much fainter magnitudes ($M_V = -2$ mag, for a total of 13 satellites) but also corrected explicitly for incompleteness\footnote{The inclusion of Canis Major as a Milky Way satellite remains controversial. Excluding it from the sample, however, does not affect the results.}.

Finally, the slope of the luminosity function for the entire Local Group, $\alpha_{\rm LG} = -1.21 \pm 0.05$ is, not surprisingly , close to that measured for the Andromeda system. This is marginally steeper than the value of $-1.1\pm 0.1$ obtained by Pritchet \& van den Bergh (1999) and Trentham et al. (2005), likely due to the discovery of new, faint members of the M31 satellite system. 

While a difference in faint end slopes between Virgo and the Local Group still stands, the gap between the two environments is not as wide as previously reported, with the two slopes now differing only at the 1.7$\sigma$ level (although inclusion of disrupted galaxies in the Virgo sample would exacerbate this difference if disruption in the Local Group has not been as effective, see \S~\ref{sec:lfucds2} and\S~\ref{sec:lfucds3}). From a theoretical perspective, this is perhaps surprising given the dramatic difference in galaxy density and morphological mix between the two environments (Mathis \& White 2002; Benson et al. 2003b); additional light on the subject might be shed by comparing the faint end slope in the Virgo core with that measured at the outskirts of the cluster.

\begin{deluxetable*}{ll|ll|ll}
\tabletypesize{\footnotesize}
\tablewidth{0pc}                        
\tablenum{7}
\tablecaption{MCMC Fitting Results for the Stellar Mass Function in the Virgo Cluster Core\label{tab7}}
\tablehead{
\multicolumn{1}{c}{No.} &
\multicolumn{1}{c}{Mass Range} &
\multicolumn{1}{|c}{${\cal M}^*$ (free) (\msun)} &
\multicolumn{1}{c}{$\alpha$} &
\multicolumn{1}{|c}{${\cal M}^*$ (fix) (\msun)} &
\multicolumn{1}{c}{$\alpha$} \\
\multicolumn{1}{c}{(1)} &
\multicolumn{1}{c}{(2)} &
\multicolumn{1}{|c}{(3)} &
\multicolumn{1}{c}{(4)} &
\multicolumn{1}{|c}{(5)} &
\multicolumn{1}{c}{(6)} 
}
\startdata
352 &  $5.6\times10^5$ $-$ $3.0\times10^{11}$ & $1.1 ^{+3.5}_{-0.9} \times10^{13}$  & $-1.350^{+0.015}_{-0.016}$ & $6.3\times10^{10}$({\it fix}) & $-1.328^{+0.015}_{-0.015}$
\enddata 
\tablecomments{MCMC fits to the stellar mass function of 352 galaxies in the Virgo core, including all galaxies brighter than the $g-$band 50\% completeness limit for the survey. The mass range spanned by the sample is listed in column 2. The ${\cal M}^*$ and faint-end slope $\alpha$ are listed in cols. 3 and 4 when both are allowed to vary, and cols. 5 and 6 when ${\cal M}^*$ is held fixed to the stellar mass corresponding to $M^* = -20.48$ mag, the ($g-$band) value measured for the field by Smith et al. (2009).}
\end{deluxetable*}

\section{The Galaxy Stellar Mass Function}
\label{sec:mfvirgo}

The multi-band imaging available from the NGVS makes it possible to calculate not just the luminosity function for galaxies in the cluster core, but also their (stellar) mass function. Given that the galaxy population in this region is dominated by old, early-type systems (Janz \& Lisker 2009; Lisker et al. 2009; Roediger et al., in preparation), we would not expect the stellar mass function slope to differ dramatically from that of the luminosity function. Nevertheless, we proceed with the analysis since a comparison between observations and the predictions of cosmological models may, in some instances, be more straightforward in terms of stellar mass. 

Galaxy luminosities have been converted to stellar masses via Bayesian modeling of their $u^*griz$ spectral energy distributions (SEDs), as described in Roediger et al. (2016). To minimize the effects of colour gradients (expected to correlate with galaxy mass), all magnitudes are measured within one effective radius.  We employed the Flexible Stellar Population Synthesis (FSPS) Simple Stellar Population (SSP) models of Conroy et al. (2009), assuming a Chabrier Initial Mass Function and exponentially-declining star formation histories.  A finite grid of 50\,000 synthetic models were generated with metalliticies ranging from 0.01 to 1.6 of the solar metallicity, star formation timescales between 0.5 and 100 Gyr$^{-1}$, and luminosity weighted ages (calculated since star formation began) between 5 and 13 Gyrs. For each galaxy, a stellar mass-to-light ratio was then estimated as the median value of the marginalized posterior probability distribution function obtained by fitting the observed SED to the grid of synthetic models, with an uncertainty equal to half the interval between the 16th and 84th percentiles (see Figure~14). We note that the ${\cal M}_*/{\cal L}$ values derived in this paper differ slightly from those presented in Grossauer et al. (2015) which were calculated using the SSP models of Bruzual \& Charlot (2003) and included star formation bursts which were assumed to have a 50\% probability of occurring at each time-step throughout the lifetime of the synthetic populations. As explained in Grossauer et al. (2015), these assumptions tend to produce ${\cal M}_*/{\cal L}$ values that are biased too low. 

\begin{figure}
\epsscale{1.0}
\plotone{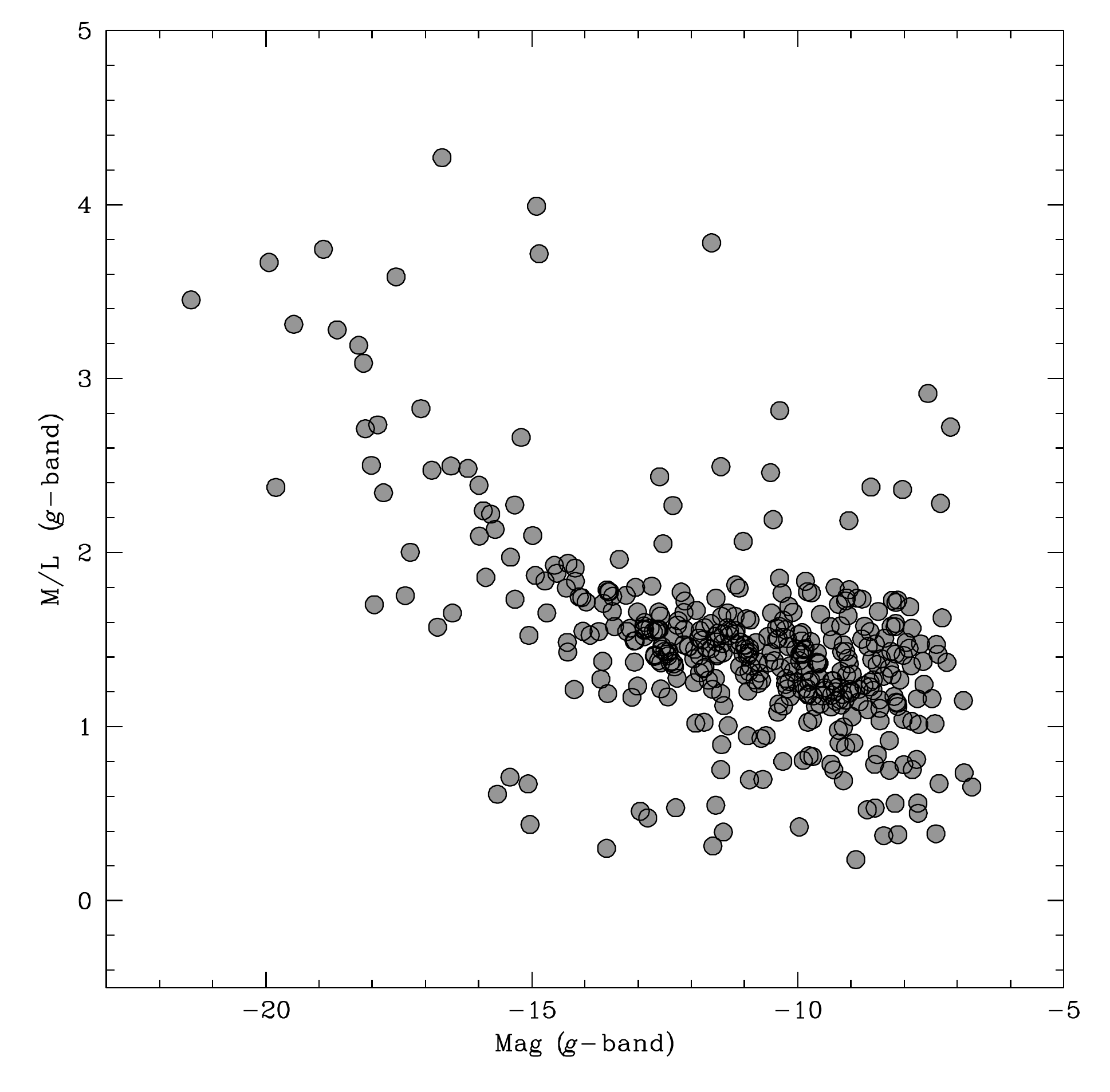}
\caption{The $g-$band stellar mass-to-light ratio of galaxies in the Virgo core region (Ferrarese et~al. 2016),  derived from SED fitting to the $u^*griz$ NGVS photometry (Roediger et~al. in preparation).}
\label{fig15}
\end{figure}

The outcome of our MCMC fitting to the resulting stellar mass function is shown in Figures 15 and 16, with parameters tabulated in Table 7. As expected, the best-fit slope of $\alpha = -1.33 \pm 0.02$ is virtually identical to that measured for the luminosity function --- a consequence of the nearly constant mass-to-light ratio of galaxies at the low-mass end of our sample.  To the best of our knowledge, this is the first measurement of the faint-end slope of the stellar mass function in the Virgo cluster that is based directly on SED modeling.

\begin{figure}
\epsscale{1.0}
\plotone{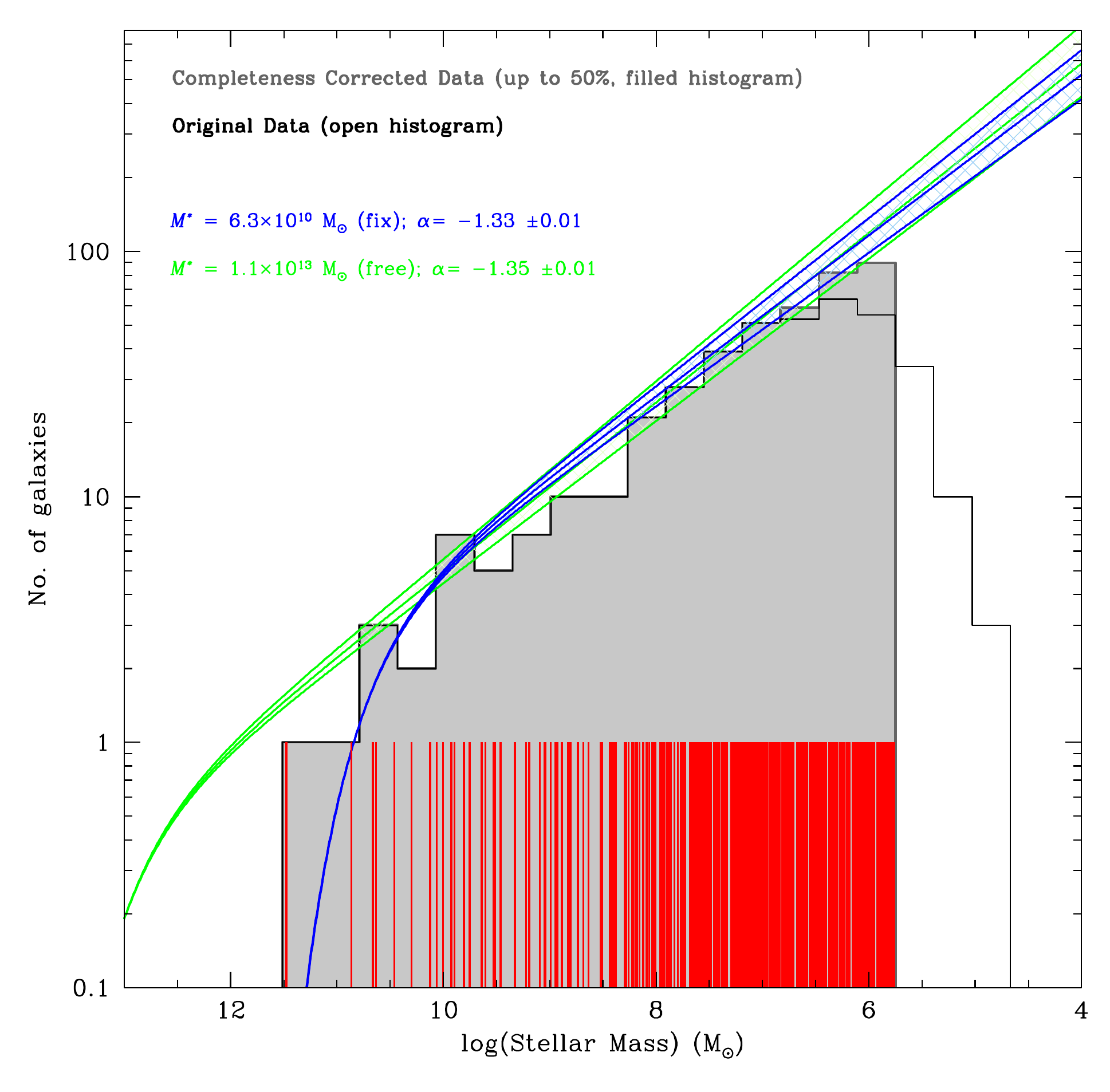}
\caption{Schechter fit to the galaxy stellar mass function in the core of the Virgo cluster. The gray histogram shows the differential mass function for galaxies brighter than $M_g = -9.13$ mag (${\cal M}^* = 5.6 \times 10^5$~\msun, see Figure 2) and corrected for incompleteness. The open black histogram shows the differential mass function constructed from the original data. Stellar masses of individual galaxies are indicated by the vertical red lines. The green line, with 1$\sigma$ confidence limits on $\alpha$, shows the best MCMC fit to the individual data when both ${\cal M}^*$ and $\alpha$ are allowed to vary. The blue line (again with 1$\sigma$ confidence limits) assumes ${\cal M}^* = 6.3 \times 10^{10}$~\msun, corresponding to $M_g^* = -20.48 $ measured in the field galaxies by Smith et al. (2009)}
\label{fig16}
\vskip .2in
\epsscale{1.15}
\plottwo{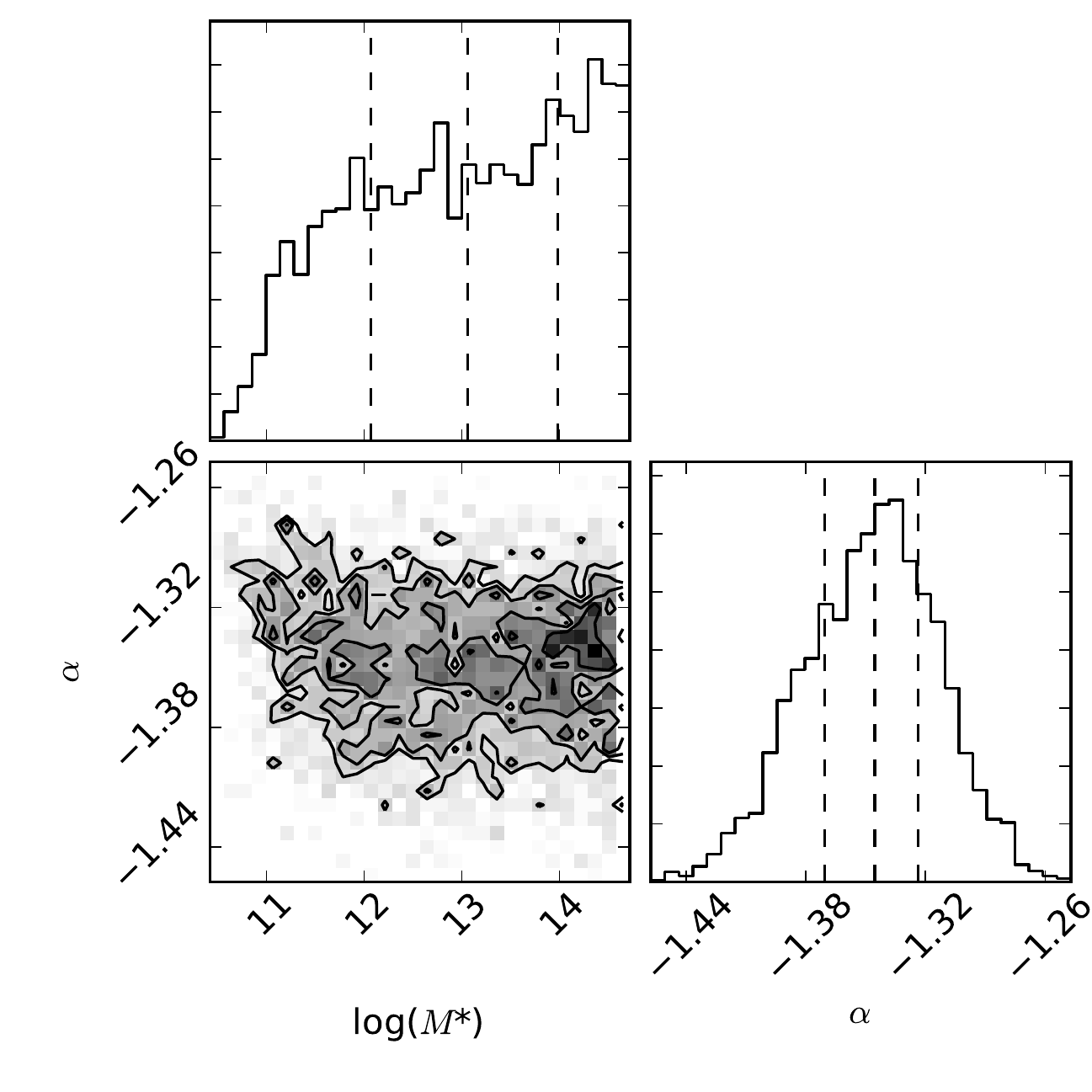}{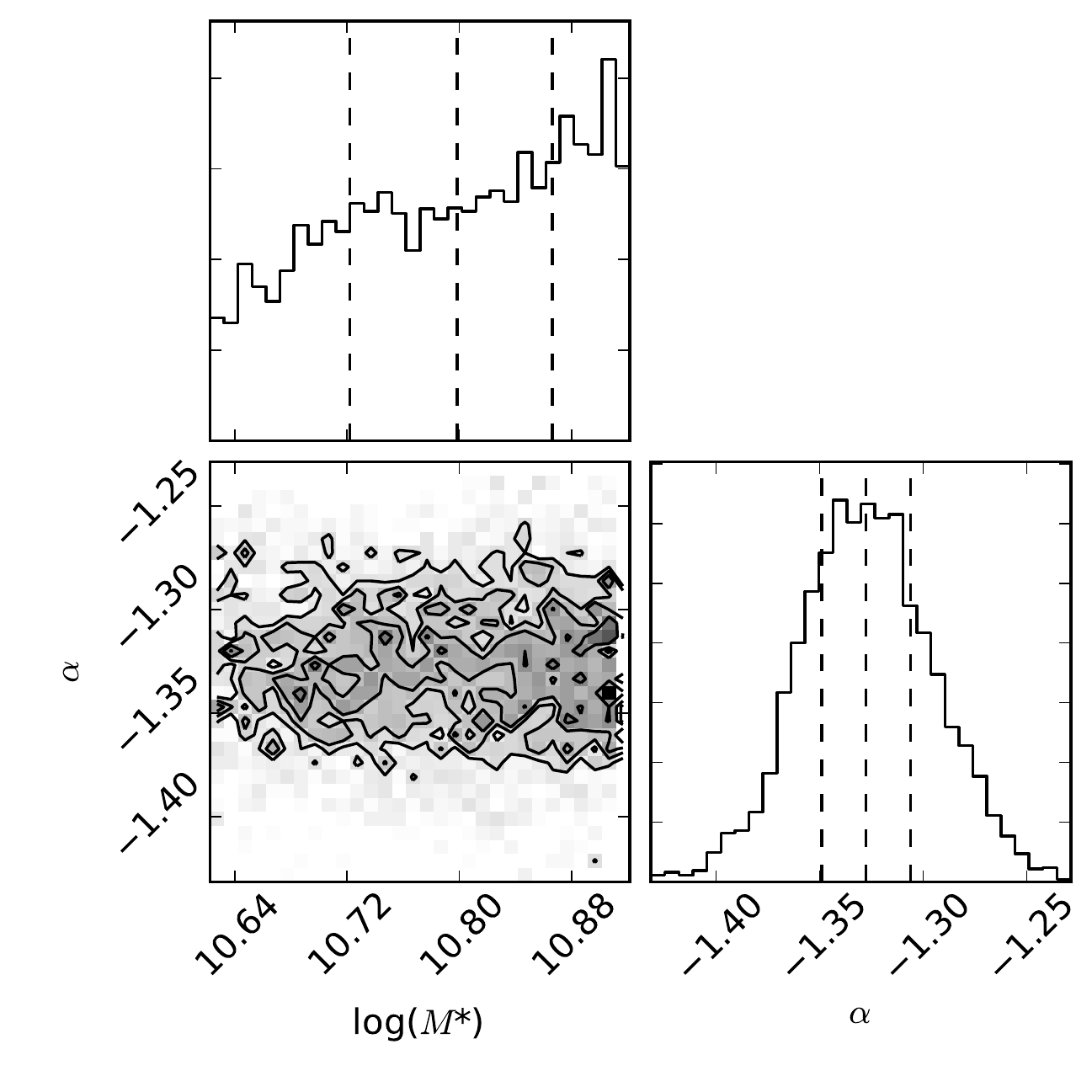}
\caption{Posterior probability density functions for ${\cal M}^*$ and $\alpha$ for galaxies in the Virgo core region when ${\cal M}^*$ is allowed to vary ({\it left panel}) or constrained within a very narrow range around ${\cal M}^* = 6.3 \times 10^{10}$ \msun~({\it right panel}). As was the case for the luminosity function shown in Figure 2, ${\cal M}^*$ is essentially unconstrained by the data, but $\alpha$ is well constrained and nearly independent of~${\cal M}^*$.}
\label{fig17}
\end{figure}

It is interesting to compare the mass function in the core of Virgo with that measured for field galaxies. Using an SDSS sample of half a million field galaxies with $z < 0.5$ and mass ${\cal M} \gtrsim 10^8$ \msun, Li \& White (2009) find that a single Schechter function  with  $\alpha = -1.155 \pm 0.008$ provides a reasonable representation of the data (fitting a triple Schechter function, which the authors note provides an even better description, leaves the faint-end slope virtually unchanged). A steeper slope is derived by Baldry et al. (2012) using a sample of over 5000 galaxies with $z < 0.06$ observed as part of the Galaxy And Mass Assembly (GAMA) survey. The authors fit a double-Schechter to their mass function; in the range $10^8 \lesssim {\cal M} < 5\times10^9$ \msun~ they find a best fitting slope of $\alpha = -1.47 \pm 0.05$. The authors further divide their sample in blue, star forming galaxies, and red, quiescent galaxies, and find that the former dominate at the low-mass end and are largely responsible for the steep slope measured for the combined sample. The luminosity function of the quiescent galaxies is significantly shallower, with only a hint of a steepening at the lowest masses. 

Regardless of these differences, the field mass function appears dissimilar from the mass function of the core of Virgo. Fitting the Virgo sample down to ${\cal M} = 10^8$ \msun~ (the approximate limit reached in the field) yields a slope $\alpha = -1.23 \pm 0.02$. Although slightly shallower than estimated using the full mass range, this slope is still steeper than measured in the field by Li \& White (2009) or by Baldry et al. (2012) for quiescent galaxies (recall that the Virgo core sample is composed almost exclusively of galaxies belonging to the red sequence). We will comment on the environmental dependence of the Virgo mass function in a future contribution using the full sample of galaxies in Virgo, reaching out to one virial radius and including a larger population and late-type systems.

\section{Summary and Conclusions}
\label{sec:summ}

We have used deep and homogenous $u^*griz$ imaging from the Next Generation Virgo Cluster Survey (Ferrarese et al. 2012) to measure the luminosity and stellar mass functions for galaxies in the core of the Virgo cluster. The analysis is based on our new catalog of 404 certain or probable cluster members found in a 3.71 deg$^2$ (0.3 Mpc$^2$) region surrounding M87, the galaxy at the dynamical center of the cluster (Ferrarese et al. 2016). Our analysis takes full advantage of the panchromatic nature, surface brightness sensitivity and exceptional spatial resolution of the NGVS data. These allow us not only to derive accurate structural and photometric parameters, but also, and perhaps most importantly, to establish robust membership criteria.

An extensive set of simulations, presented in a companion paper (Ferrarese et al. 2016), shows our NGVS catalog to be 50\% complete at a $g-$band magnitude of $M_g \simeq -9.13$, which we have adopted as the limiting magnitude for our measurement of the luminosity function. Applying this cutoff reduces our sample from 404 to 352 galaxies (in the $g-$band), 30\% of which are new detections. The NGVS completeness limit is $2.2$ mag fainter than that of Sandage et al. (1985) who first characterized Virgo's luminosity function using the VCC catalog of Binggeli et al. (1985). Additionally, the NGVS completeness limit is between $1.2$ and $3.8$ mag fainter than that of galaxy samples previously used to measure the luminosity function in this benchmark cluster.   

We have employed a Bayesian approach that fits a Schechter function to individual datapoints and their errors, thus avoiding the loss of information that can result from collapsing the data onto a histogram representation. The faint-end slope $\alpha$ has always been treated as a free parameter, while separate fits were performed by first allowing $M^*$ to vary and then setting it equal to the value measured for field galaxies from SDSS data (equivalent to $M_g = -20.48$ mag, Smith et al. 2009). We find that although $M^*$ is not well constrained by the data, owing to the small number of bright galaxies within our sample, our measurement of $\alpha$ is quite robust and insensitive to the precise choice of $M^*$. Using the same Bayesian  methodology, we have also provided new measurements for the galaxy luminosity function for the Milky Way, M31 and Local Group satellites. 

Our main results are as follows:

\begin{itemize}

\item The faint-end slope of the luminosity function in the core of Virgo is $\alpha = -1.33 \pm 0.02$, which is significantly shallower than the slope of the $\Lambda$CDM halo mass function ($\alpha \sim -1.9$, Springel et al 2008; Klypin et al. 1999; Moore et al. 1999; Han et al. 2015). This measurement refers to the $g-$band, but the values measured in $u^*riz$, down to the 50\% completeness limit appropriate to each band, are all consistent with the $g-$band measurement to within 1$\sigma$. Within the core of Virgo, and down to the 50\% completeness limit of our survey, $g=-9.13$ mag, the NGVS data firmly rule out slopes as steep as $\alpha \sim 1.7 - 2.2$ (Impey et al. 1988; Phillipps et al. 1998) or as shallow as $\alpha \sim -1.0$ (Trentham \& Tully 2002). 

\item There may be some evidence for a dip in the luminosity function at $M_g \sim - 17$ mag and, possibly, a steepening of the luminosity function fainter than $M_g \sim -14.4$ mag. These claims have been made before (e.g., Sandage et al. 1985; Trentham \& Tully 2002; Trentham \& Hodgkin 2002; Rines \& Geller 2008; McDonald et al. 2009), although they are only marginally significant based on NGVS data. The significance of these features, as well as the question of whether the luminosity function depends on location within the cluster, will be revisited in a future paper that analyses the full sample of galaxies detected in the NGVS.

\item The core region contains 92 known ultra compact dwarfs (UCDs). Under the assumption that UCDs are the surviving nuclei of tidally disrupted nucleated galaxies (an assumption we will adopt for all conclusions involving UCDs; e.g., Bekki et al. 2003), we have explored their possible impact on the luminosity function by adding them to the sample of ``surviving'' galaxies observed in the core region. While their inclusion definitely steepens the luminosity function, it does not increase the slope of the observed luminosity function to the point where it equals that of the subhalo mass function predicted by $\Lambda$CDM structure formation models if the efficiency of galaxy formation where constant across a wide range of scales. Simply adding the UCD magnitudes directly from the catalog of Liu et al. (2015a) steepens the slope to $\alpha = -1.42 \pm 0.02$, while adding the magnitudes of the UCD {\it progenitors} (estimated from the relation of nuclear-to-total galaxy magnitude from S\'anchez-Janssen et al. 2016b) increases the slope to $\alpha = -1.53 \pm 0.07$. If we also correct for the expected number of disrupted {\it non-nucleated} galaxies (and assume that the ratio of nucleated to non-nucleated disrupted galaxies is the same as observed for existing galaxies), then the slope steepens further, to $\alpha = -1.60 \pm 0.06$. Under these assumptions, the observed ratio of UCDs to galaxies in the Virgo core suggests that, for every halo that hosts a galaxy that has survived to the present day, 2 to 3 halos have been tidally disrupted. 

\item The core region contains $\sim$ 11\,000 globular clusters, of which $\sim$ 7300 belonging to the blue population. If the UCD progenitors hosted populations of globular clusters having properties (in particular, specific frequency) similar to those of present-day galaxies --- and if the nucleation to non-nucleation fraction among disrupted galaxies is similar to that of surviving galaxies --- then we estimate the disrupted galaxies to have contributed $240^{+49}_{-24}$ UCDs and $2643^{+756}_{-198}$ globular clusters to the Virgo core. The disrupted {\it nucleated} galaxies contribute all of the UCDs (by definition) and $\sim 77$\% of these globular clusters (due to the fact that  the most luminous progenitor galaxies, which contain the largest number of globular clusters, are nearly always nucleated). In this picture, we estimate that as many as one third of the (mostly blue) globular clusters in the core of Virgo could have originated from disrupted satellites.

\item Only a fraction of the $\sim 240$ predicted UCDs would be easily recognizable as such: i.e., the census of UCDs in the M87 region includes only 92 objects (Liu et~al. 2015a; see also Zhang et al. 2015) down to $M_g \sim -9.6$. The remaining $\sim 150$ UCDs, which would be expected to have magnitudes in the range $-9.6 \leq g \leq -6.7$, would be classified as globular clusters. 

\item In this scenario, we estimate that the total amount of luminosity contributed by disrupted galaxies in the core region could amount to $\sim$ 85\% of the luminosity of M87, the supergiant elliptical galaxy at the dynamical center of Virgo. Alternatively, the luminosity associated with disrupted galaxies corresponds to $\sim$ 40\% of the combined luminosity of all member galaxies in the Virgo core region. All in all, these calculations suggest that the $\sim$ 400 galaxies we detect in the core region may represent as little as a quarter of the initial population of (surviving plus disrupted) galaxies in this region.

\item We present new luminosity function measurements for the M31, Milky Way and the Local Group satellites using updated galaxy catalogs. These measurements have been computed using the same Bayesian approach adopted for Virgo (although we apply no completeness corrections for the Local Group samples). Down to the completeness limit of the NGVS ($M_g = -9.13$), the luminosity function of the M31 system has slope $\alpha_{\rm M31} = -1.23 \pm 0.07$. The luminosity function for the Milky Way has a slightly shallower faint-end slope, $\alpha_{\rm MW} =  -1.03 \pm 0.1$, but this would likely steepen if completeness corrections --- which are more severe for the Milky Way rather than for M31 --- are taken into account. Overall, the luminosity function for Local Group galaxies is found to be $\alpha_{\rm LG} = -1.21 \pm 0.05$.
This is, somewhat surprisingly, only slightly shallower than that measured in the core of the Virgo cluster (a 1.7$\sigma$ difference).

\item We have derived stellar masses for the Virgo core galaxies by fitting their $u^*griz$ spectral energy distributions. The resulting stellar mass function, measured over the range $5.6\times10^5 \lesssim {\cal M} \lesssim 3.0\times 10^{11}$ \msun, has a faint-end slope of $\alpha = -1.32 \pm 0.02$, which is nearly identical to the luminosity functions measured in the various NGVS bandpasses.

\end{itemize}

\acknowledgments

We wish to thank the anonymous referee for the very thorough reading of the paper, and the many helpful and constructive comments. This work was supported in part by the Canadian Advanced Network for Astronomical Research (CANFAR) which has been made possible by funding from CANARIE under the Network-Enabled Platforms program. This research also used the facilities of the Canadian Astronomy Data Centre operated by the National Research Council of Canada with the support of the Canadian Space Agency. The authors further acknowledge use of the NASA/IPAC Extragalactic Database (NED) which is operated by the Jet Propulsion Laboratory, California Institute of Technology, under contract with the National Aeronautics and Space Administration, and the HyperLeda database ({\tt http://leda.univ-lyon1.fr}). S.C. and J.T. acknowledge support from the Natural Science and
Engineering Research Council of Canada through a Research Discovery Grant. E.W.P. acknowledges support from the National Natural Science Foundation of China under Grant Nos. 11173003 and 11573002, and from the Strategic Priority Research Program, `The Emergence of Cosmological Structures', of the Chinese Academy of Sciences, Grant No. XDB09000105. This work is supported in part by the French Agence Nationale de la Recherche grant VIRAGE (ANR10-BLANC-0506-01). C.L. acknowledges the National Key Basic Research Program of China (2015CB857002), NSFC grants 11203017, 11125313 and 11473020. E.T. and R.G. acknowledge NSF grants AST-1010039 and AST-1412504.

Facilities: Canada-France-Hawaii Telescope (CFHT)


\begin{thebibliography}{}
\bibitem[Adami et al.(2007)]{2007A&A...472..749A} Adami, C., Picat, J.~P., Durret, F., et al.\ 2007, \aap, 472, 749 
\bibitem[Agulli et al.(2014)]{2014MNRAS.444L..34A} Agulli, I., Aguerri, J.~A.~L., S{\'a}nchez-Janssen, R., et al.\ 2014, \mnras, 444, L34 
\bibitem[Baldry et al.(2012)]{2012MNRAS.421..621B} Baldry, I.~K., Driver, S.~P., Loveday, J., et al.\ 2012, \mnras, 421, 621 
\bibitem[Beasley et al.(2002)]{2002MNRAS.333..383B} Beasley, M.~A., Baugh, C.~M., Forbes, D.~A., Sharples, R.~M., \& Frenk, C.~S.\ 2002, \mnras, 333, 383 
\bibitem[Bechtol et al.(2015)]{2015ApJ...807...50B} Bechtol, K., Drlica-Wagner, A., Balbinot, E., et al.\ 2015, \apj, 807, 50 
\bibitem[()]{} Bellazzini, M., Gennari, N., Ferraro, F.~R., \& Sollima, A. 2004, MNRAS 354, 708             
\bibitem[()]{} Bellazzini, M., Ibata, R., Monaco, L., et al. 2004, MNRAS 354, 1263              
\bibitem[()]{} Bellazzini, M., Gennari, N., \& Ferraro, F.~R. 2005, MNRAS 360, 185               
\bibitem[()]{} Bellazzini, M., Ibata, R., Martin, N., et al. 2006, MNRAS 366, 865              
\bibitem[Bekki et al.(2001)]{2001ApJ...557L..39B} Bekki, K., Couch, W.~J., Drinkwater, M.~J., \& Gregg, M.~D.\ 2001, \apjl, 557, L39
\bibitem[Bekki et al.(2003)]{2003MNRAS.344..399B} Bekki, K., Couch, W.~J., Drinkwater, M.~J., \& Shioya, Y.\ 2003, \mnras, 344, 399 
\bibitem[Benson et al.(2002)]{2002MNRAS.333..156B} Benson, A.~J., Lacey, C.~G., Baugh, C.~M., Cole, S., \& Frenk, C.~S.\ 2002, \mnras, 333, 156 
\bibitem[Benson et al.(2003)]{2003ApJ...599...38B} Benson, A.~J., Bower, R.~G., Frenk, C.~S., Lacey, C.~G., Baugh, C.~M., \& Cole, S.\ 2003a, \apj, 599, 38 
\bibitem[()]{}Benson A. J., Frenk C. S., Baugh C. M., Cole S., Lacey C. G., 2003b, MNRAS, 343, 679
\bibitem[()]{} Bernard, E.~J., Monelli, M., Gallart, C., et al. 2009, ApJ 699, 1742              
\bibitem[()]{} Bernard, E.~J., Monelli, M., Gallart, C., et al. 2010, ApJ 712, 1259              
\bibitem[Binggeli et al.(1985)]{1985AJ.....90.1681B} Binggeli, B., Sandage, A., \& Tammann, G.~A.\ 1985, \aj, 90, 1681
\bibitem[Binggeli et al.(1987)]{1987AJ.....94..251B} Binggeli, B., Tammann, G.~A., \& Sandage, A.\ 1987, \aj, 94, 251 
\bibitem[Binggeli \& Cameron(1991)]{1991A&A...252...27B} Binggeli, B., \& Cameron, L.~M.\ 1991, \aap, 252, 27 
\bibitem[Blakeslee et al.(2009)]{2009ApJ...694..556B} Blakeslee, J.~P., Jord{\'a}n, A., Mei, S., et al.\ 2009, \apj, 694, 556 
\bibitem[Blanton et al.(2003)]{2003ApJ...592..819B} Blanton, M.~R., Hogg, D.~W., Bahcall, N.~A., et al.\ 2003, \apj, 592, 819 
\bibitem[Blanton et al.(2005)]{2005ApJ...631..208B} Blanton, M.~R., Lupton, R.~H., Schlegel, D.~J., et al.\ 2005, \apj, 631, 208 
\bibitem[Boselli et al.(2016)]{2014A&A...570A..69B} Boselli, A., Voyer, E., Boissier, S., et al.\ 2016, \aap, 585, 2 
\bibitem[Bruzual \& Charlot(2003)]{2003MNRAS.344.1000B} Bruzual, G., \& Charlot, S.\ 2003, \mnras, 344, 1000 
\bibitem[Bullock et al.(2000)]{2000ApJ...539..517B} Bullock, J.~S., Kravtsov, A.~V., \& Weinberg, D.~H.\ 2000, \apj, 539, 517 
\bibitem[Castro-Rodrigu{\'e}z et al.(2009)]{2009A&A...507..621C} Castro-Rodrigu{\'e}z, N., Arnaboldi, M., Aguerri, J.~A.~L., et al.\ 2009, \aap, 507, 621 
\bibitem[Chiboucas et al.(2009)]{2009AJ....137.3009C} Chiboucas, K., Karachentsev, I.~D., \& Tully, R.~B.\ 2009, \aj, 137, 3009 
\bibitem[Chiboucas et al.(2013)]{2013AJ....146..126C} Chiboucas, K., Jacobs, B.~A., Tully, R.~B., \& Karachentsev, I.~D.\ 2013, \aj, 146, 126 
\bibitem[Chilingarian et al.(2008)]{2008MNRAS.390..906C} Chilingarian, I.~V., Cayatte, V., \& Bergond, G.\ 2008, \mnras, 390, 906 
\bibitem[()]{} Clementini, G., Gratton, R., Bragaglia, A., et al. 2003, AJ 125, 1309              
\bibitem[()]{} Cole, A.~A., Skillman, E.~D., Tolstoy, E., et al. 2007, ApJ 659, L17              
\bibitem[()]{}Conn, A.~R., Ibata, R.~A., Lewis, G.~F., et al.\ 2012, \apj, 758, 11 
\bibitem[Conroy et al.(2009)]{2009ApJ...699..486C} Conroy, C., Gunn, J.~E., \& White, M.\ 2009, \apj, 699, 486 
\bibitem[Cooper et al.(2015)]{2015MNRAS.451.2703C} Cooper, A.~P., Gao, L., Guo, Q., et al.\ 2015, \mnras, 451, 2703 
\bibitem[C{\^o}t{\'e} et al.(1998)]{1998ApJ...501..554C} C{\^o}t{\'e}, P., Marzke, R.~O., \& West, M.~J.\ 1998, \apj, 501, 554 
\bibitem[C{\^o}t{\'e} et al.(2002)]{2002ApJ...567..853C} C{\^o}t{\'e}, P., West, M.~J., \& Marzke, R.~O.\ 2002, \apj, 567, 853 
\bibitem[C{\^o}t{\'e} et al.(2006)]{2006ApJS..165...57C} C{\^o}t{\'e}, P., et al.\ 2006, \apjs, 165, 57 
\bibitem[()]{} C{\^o}t{\'e}, P., et al. 2016, ApJ, to be submitted February 2016
\bibitem[Croton(2006)]{2006MNRAS.369.1808C} Croton, D.~J.\ 2006, \mnras, 369, 1808 
\bibitem[den Brok et al.(2014)]{2014MNRAS.445.2385D} den Brok, M., Peletier, R.~F., Seth, A., et al.\ 2014, \mnras, 445, 2385 
\bibitem[De Propris et al.(2003)]{2003MNRAS.342..725D} De Propris, R., Colless, M., Driver, S.~P., et al.\ 2003, \mnras, 342, 725 
\bibitem[()]{} de Vaucouleurs, G., de Vaucouleurs, A., Corwin, Jr., et al. 1991, Third Reference Catalog of Bright Galaxies. Volume 1-3, XII, 2069 pp. 7 Springer-Verlag Berlin Heidelberg
\bibitem[Diehl et al.(2014)]{2014SPIE.9149E..0VD} Diehl, H.~T., Abbott, T.~M.~C., Annis, J., et al.\ 2014, \procspie, 9149, 91490V 
\bibitem[Dolag et al.(2010)]{2010MNRAS.405.1544D} Dolag, K., Murante, G., \& Borgani, S.\ 2010, \mnras, 405, 1544 
\bibitem[()]{} Dolphin, A.~E., Saha, A., Claver, J., et al. 2002, AJ 123, 3154              
\bibitem[Drinkwater et al.(1999)]{1999ApJ...511L..97D} Drinkwater, M.~J., Phillipps, S., Gregg, M.~D., et al.\ 1999, \apjl, 511, L97 
\bibitem[Drinkwater et al.(2003)]{2003Natur.423..519D} Drinkwater, M.~J., Gregg, M.~D., Hilker, M., Bekki, K., Couch, W.~J., Ferguson, H.~C., Jones,  J.~B., \& Phillipps, S.\ 2003, \nat, 423, 519 
\bibitem[()]{}Drinkwater, M. J., Gregg, M. D., Couch, W. J., et al. 2004, PASA, 21, 375
\bibitem[Drlica-Wagner et al.(2015)]{2015ApJ...813..109D} Drlica-Wagner, A., Bechtol, K., Rykoff, E.~S., et al.\ 2015, \apj, 813, 109 
\bibitem[Durrell et al.(2002)]{2002ApJ...570..119D} Durrell, P.~R., Ciardullo, R., Feldmeier, J.~J., Jacoby, G.~H., \& Sigurdsson, S.\ 2002, \apj, 570, 119 
\bibitem[Durrell et al.(2014)]{2014ApJ...794..103D} Durrell, P.~R., C{\^o}t{\'e}, P., Peng, E.~W., et al.\ 2014, \apj, 794, 103
\bibitem[Feldmeier et al.(2004)]{2004ApJ...615..196F} Feldmeier, J.~J., Ciardullo, R., Jacoby, G.~H., \& Durrell, P.~R.\ 2004, \apj, 615, 196 
\bibitem[()]{} Ferrarese, L. et al. 2015, ApJ, submitted
\bibitem[Ferrarese et al.(2012)]{2012ApJS..200....4F} Ferrarese, L., C{\^o}t{\'e}, P., Cuillandre, J.-C., et al.\ 2012, \apjs, 200, 4
\bibitem[Foreman-Mackey et al.(2013)]{2013PASP..125..306F} Foreman-Mackey, D., Hogg, D.~W., Lang, D., \& Goodman, J.\ 2013, \pasp, 125, 306 
\bibitem[Frank et al.(2011)]{2011MNRAS.414L..70F} Frank, M.~J., Hilker, M., Mieske, S., et al.\ 2011, \mnras, 414, L70 
\bibitem[()]{} Gao L., Navarro J. F., Frenk C. S., Jenkins A., Springel V., White S. D. M., 2012, MNRAS, 425, 2169
\bibitem[Georgiev et al.(2010)]{2010MNRAS.406.1967G} Georgiev, I.~Y., Puzia, T.~H., Goudfrooij, P., \& Hilker, M.\ 2010, \mnras, 406, 1967 
\bibitem[()]{} Gieren, W., Pietrzynski, G., Nalewajko, K., et al. 2006, ApJ 647, 1056              
\bibitem[Goerdt et al.(2008)]{2008MNRAS.385.2136G} Goerdt, T., Moore, B., Kazantzidis, S., et al.\ 2008, \mnras, 385, 2136 
\bibitem[Gonz{\'a}lez et al.(2006)]{2006A&A...445...51G} Gonz{\'a}lez, R.~E., Lares, M., Lambas, D.~G., \& Valotto, C.\ 2006, \aap, 445, 51 
\bibitem[Gonzalez et al.(2007)]{2007ApJ...666..147G} Gonzalez, A.~H., Zaritsky, D., \& Zabludoff, A.~I.\ 2007, \apj, 666, 147 
\bibitem[Governato et al.(2007)]{2007MNRAS.374.1479G} Governato, F., Willman, B., Mayer, L., et al.\ 2007, \mnras, 374, 1479 
\bibitem[()]{} Grillmair, C.~J., Lauer, T.~R., Worthey, G., et al. 1996, AJ 112, 1975              
\bibitem[Grossauer et al.(2015)]{2015ApJ...807...88G} Grossauer, J., Taylor, J.~E., Ferrarese, L., et al.\ 2015, \apj, 807, 88
\bibitem[Han et al.(2015)]{2015arXiv150902175H} Han, J., Cole, S., Frenk, C.~S., \& Jing, Y.\ 2015, arXiv:1509.02175 
\bibitem[Harris et al.(2013)]{2013ApJ...772...82H} Harris, W.~E., Harris, G.~L.~H., \& Alessi, M.\ 2013, \apj, 772, 82                  
\bibitem[Ha{\c s}egan et al.(2005)]{2005ApJ...627..203H} Ha{\c s}egan, M., Jord{\'a}n, A., C{\^o}t{\'e}, P., et al.\ 2005, \apj, 627, 203 
\bibitem[()]{} Hidalgo, S.~L., Aparicio, A., Mart\'inez-Delgado, D., \& Gallart, C. 2009, ApJ 705, 704             
\bibitem[Hilker et al.(1999)]{1999A&AS..134...75H} Hilker, M., Infante, L., Vieira, G., Kissler-Patig, M., \& Richtler, T.\ 1999, \aaps, 134, 75 
\bibitem[()]{} Huchra, J.~P., Vogeley, M.~S., \& Geller, M.~J. 1999, ApJS 121, 287               
\bibitem[()]{} Ibata, R.~A., Gilmore, G., \& Irwin, M.~J. 1994, Nature 370, 194               
\bibitem[Impey et al.(1988)]{1988ApJ...330..634I} Impey, C., Bothun, G., \& Malin, D.\ 1988, \apj, 330, 634 
\bibitem[Irwin(1994)]{1994ESOC...49...27I} Irwin, M.~J.\ 1994, European Southern Observatory Conference and Workshop Proceedings, 49, 27 
\bibitem[()]{} Irwin, M. \& Hatzidimitriou, D. 1995, MNRAS 277, 1354                 
\bibitem[()]{} Izenman, A. J. 1991, in Recent Developments in Nonparametric Density Estimation. Journal of the American Statistical Association, 86, 205
\bibitem[Janz \& Lisker(2009)]{2009ApJ...696L.102J} Janz, J., \& Lisker, T.\ 2009, \apjl, 696, L102 
\bibitem[Janz et al.(2015)]{2015MNRAS.449.1716J} Janz, J., Forbes, D.~A., Norris, M.~A., et al.\ 2015, \mnras, 449, 1716 
\bibitem[Jord{\'a}n et al.(2004)]{2004AJ....127...24J} Jord{\'a}n, A., C{\^o}t{\'e}, P., West, M.~J., et al.\ 2004, \aj, 127, 24 
\bibitem[Kaiser et al.(2010)]{2010SPIE.7733E..0EK} Kaiser, N., Burgett, W., Chambers, K., et al.\ 2010, \procspie, 7733, 77330E 
\bibitem[()]{} Kalirai, J.~S., Beaton, R.~L., Geha, M.~C., et al. 2010, ApJ 711, 671              
\bibitem[Karachentsev et al.(2002)]{2002A&A...385...21K} Karachentsev, I.~D., Sharina, M.~E., Dolphin, A.~E., et al.\ 2002, \aap, 385, 21 
\bibitem[Kim \& Jerjen(2015)]{2015ApJ...808L..39K} Kim, D., \& Jerjen, H.\ 2015, \apjl, 808, L39 
\bibitem[Kim et al.(2014)]{2014ApJS..215...22K} Kim, S., Rey, S.-C., Jerjen, H., et al.\ 2014, \apjs, 215, 22 
\bibitem[Kim et al.(2015)]{2015ApJ...804L..44K} Kim, D., Jerjen, H., Mackey, D., Da Costa, G.~S., \& Milone, A.~P.\ 2015b, \apjl, 804, L44 
\bibitem[Kim et al.(2015)]{2015ApJ...803...63K} Kim, D., Jerjen, H., Milone, A.~P., Mackey, D., \& Da Costa, G.~S.\ 2015a, \apj, 803, 63 
\bibitem[Klypin et al.(1999)]{1999ApJ...522...82K} Klypin, A., Kravtsov, A.~V., Valenzuela, O., \& Prada, F.\ 1999, \apj, 522, 82 
\bibitem[Klypin et al.(2015)]{2015MNRAS.454.1798K} Klypin, A., Karachentsev, I., Makarov, D., \& Nasonova, O.\ 2015, \mnras, 454, 1798 
\bibitem[Koda et al.(2015)]{2015ApJ...807L...2K} Koda, J., Yagi, M., Yamanoi, H., \& Komiyama, Y.\ 2015, \apjl, 807, L2 
\bibitem[Koposov et al.(2008)]{2008ApJ...686..279K} Koposov, S., Belokurov, V., Evans, N.~W., et al.\ 2008, \apj, 686, 279 
\bibitem[Koposov et al.(2015)]{2015ApJ...805..130K} Koposov, S.~E., Belokurov, V., Torrealba, G., \& Evans, N.~W.\ 2015, \apj, 805, 130 
\bibitem[Krusch et al.(2006)]{2006A&A...459..759K} Krusch, E., Rosenbaum, D., Dettmar, R.-J., et al.\ 2006, \aap, 459, 759 
\bibitem[Laevens et al.(2015)]{2015ApJ...813...44L} Laevens, B.~P.~M., Martin, N.~F., Bernard, E.~J., et al.\ 2015b, \apj, 813, 44 
\bibitem[Laevens et al.(2015)]{2015ApJ...802L..18L} Laevens, B.~P.~M., Martin, N.~F., Ibata, R.~A., et al.\ 2015a, \apjl, 802, L18 
\bibitem[()]{} Lee, M.~G. 1995, AJ 110, 1129                    
\bibitem[Li \& White(2009)]{2009MNRAS.398.2177L} Li, C., \& White, S.~D.~M.\ 2009, \mnras, 398, 2177 
\bibitem[Lieder et al.(2012)]{2012A&A...538A..69L} Lieder, S., Lisker, T., Hilker, M., Misgeld, I., \& Durrell, P.\ 2012, \aap, 538, A69 
\bibitem[Lieder et al.(2013)]{2013A&A...559A..76L} Lieder, S., Mieske, S., S{\'a}nchez-Janssen, R., et al.\ 2013, \aap, 559, A76 
\bibitem[()]{} Liu, C., Peng, E.~W., C\^ot\'e, P., Ferrarese, L., Jord\'an, A., , Mihos, J.C., Zhang, H.-X., et al. \ 2015a, \apj, 812, 34
\bibitem[Liu et al.(2015)]{2015ApJ...812L...2L} Liu, C., Peng, E.~W., Toloba, E., et al.\ 2015b, \apjl, 812, L2 
\bibitem[Lisker et al.(2007)]{2007ApJ...660.1186L} Lisker, T., Grebel, E.~K., Binggeli, B., \& Glatt, K.\ 2007, \apj, 660, 1186 
\bibitem[Lisker et al.(2009)]{2009ApJ...706L.124L} Lisker, T., Janz, J., Hensler, G., et al.\ 2009, \apjl, 706, L124 
\bibitem[Loveday(1997)]{1997ApJ...489...29L} Loveday, J.\ 1997, \apj, 489, 29 
\bibitem[Mahdavi et al.(2005)]{2005AJ....130.1502M} Mahdavi, A., Trentham, N., \& Tully, R.~B.\ 2005, \aj, 130, 1502 
\bibitem[()]{} Majewski, S.~R., Skrutskie, M.~F., Weinberg, M.~D., \& Ostheimer, J.~C. 2003, ApJ 599, 1082             
\bibitem[()]{} Martin, N.~F., McConnachie, A.~W., Irwin, M., et al. 2009, ApJ 705, 758              
\bibitem[()]{}Martin, N.~F., Slater, C.~T., Schlafly, E.~F., et al.\ 2013a, \apj, 772, 15 
\bibitem[()]{}Martin, N.~F., Schlafly, E.~F., Slater, C.~T., et al.\ 2013b, \apjl, 779, L10
\bibitem[Martin et al.(2015)]{2015ApJ...804L...5M} Martin, N.~F., Nidever, D.~L., Besla, G., et al.\ 2015, \apjl, 804, L5 
\bibitem[()]{} Mateo, M., Olszewski, E.~W., \& Morrison, H.~L. 1998, ApJ 508, L55               
\bibitem[Mathis \& White(2002)]{2002MNRAS.337.1193M} Mathis, H., \& White, S.~D.~M.\ 2002, \mnras, 337, 1193 
\bibitem[()]{} McConnachie, A.~W., Irwin, M.~J., Ferguson, A.~M.~N., et al. 2005, MNRAS 356, 979              
\bibitem[()]{} McConnachie, A.~W. \& Irwin, M.~J. 2006, MNRAS 365, 1263                 
\bibitem[()]{} McConnachie, A.~W., Arimoto, N., Irwin, M., \& Tolstoy, E. 2006, MNRAS 373, 715             
\bibitem[McConnachie(2012)]{2012AJ....144....4M} McConnachie, A.~W.\ 2012, \aj, 144, 4 
\bibitem[McConnachie et al.(2009)]{2009Natur.461...66M} McConnachie, A.~W., Irwin, M.~J., Ibata, R.~A., et al.\ 2009, \nat, 461, 66 
\bibitem[McDonald et al.(2009)]{2009MNRAS.394.2022M} McDonald, M., Courteau, S., \& Tully, R.~B.\ 2009, \mnras, 394, 2022 
\bibitem[McLaughlin(1999)]{1999ApJ...512L...9M} McLaughlin, D.~E.\ 1999, \apjl, 512, L9 
\bibitem[Mei et al.(2007)]{2007ApJ...655..144M} Mei, S., Blakeslee, J.~P., C{\^o}t{\'e}, P., et al.\ 2007, \apj, 655, 144 
\bibitem[Mieske et al.(2008)]{2008A&A...487..921M} Mieske, S., Hilker, M., Jord{\'a}n, A., et al.\ 2008, \aap, 487, 921 
\bibitem[Mihos et al.(2005)]{2005ApJ...631L..41M} Mihos, J.~C., Harding, P., Feldmeier, J., \& Morrison, H.\ 2005, \apjl, 631, L41 
\bibitem[Mihos et al.(2015)]{2015ApJ...809L..21M} Mihos, J.~C., Durrell, P.~R., Ferrarese, L., et al.\ 2015, \apjl, 809, L21 
\bibitem[()]{} Monaco, L., Bellazzini, M., Ferraro, F.~R., \& Pancino, E. 2004, MNRAS 353, 874             
\bibitem[Moore et al.(1999)]{1999ApJ...524L..19M} Moore, B., Ghigna, S., Governato, F., Lake, G., Quinn, T., Stadel, J., \& Tozzi, P.\ 1999, \apjl, 524, L19 
\bibitem[Mu{\~n}oz et al.(2014)]{2014ApJS..210....4M} Mu{\~n}oz, R.~P., Puzia, T.~H., Lan{\c c}on, A., et al.\ 2014, \apjs, 210, 4 
\bibitem[Murante et al.(2007)]{2007MNRAS.377....2M} Murante, G., Giovalli, M., Gerhard, O., Arnaboldi, M., Borgani, S., \& Dolag, K.\ 2007, \mnras, 377, 2 
\bibitem[Okamoto et al.(2008)]{2008MNRAS.390..920O} Okamoto, T., Gao, L., \& Theuns, T.\ 2008, \mnras, 390, 920 
\bibitem[Paudel et al.(2010)]{2010MNRAS.405..800P} Paudel, S., Lisker, T., Kuntschner, H., Grebel, E.~K., \& Glatt, K.\ 2010, \mnras, 405, 800
\bibitem[Peng et al.(2006)]{2006ApJ...639...95P} Peng, E.~W., Jord{\'a}n, A., C{\^o}t{\'e}, P., et al.\ 2006, \apj, 639, 95 
\bibitem[Peng et al.(2008)]{2008ApJ...681..197P} Peng, E.~W., Jord{\'a}n, A., C{\^o}t{\'e}, P., et al.\ 2008, \apj, 681, 197 
\bibitem[Pfeffer \& Baumgardt(2013)]{2013MNRAS.433.1997P} Pfeffer, J., \& Baumgardt, H.\ 2013, \mnras, 433, 1997 
\bibitem[Phillipps et al.(1998)]{1998ApJ...493L..59P} Phillipps, S., Parker, Q.~A., Schwartzenberg, J.~M., \& Jones, J.~B.\ 1998, \apjl, 493, L59 
\bibitem[Phillipps et al.(2001)]{2001ApJ...560..201P} Phillipps, S., Drinkwater, M.~J., Gregg, M.~D., \& Jones, J.~B.\ 2001, \apj, 560, 201 
\bibitem[()]{} Pietrzynski, G., Gieren, W., Szewczyk, O., et al. 2008 AJ 135, 1993              
\bibitem[()]{} Pietrzynski, G., Gorski, M., Gieren, W., et al. 2009, AJ 138, 459              
\bibitem[Popesso et al.(2005)]{2005A&A...433..415P} Popesso, P., B{\"o}hringer, H., Romaniello, M., \& Voges, W.\ 2005, \aap, 433, 415 
\bibitem[Popesso et al.(2006)]{2006A&A...445...29P} Popesso, P., Biviano, A., B{\"o}hringer, H., \& Romaniello, M.\ 2006, \aap, 445, 29 
\bibitem[Pritchet \& van den Bergh(1999)]{1999AJ....118..883P} Pritchet, C.~J., \& van den Bergh, S.\ 1999, \aj, 118, 883 
\bibitem[Purcell et al.(2007)]{2007ApJ...666...20P} Purcell, C.~W., Bullock, J.~S., \& Zentner, A.~R.\ 2007, \apj, 666, 20 
\bibitem[Rees \& Ostriker(1977)]{1977MNRAS.179..541R} Rees, M.~J., \& Ostriker, J.~P.\ 1977, \mnras, 179, 541 
\bibitem[()]{} Richardson, J.~C., Irwin, M.~J., McConnachie, A.~W., et al. 2011, ApJ 732, 76              
\bibitem[Rines \& Geller(2008)]{2008AJ....135.1837R} Rines, K., \& Geller, M.~J.\ 2008, \aj, 135, 1837 
\bibitem[()]{} Roediger, J., et al. 2016, ApJ, to be submitted February 2016
\bibitem[Rudick et al.(2010)]{2010ApJ...720..569R} Rudick, C.~S., Mihos, J.~C., Harding, P., et al.\ 2010, \apj, 720, 569 
\bibitem[Sabatini et al.(2003)]{2003MNRAS.341..981S} Sabatini, S., Davies, J., Scaramella, R., et al.\ 2003, \mnras, 341, 981 
\bibitem[Sabatini et al.(2005)]{2005MNRAS.357..819S} Sabatini, S., Davies, J., van Driel, W., et al.\ 2005, \mnras, 357, 819 
\bibitem[()]{} S\'anchez-Janssen R., et al. 2016a, ApJ, submitted
\bibitem[()]{} S\'anchez-Janssen R., et al. 2016b, ApJ, to be submitted February 2016
\bibitem[Sandage et al.(1985)]{1985AJ.....90.1759S} Sandage, A., Binggeli, B., \& Tammann, G.~A.\ 1985, \aj, 90, 1759 
\bibitem[()]{} Sanna, N., Bono, G., Stetson, P.~B., et al. 2010, ApJ 722, L244              
\bibitem[()]{} Saviane, I., Held, E.~V., \& Piotto, G. 1996, A\&A 315, 40               
\bibitem[Seth et al.(2014)]{2014Natur.513..398S} Seth, A.~C., van den Bosch, R., Mieske, S., et al.\ 2014, \nat, 513, 398 
\bibitem[Schechter(1976)]{1976ApJ...203..297S} Schechter, P.\ 1976, \apj, 203, 297 
\bibitem[Schlegel et al.(1998)]{1998ApJ...500..525S} Schlegel, D.~J., Finkbeiner, D.~P., \& Davis, M.\ 1998, \apj, 500, 525
\bibitem[Smith et al.(2009)]{2009MNRAS.397..868S} Smith, A.~J., Loveday, J., \& Cross, N.~J.~G.\ 2009, \mnras, 397, 868 
\bibitem[Somerville(2002)]{2002ApJ...572L..23S} Somerville, R.~S.\ 2002, \apjl, 572, L23 
\bibitem[Springel et al.(2008)]{2008MNRAS.391.1685S} Springel, V., et al.\ 2008, \mnras, 391, 1685 
\bibitem[Tollerud et al.(2008)]{2008ApJ...688..277T} Tollerud, E.~J., Bullock, J.~S., Strigari, L.~E., \& Willman, B.\ 2008, \apj, 688, 277 
\bibitem[Tonini(2013)]{2013ApJ...762...39T} Tonini, C.\ 2013, \apj, 762, 39 
\bibitem[Trentham \& Hodgkin(2002)]{2002MNRAS.333..423T} Trentham, N., \& Hodgkin, S.\ 2002, \mnras, 333, 423 
\bibitem[()]{}Trentham, N., \& Tully, R. B. 2002, MNRAS, 335, 712
\bibitem[Trentham(1998)]{1998MNRAS.294..193T} Trentham, N.\ 1998, \mnras, 294, 193 
\bibitem[Trentham et al.(2005)]{2005MNRAS.357..783T} Trentham, N., Sampson, L., \& Banerji, M.\ 2005, \mnras, 357, 783 
\bibitem[Trentham et al.(2006)]{2006MNRAS.369.1375T} Trentham, N., Tully, R.~B., \& Mahdavi, A.\ 2006, \mnras, 369, 1375 
\bibitem[Turner et al.(2012)]{2012ApJS..203....5T} Turner, M.~L., C{\^o}t{\'e}, P., Ferrarese, L., et al.\ 2012, \apjs, 203, 5 
\bibitem[Tully et al.(2002)]{2002ApJ...569..573T} Tully, R.~B., Somerville, R.~S., Trentham, N., \& Verheijen, M.~A.~W.\ 2002, \apj, 569, 573
\bibitem[()]{} Tully, R.~B., Rizzi, L., Dolphin, A.~E., et al. 2006, AJ 132, 729              
\bibitem[()]{} Udalski, A., Szymanski, M., Kubiak, M., et al. 1999, Acta Astronomica 49, 201             
\bibitem[van Dokkum et al.(2015)]{2015ApJ...798L..45V} van Dokkum, P.~G., Abraham, R., Merritt, A., et al.\ 2015, \apjl, 798, L45 
\bibitem[()]{} Vansevi{\v c}ius, V., Arimoto, N., Hasegawa, T., et al. 2004, ApJ 611, L93              
\bibitem[Villegas et al.(2010)]{2010ApJ...717..603V} Villegas, D., Jord{\'a}n, A., Peng, E.~W., et al.\ 2010, \apj, 717, 603 
\bibitem[()]{} Walker, M.~G., Mateo, M., \& Olszewski, E.~W. 2009, AJ 137, 3100               
\bibitem[White \& Rees(1978)]{1978MNRAS.183..341W} White, S.~D.~M., \& Rees, M.~J.\ 1978, \mnras, 183, 341 
\bibitem[Yagi et al.(2002)]{2002AJ....123...87Y} Yagi, M., Kashikawa, N., Sekiguchi, M., et al.\ 2002, \aj, 123, 87 
\bibitem[Yamanoi et al.(2012)]{2012AJ....144...40Y} Yamanoi, H., Komiyama, Y., Yagi, M., et al.\ 2012, \aj, 144, 40 
\bibitem[York et al.(2000)]{2000AJ....120.1579Y} York, D.~G., Adelman, J., Anderson, J.~E., Jr., et al.\ 2000, \aj, 120, 1579 
\bibitem[()]{} van de Rydt, F., Demers, S., \& Kunkel, W.~E. 1991, AJ 102, 130             
\bibitem[Zhang et al.(2015)]{2015ApJ...802...30Z} Zhang, H.-X., Peng, E.~W., C{\^o}t{\'e}, P., et al.\ 2015, \apj, 802, 30
\bibitem[Zibetti(2008)]{2008IAUS..244..176Z} Zibetti, S.\ 2008, IAU Symposium, 244, 176 
\end{thebibliography}
\end{document}